\documentclass[reprint, aps, prd, twocolumn, nofootinbib,superscriptaddress]{revtex4-2}

\pdfoutput=1

\usepackage[T1]{fontenc} 
\usepackage{amsmath}
\usepackage{amsfonts}
\usepackage{amsthm}
\usepackage{amssymb}
\usepackage{latexsym, array,multirow,  verbatim, enumerate}
\usepackage{slashed}
\usepackage{booktabs}
\usepackage{tabularx}
\usepackage{cancel}
\usepackage{dcolumn}
\usepackage{bm}
\usepackage{graphicx} 
\usepackage{pifont}
\usepackage{empheq}
\usepackage[utf8]{inputenc}
\usepackage{bm}
\usepackage{blkarray}
\usepackage{float}
\usepackage{multirow}
\usepackage{physics}
\graphicspath{{./Figures/}}
\usepackage{hyperref}
\usepackage{natbib}

\newcommand{\solarmass}{\textup{M}_\odot}
\newcommand{\solarrad}{\textup{R}_\odot}

\begin{document}


\title{\boldmath Q-balls in the sky }
\author{Arhum Ansari}
\email{ansari.arhum@students.iiserpune.ac.in}
\author{Lalit Singh Bhandari}
\email{bhandari.lalitsingh@students.iiserpune.ac.in}
\author{Arun M. Thalapillil}
\email{thalapillil@iiserpune.ac.in}

\affiliation{ Department of Physics,\\ Indian Institute of Science Education and Research Pune,\\ Pune 411008, India}

\date{\today}


\begin{abstract}
There may exist extended configurations in the dark matter sector that are analogues of structures in the visible sector. In this work, we explore non-topological solitonic configurations, specifically Q-balls, and study when they may form macroscopic astrophysical structures and what their distinct characteristics might be. We study in some detail theoretical bounds on their sizes and constraints on the underlying parameters, based on criteria for an astrophysical Q-ball's existence, gravitational stability and viability of solutions. Following this path, one is able to obtain novel limits on astrophysical Q-ball sizes and their underlying parameters. We also explore the gravitational lensing features of different astrophysical Q-ball profiles, which are more general than the simple thin-wall limit. It is seen that the magnification characteristics may be very distinct, depending on the actual details of the solution, even for astrophysical Q-balls having the same size and mass. Assuming that such astrophysical Q-balls may form a small component of the dark matter in the universe, we place limits on this fraction from the gravitational microlensing surveys EROS-2, OGLE-IV, HSC-Subaru and the proposed future survey WFIRST. Exploring various astrophysical Q-ball profiles and sizes, it is found that while for most intermediate masses that we consider, the dark matter fraction comprising astrophysical Q-balls is at most sub-percent, for other masses it may be significantly higher.
\end{abstract}

\maketitle

\section{Introduction}\label{sec:intro}

The true nature of dark matter is among the foremost questions in physics today (See\,\cite{Bertone:2004pz,Feng:2010gw,Bergstrom:2009ib,Steffen:2008qp},\cite{ParticleDataGroup:2022pth}(Ch.27) and references therein, for instance). The present evidence suggests that this sector interacts only feebly with visible matter. Given the presence of such an extraordinary dark sector, one is led to speculate if this domain furnishes, apart from some plethora of dark elementary particles, bound states similar to those in the visible sector. This may, for instance, include field-theoretic, extended, solitonic objects as well as gravitationally bound astrophysical objects that are similar to stars. Analogues in the visible sector which are electromagnetically inert, for instance, massive astrophysical compact halo objects (MACHOs) and dark clouds (see, for instance,\,\cite{Henriksen:1994ep,Gerhard:1995ff} and related references), have long been investigated\,\cite{Bertone:2016nfn} with the speculation that they may form at least some fraction of the missing matter in the universe.

The constituents of the dark sector could also possibly form a rich set of structures. Potential dark matter structures that have been speculated range from galactic or sub-galactic scales all the way to almost particle-like bound states (See, for example,\,\cite{Erickcek:2011us,Fan:2013tia,Fan:2013yva,Barenboim:2013gya,Fan2014,Wise:2014jva,Graham:2015rva, Fairbairn:2017dmf,Gresham:2017cvl, Dror:2017gjq,Grabowska:2018lnd,Curtin:2019ngc,Fox:2023aat} ). Particularly intriguing in recent years have been bosonic field configurations, either forming expansive diffuse structures or exotic astrophysical compact objects, that mimic ordinary neutron stars or white dwarfs (for instance, see\,\cite{Ruffini:1969qy, Kling:2017mif, Kling:2017hjm, Cardoso:2019rvt, Cardoso:2021ehg, Visinelli:2021uve,Boskovic:2021nfs,DelGrosso:2023trq} and related references).

There have been seminal works on various astrophysical aspects of extended dark matter structures and star-like objects---for instance, looking at their gravitational lensing characteristics\,\cite{Croon:2020wpr,Bai:2020jfm,Croon:2020ouk,Fujikura:2021omw}, signatures from baryonic accretion\,\cite{Curtin:2019ngc,Bai:2020jfm},  gravitational wave signatures\,\cite{Giudice:2016zpa,Croon:2022tmr} and so on (For a more comprehensive survey and summary of the existing literature, in these contexts, we kindly refer the interested reader to some of the excellent current reviews and white papers\,\cite{Marsh:2022dqv,Visinelli:2021uve,Boskovic:2021nfs,Drlica-Wagner:2022lbd,Green:2022hhj}). Additionally, there have been various studies of solitonic objects composed of fields that serve as potential candidates for dark matter\,\cite{Frieman:1988ut,Kusenko:1997ad,Kusenko:1997si,Lee:1991ax,Rosen:1968mfz,Rosen:1968zwl,Coleman:1985ki,Arias:2014tka,Vachaspati:2009jx,Long:2014lxa,Long:2014mxa,Hyde:2013fia,Hindmarsh:2014zba,Hartmann:2009ki,Forgacs:2016iva,Forgacs:2019tbn,Brihaye:2009fs,Babeanu:2011ie}.

In this work we focus on non-topological solitons\,\cite{Frieman:1988ut,Kusenko:1997si,Kusenko:1997ad,Lee:1991ax} that could form astrophysical structures. Specifically, we would like to study Q-balls\,\cite{Rosen:1968mfz,Rosen:1968zwl,Coleman:1985ki} in an astrophysical context. These are bound configurations of bosonic fields, owing their existence to a conserved charge ($Q$)\,\cite{Rosen:1968mfz,Rosen:1968zwl,Coleman:1985ki}, hence the `Q' in their nomenclature. Their existence and stability are predicated on unique characteristics in the theory and for large charges, their energy and volume scale linearly with it. It is the latter scaling that makes such objects behave almost like homogeneous lumps of exotic matter-- Q-matter---with `Q' playing the role of a putative particle number. Such objects have been studied in various contexts since their conception---decay of such objects when couplings to other fields are introduced\,\cite{Cohen:1986ct}, their small charge limits\,\cite{Kusenko:1997ad}, as dark matter candidates\,\cite{Kusenko:1997si,Kusenko:2001vu,Troitsky:2015mda}, in the context of supersymmetry and baryogenesis\,\cite{Enqvist:1997si}, cosmological phase transitions\,\cite{Krylov:2013qe}, spinning Q-balls\,\cite{Kleihaus:2005me}, charge-swapping Q-balls\,\cite{Copeland:2014qra, Xie:2021glp,Hou:2022jcd}, gravitational waves during Q-ball formation from condensate fragmentation\,\cite{Zhou:2015yfa} and so forth. Interest in such objects has persisted, and they have also been subjects of many dedicated recent studies, for instance, in the context of generating lepton asymmetry and enhanced gravitational waves during Q-ball decay\,\cite{Kasuya:2022cko}, Q-ball superradiance\,\cite{Saffin:2022tub}, stress stability criteria in gauged Q-balls\,\cite{Loiko:2022noq}, fermion scattering solutions for one-dimensional Q-balls\,\cite{Loginov:2022okj}, topological and non-topological charged Q-monopole-balls\,\cite{Bai:2021mzu}, and many others.

There exist in the literature few analytical solutions and approximations for particular potentials\,\cite{Rosen:1969ay,Theodorakis:2000bz,Gulamov:2013ema,PaccettiCorreia:2001wtt,MacKenzie:2001av,Ioannidou:2003ev,Ioannidou:2005hk,Ioannidou:2004vr}, that accommodate Q-ball solutions, but the analytic tractability of general scenarios has largely been challenging. Recently, there has been some progress\,\cite{Heeck:2020bau,Heeck:2021zvk,Heeck:2022iky} in obtaining better analytic approximations to Q-ball profiles and associated quantities. 

We hope to leverage this recent progress\,\cite{Heeck:2020bau,Heeck:2021zvk,Heeck:2022iky} in the analytic approximations for Q-ball profiles by adapting it to study aspects of Q-balls when they may form astrophysically viable objects. Particularly, in this context, we explore a broader set of solutions than the strict thin-wall limit\,\cite{Coleman:1985ki}. We theoretically study when the flat spacetime Q-ball solutions may also be considered as good approximations in the astrophysical context. We broach this question by considering Jeans' criterion and probing the non-gravitational limit of such macroscopic Q-balls. Similarly, for astrophysically viable Q-balls, from a careful study of the mass density profiles and gravitational lensing features, we point out distinct characteristics of various solutions. Through microlensing survey constraints, we also bound the fraction of total dark matter that may be in the form of astrophysical Q-balls. For this latter part, we will partly modify and adapt few of the methodologies in\,\cite{Croon:2020wpr,Bai:2020jfm}. The actual cosmological origins and formation of astrophysical Q-balls is an intriguing topic of study (see, for instance,\,\cite{Kusenko:1997si, Krylov:2013qe, Dine:2003ax,Enqvist:2003gh,Troitsky:2015mda} and related references). For the aspects we focus on in this study, we will nevertheless be largely agnostic to any specific formation mechanism.

In Sec.\,\ref{sec:qballs}, we briefly review the theoretical framework for Q-balls, fixing conventions and notations along the way. Then, in Sec.\,\ref{sec:microlns}, we summarise aspects of gravitational lensing, specifically microlensing, that we will subsequently utilise to study astrophysical Q-balls. Sec.\,\ref{sec:results} contains the main results of the study. Here, we present our theoretical discussions pertaining to the astrophysical viability of Q-ball solutions and their unique gravitational lensing characteristics, as well as microlensing constraints. We summarise and conclude in Sec.\,\ref{sec:summary}.

\section{Non-topological solitons with a conserved charge}
\label{sec:qballs}

Q-balls\,\cite{Rosen:1968mfz,Rosen:1968zwl,Coleman:1985ki} are non-topological solitons that may occur in theories with a conserved global charge. They arise as stable, localised configurations in generic scalar field theories whose potential satisfies specific conditions. 

For the purposes of our present discussion, consider a theory with a conserved $U(1)$ charge, described by a Lagrangian density given by
\begin{equation}\label{eq:lagcompscalar}
\mathcal{L}=\partial_\mu\Phi \partial^\mu\Phi^\dagger-U(\Phi^\dagger \Phi)\;.
\end{equation}   
 The stable vacuum is assumed to be at $\lvert \Phi \rvert=0$, and without loss of generality, the potential is assumed to be vanishing in the vacuum.
Q-ball solutions will exist for all potentials $U(\Phi^\dagger \Phi)$ when $U(\Phi^\dagger \Phi)/\Phi^\dagger\Phi$ has a minimum, without loss of generality, at a real, positive field value $\Phi_\text{\tiny{Q}}$ satisfying
\begin{equation}\label{eq:qballexist}
0\leqslant\frac{U(\Phi_\text{\tiny Q})}{\Phi^2_\text{\tiny Q}}\equiv \omega^2_\text{\tiny Q} <m^2\;.
\end{equation}
Here, the mass of the fundamental scalar quanta $\Phi$ is defined as $m^2\equiv \frac{d^2U(\Phi^\dagger\Phi)}{d\Phi^\dagger d\Phi}\big\vert_{\Phi=0}$.  The theory defined by Eq.\,(\ref{eq:lagcompscalar}) is invariant under a global $U(1)$ symmetry with the associated conserved charge for Q-balls in the local $\Phi_\text{\tiny{Q}}$ minimum given by
\begin{equation}\label{eq:Qcharge}
Q\equiv i\int d^3x(\Phi\dot{\Phi}^\dagger-\Phi^\dagger\dot{\Phi}) \;.
\end{equation}
The condition for the existence of Q-balls given in Eq.\,(\ref{eq:qballexist}) is equivalent, for large $Q$-charges, to a requirement that the Q-ball configuration is more energetically favourable than a collection of scalar particles with the same total charge.

The Hamiltonian for the theory is
\begin{equation}\label{eq:energy}
H= \int d^3x \left(\vert\dot{\Phi}\vert^2+\abs{\nabla\Phi}^2+U(\abs{\Phi})\right)\;,
\end{equation}
and to study the existence and possibility of objects with a fixed charge $Q$, we may analyse the functional\,\cite{Kusenko:1997ad}
\begin{equation}\label{eq:functional}
\mathcal{F}= H+\omega \left(Q-i\int d^3x(\Phi\dot{\Phi}^\dagger-\Phi^\dagger\dot{\Phi})\right)\;.
\end{equation} 
Here, $\omega$ is a Lagrange multiplier. Minimising the functional in Eq.\,(\ref{eq:functional}), it may be shown that the minimum energy configuration, subject to a fixed Q-charge, will have a time dependence that goes like
\begin{equation}\label{eq:phitdepen}
\Phi(r,t)=\phi(r) e^{i\omega t}\;.
\end{equation}
The above time dependence gives a stationary solution that helps evade Derrick's theorem\,\cite{Derrick:1964ww}. With this time dependence, the equation of motion now becomes
\begin{equation}\label{eq:eom}
\phi''+\frac{2}{r}\phi'=-\frac{1}{2}\frac{\partial}{\partial\phi}\left[\omega^2\phi^2-U(\phi)\right]\;,
\end{equation} 
and the total charge, energy and equation of motion may be written as
\begin{equation}\label{eq:Qchargephi}
Q= 8\pi \omega \int dr r^2\phi(r)^2\;,
\end{equation}
\begin{eqnarray}\label{eq:energyphi}
\nonumber E&=&4\pi\int drr^2\left[\phi'^2+\phi^2\omega^2+U(\phi)\right]\nonumber  \nonumber  \;,\\ 
&=& \omega Q+\frac{8\pi}{3}\int drr^2\phi'^2 \; .
\end{eqnarray}
At this juncture, we would like to re-emphasize that, in this study, by Q-ball solutions, we mean the solution one obtains strictly from Eq.\,\eqref{eq:eom}. This distinction is being made to distinguish Q-balls from Q-stars, where the effects of gravity may have to be taken into account by solving a Gross-Pitaevskii-Poisson equation. We will also comment on some of these aspects later in Sec.\,\ref{sec:thbounds}.

From above, we identify the corresponding charge and energy density profiles for a given field profile to be
\begin{equation}
\rho_{\text{\tiny Q}}^{\text{\tiny C}}(r)=2\omega\phi(r)^2\;,
\end{equation}
\begin{equation}\label{eq:fullrhoprofile}
\rho_{\text{\tiny Q}}^{\text{\tiny M}}(r)=2\omega^2\phi(r)^2+\frac{2}{3}\phi(r)'^2\;.
\end{equation}
Also, differentiating Eq.\,(\ref{eq:energyphi}) it is seen that\,\cite{PaccettiCorreia:2001wtt}
\begin{equation}\label{eq:debydq}
\dv{E}{\omega}=\omega\dv{Q}{\omega}\;.
\end{equation}
This is an exact relation that shows that $\omega$ may be interpreted as a chemical potential. When $dE/dQ>0$, the Q-ball configuration is expected to be unstable or metastable, with it being energetically favourable for the Q-ball to shed fundamental $\Phi$ quanta.

To make the study more concrete, let us specifically look at the simplest potential that allows for Q-ball solutions. Stable quartic potentials do not yield Q-ball configurations. Confining to simple field representations, the simplest potential that contains Q-ball-like solutions is the non-renormalizable sextic potential
\begin{equation}\label{eq:qballsexticpot}
U(\phi)=m^2\phi^2+\lambda\phi^4+\zeta\phi^6\;.
\end{equation}
Apart from being the simplest potential where Q-ball configurations may be studied, thereby acting as a good proxy for investigating salient features, it is also a potential where so far no exact solutions have been found. Therefore, unlike few other special potentials where exact Q-ball solutions have been found\,\cite{Rosen:1969ay,Theodorakis:2000bz,Gulamov:2013ema,PaccettiCorreia:2001wtt,MacKenzie:2001av,Ioannidou:2003ev,Ioannidou:2005hk,Ioannidou:2004vr}, in this case, improved analytic solutions and studies may be even more pertinent\,\cite{Heeck:2020bau}. Moreover, from an effective field theory perspective, the sextic potential may arise as a low-energy approximation to a UV-complete framework in the dark sector\,\cite{Heeck:2022iky}. All of these reasons make it relevant to study the sextic potential as a stereotypical potential in the dark sector that may lead to astrophysical Q-balls.

The second term with $\lambda<0$ provides the attractive interaction, whereas the third term is repulsive in nature and can balance the attractive interaction, which hints towards forming stable bound states like Q-balls. When $\lambda<0$, Q-ball solutions may be found, with the model having its global minima at $\phi=0$ and a local minimum for $U(\Phi^\dagger \Phi)/\Phi^\dagger\Phi$ at 
\begin{equation}\label{eq:Qminima}
\phi=\phi_\text{\tiny Q}\equiv \sqrt{\frac{\lvert \lambda \rvert}{2\zeta}}\; .
\end{equation}
In the latter minima, the ground state quanta may be Q-ball configurations, subject to the fulfilment of the required conditions. From Eq.\,(\ref{eq:qballexist}) we  have for this potential\,\cite{Heeck:2020bau}
\begin{equation}\label{eq:omegaqexp}
\omega_\text{\tiny Q}=m\sqrt{1-\frac{\lambda^2}{4m^2\zeta }} \; .
\end{equation}
It is clear from above that $\omega_\text{\tiny Q}<m$ is satisfied, and the existence of Q-balls is possible for
\begin{equation}\label{eq:lambda_zeta_rel}
0<\lambda^2\leqslant 4m^2\zeta \;.
\end{equation}
Note that these restrictions on the Lagrangian parameters, among other features, make diffuse Q-ball structures potentially more distinct than generic bosonic field configurations like bose stars.

Analysing the form of the effective potential $\left(\omega^2\phi^2-U(\phi)\right)$ appearing in the equation of motion Eq.\,(\ref{eq:eom}), as a function of $\omega$, one finds that suitable Q-ball solutions only occur for $\omega>\omega_\text{\tiny Q}$ and $\omega<m$. This, in combination with the necessary condition Eq.\,(\ref{eq:qballexist}), then implies that we must have
\begin{equation}\label{eq:exist_cond}
\omega_\text{\tiny Q}<\omega<m\;,
\end{equation}
for viable Q-ball solutions.

In the lower limit of the viable $\omega$ region, $\omega\rightarrow \omega_{\text{\tiny Q}}$, one recovers the original Q-ball solution of Coleman\,\cite{Coleman:1985ki}. In the limit $\omega\rightarrow \omega_{\text{\tiny Q}}$, the actual field profile obtained by solving Eq.\,(\ref{eq:eom}) is very well approximated by\,\cite{Coleman:1985ki} 
\begin{equation}\label{eq:kzero}
\phi^{\text{\tiny TW}}(r)=
\begin{cases}
\phi_\text{\tiny Q}& r\leqslant R_\text{\tiny Q}\;,\\
0& r>R_\text{\tiny Q}\;,
\end{cases}
\end{equation}
where $R_\text{\tiny Q}$ is defined implicitly in terms of the actual field profile through
\begin{equation}\label{eq:RdefTW}
R_\text{\tiny Q}=\left(\frac{3\omega}{4 \pi \omega_{\text{\tiny Q}}  \phi_{\text{\tiny Q}}^2 }\int_{0}^{\infty}d^3r\phi(r)^2\right)^{\frac{1}{3}} \;.
\end{equation}
We will refer to these solutions as thin-wall (TW) Q-balls.

Recently, better theoretical approximations for Q-ball solutions have been found, enabling one to explore more of the $\omega$-parameter space regions, encompassing $\omega_\text{\tiny Q}<\omega<m$. In particular, this allows studies beyond the simplest step-function profiles of the exact thin-wall limit Q-balls. Following methodologies developed in\,\cite{Heeck:2020bau}, better analytic profiles may be obtained for the interior, exterior and transition regions of the Q-ball. Remarkably, for large Q-balls, the full analytic profile may be approximated by a single function\,\cite{Heeck:2020bau} of the form
\begin{equation}\label{global_transition_profile}
\phi(r)=\frac{\phi_*}{\sqrt{1+2\exp\left[2\sqrt{m^2-\omega_\text{\tiny Q}^2}\,(r-R_\text{\tiny Q})\right]~}}\;,
\end{equation}
where
 \begin{equation}\label{eq:phiplus}
\phi_*^2=\frac{\phi_\text{\tiny Q}^2}{3}\left[2+\sqrt{1+3\left(\frac{\omega^2-\omega_\text{\tiny Q}^2}{m^2-\omega_\text{\tiny Q}^2}\right)}~\right]\;.
\end{equation}

The radius $R_\text{\tiny Q}$ in the expression above is defined implicitly through
\begin{equation}\label{eq:qball_radii}
\frac{d^2\phi(r)}{dr^2} \Big\vert_{r=R_\text{\tiny Q}}=0 \; ,
\end{equation}
using the full field profile approximation. When $\omega$ is very near the $\omega_{\text{\tiny Q}}$ lower limit, the above definition starts coinciding with the thin-wall definition Eq.\,(\ref{eq:RdefTW}) and the field profile also starts coinciding with that given in Eq.\,\eqref{eq:kzero}. To leading order, Eq.\,\eqref{eq:qball_radii} gives large Q-ball radii to be\,\cite{Heeck:2020bau}
\begin{equation}\label{eq:Qball_radius}
R_\text{\tiny Q}=\frac{\sqrt{m^2-\omega_{\text{\tiny Q}}^2}}{~\omega^2-\omega_{\text{\tiny Q}}^2}\;.
\end{equation}
To distinguish the improved, more general field profile, given in Eq.\,\eqref{global_transition_profile}, from a limit of it given by Eq.\,\eqref{eq:kzero} (i.e. the TW solution limit), we will refer to the former as beyond-thin-wall (BTW) Q-balls. 

For large Q-balls, the approximate BTW Q-ball profile given by Eq.\,\eqref{global_transition_profile} is in agreement with the relation given by Eq.\,\eqref{eq:debydq}\,\cite{Heeck:2020bau}. $dE/dQ<0$ and $E<m\, Q$ ensure the field-theoretic stability of Q-balls, against decay into free scalar quanta. These will restrict the viable Lagrangian parameters in turn. We will, therefore, also impose these field-theoretic stability constraints along with the gravitational stability criteria later in Sec.\,\ref{sec:thbounds}, while investigating astrophysically viable Q-balls.

If we expand the domain of symmetry, one may make the $U(1)$ invariance of the Lagrangian, given in Eq.\eqref{eq:lagcompscalar}, local. This gives rise to gauged Q-balls\,\cite{Lee:1988ag}, which imposes some restrictions on the charge and radius of the Q-balls. However, it has been shown that gauged Q-ball profile can still be described by the expression given in Eq.\eqref{global_transition_profile} with the following identification of the Lagrange multiplier\,\cite{Heeck:2021zvk}
\begin{equation}\label{eq:mapgauge}
	\omega_{\text{\tiny G}}=\omega g \phi_{\text{\tiny Q}} R_\text{\tiny Q}\coth(g\phi_{\text{\tiny Q}} R_\text{\tiny Q})\;.
\end{equation}
In the above expression, $\omega_{\text{\tiny G}}$ and $\omega$ are the Lagrange multipliers for gauged and global Q-balls, respectively, and $g$ is the coupling constant between the scalar field and the gauge field. Owing to this mapping, some of the analyses for global Q-balls may be translated for gauged Q-balls. We leave a detailed analyses for future work.

One may also look at more generalised potentials given that the particular polynomial potential given in Eq.\eqref{eq:qballsexticpot}, the simplest one which accommodates Q-balls, may be considered as an effective potential coming from some fundamental renormalisable theory. Most renormalisable theories in the low energy limit give rise to polynomial potential with arbitrary exponents whose first few terms can be of the form,
\begin{equation}\label{eq:polypot}
	U(\phi)=m^2\phi^2+\lambda\phi^p+\zeta\phi^q\;,
\end{equation}
with arbitrary powers $p$ and $q$. General polynomial potentials of the above form were studied in detail recently\,\cite{Heeck:2022iky}. It is seen that global properties like total charge and total energy have only a mild dependence on the polynomial powers $p$ and $q$\,\cite{Heeck:2022iky}. From such indications, it may be speculated that the sextic potential in \eqref{eq:qballsexticpot} may capture some of the more salient features of generic Q-ball configurations while being analytically more tractable.

In the next section, we briefly review the theoretical framework for gravitational lensing while defining and clarifying relevant ideas and notations that will be useful to us in our study.

\section{Gravitational microlensing by extended structures}\label{sec:microlns}

If there is an extended mass distribution in the path of light emanating from astrophysical or cosmological sources, they may be deflected or distorted. This broad effect is termed gravitational lensing\,\cite{1936Sci....84..506E}. 
\begin{figure}[h!]
	\centering
\includegraphics[scale=0.4]{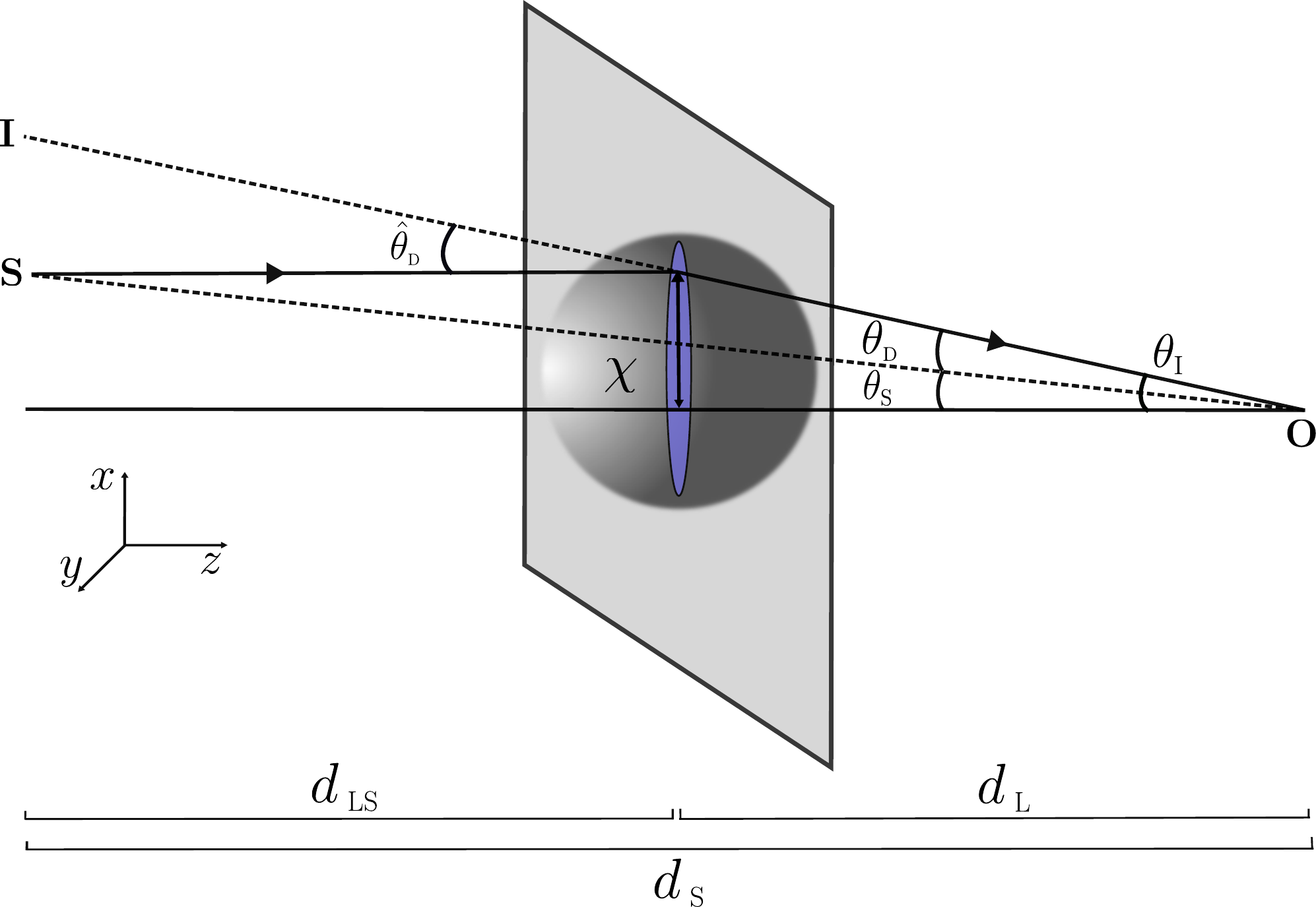}
\caption{A diagrammatic representation of a lens, observer and source lensing system showing the various relevant quantities involved. The directed solid line illustrates the path of a null ray. The lens plane is shown at the centre, along with a spherical lens.}
\label{fig:gr_lens_fig}
\end{figure}
 
Consider Fig.\ref{fig:gr_lens_fig} illustrating a typical gravitational lensing system. In the diagram, the light source (S), observer (O) and image position (I) are shown along with the respective source angular position ($\theta_S$), image angular positions ($\theta_I$) and deflection angle ($\hat{\theta}_D$). $d_{\text{\tiny S}}$ denotes the angular diameter distance of the source from the observer, $d_{\text{\tiny L}}$ the angular diameter distance of the lensing mass from the observer and $d_{\text{\tiny LS}}$ is the angular diameter distance from the source to the lensing mass. $\chi$ denotes the transverse distance to the null ray, and z-distances are measured perpendicular to the lens plane. In almost all cases of interest, and in particular, in the case of interest to us, the deflection angles involved are small, and the lens may be approximated to be a planar mass distribution by projecting the mass of the lens on a sheet orthogonal to the line of sight.

For point-like objects, meaning their intrinsic size is negligible compared to the other characteristic length scales in the system, the deflection angle of light rays sufficiently far from an object is directly related to the transverse gradient of the configuration's Newtonian gravitational potential ($\varphi^{\text{\tiny{G}}}$). More precisely, for a total configuration mass $M$, the point lens approximation is valid when the size of the lens is much smaller than the corresponding Einstein radius of the system defined by\,\cite{1936Sci....84..506E}
\begin{equation}\label{eq:ein_rad}
R_{\text{\tiny E}}=\left(\frac{4 G_{\text{\tiny N}} M}{c^2}\frac{d_{\text{\tiny L}} d_{\text{\tiny LS}}}{d_{\text{\tiny S}}}\right)^{\frac{1}{2}} \; .
\end{equation}
Here $G_{\text{\tiny N}}$ is the Newton's gravitational constant. This is, for instance, relevant in many gravitational lensing studies involving compact astrophysical objects like black holes or neutron stars. More pertinently, for our purposes, this is also applicable to small Q-balls. For a total lens mass $M$, the deflection angle would for instance be given by\,\cite{Narayan:1996ba}
\begin{equation}
\hat{\theta}_{\text{\tiny D}}=\frac{2}{c^2}\int \nabla_{\chi}\varphi^{\text{\tiny{G}}} ~dz=\frac{4G_{\text{\tiny N}} M}{c^2 |\vec{\chi}|}\hat{\chi} \; .
\end{equation}

For extended mass distributions, like the Q-ball configurations we are interested in, one essentially adds up the contributions due to the individual mass elements. If $\chi$ denotes the transverse distance (with respect to some origin) of a light ray passing through an extended mass distribution and $\chi'$ denotes the position of an individual mass element of the distribution, then the deflection angle $\hat{\theta}_{\text{\tiny D}}(\chi)$ is given by
\begin{eqnarray}\label{eq:deflec_angle_extended}
	\hat{\theta}_{\text{\tiny D}}(\chi)&=&\frac{4G_{\text{\tiny N}}}{c^2}\int\frac{(\vec{\chi}-\vec{\chi}')dM'}{\abs{\vec{\chi}-\vec{\chi}'}^2}\;,\nonumber\\
	&=&\frac{4G_{\text{\tiny N}}}{c^2}\int d^2\chi'\frac{(\vec{\chi}-\vec{\chi}')\sigma(\chi^\prime)}{\abs{\vec{\chi}-\vec{\chi}'}^2} \;.
\end{eqnarray}
Here, for a density distribution $\rho(\chi,z)$, we have defined the planar density distribution $\sigma(\chi)=\int_{-\infty}^\infty \rho(\chi,z)dz$. A special case arises for spherically symmetric lenses. In that case, we can shift the origin to the centre, and Eq.\eqref{eq:deflec_angle_extended} becomes\,\cite{Narayan:1996ba}
\begin{eqnarray}\label{eq:deflection_spherically_symm}
\hat{\theta}_{\text{\tiny D}}(\chi)&=&\frac{4G_{\text{\tiny N}}}{c^2 |\vec{\chi}|}\int_0^\chi d^2\chi'\sigma(\chi')\;,\nonumber\\
&=&\frac{4G_{\text{\tiny N}}\widetilde{M}(\chi)}{c^2 \chi}\;,
\end{eqnarray}
where 
\begin{equation}\label{eq:mproj}
\widetilde{M}(\chi)=\int_0^\chi d^2\chi'\sigma(\chi')=2\pi\int_0^{\chi} d\chi' \chi'\sigma(\chi')\;.
\end{equation}

The angular source position and angular image positions are related via the lens equation
\begin{equation}\label{eq:temp_lens_eq}
\theta_{\text{\tiny S}}=\theta_{\text{\tiny I}}-\theta_{\text{\tiny D}}(\theta_{\text{\tiny I}})\;,
\end{equation}
where 
\begin{equation}\label{eq:alpha_alphap_rel}
\theta_\text{\tiny D}=\frac{d_{\text{\tiny LS}}}{d_{\text{\tiny S}}}\hat{\theta}_{\text{\tiny D}}\;.
\end{equation}
Using Eq.\,(\ref{eq:deflection_spherically_symm}) with the identification $\chi=d_{\text{\tiny L}}\theta_{\text{\tiny I}}$, the lens equation may be written as
\begin{equation}\label{eq:lens_eq}
\theta_{\text{\tiny S}}=\theta_{\text{\tiny I}}-\frac{d_{\text{\tiny LS}}}{d_{\text{\tiny S}}d_{\text{\tiny L}}}\frac{4G_{\text{\tiny N}}\widetilde{M}(\theta_{\text{\tiny I}})}{c^2\theta_{\text{\tiny I}}}\;.
\end{equation}
For a general extended mass distribution, given a source position $\theta_{\text{\tiny S}}$, we solve the above lens equation to find the image positions $\theta_{\text{\tiny I}}$. 

When $\theta_{\text{\tiny S}}=0$---meaning the source, lens and observer are aligned along the line of sight---the angular image position, for a point lens of total mass $M$, is defined to be at the point-like Einstein angle ($\theta_{\text{\tiny E}}$). This is related to Eq.\,(\ref{eq:ein_rad}) through 
\begin{equation}\label{eq:einstein_angle}
\theta_{\text{\tiny E}}=\frac{R_{\text{\tiny E}}}{d_{\text{\tiny L}} }\;.
\end{equation}
We will use $R_{\text{\tiny E}}$ and $\theta_{\text{\tiny E}}$, when required, to normalise the Q-ball radius and relevant lensing angles.

By Liouville's theorem, the surface brightness of the source is conserved under gravitational lensing. Under gravitational microlensing, as a Q-ball configuration gradually moves across a light source, say a background galaxy, there will be a waxing and waning of its measured brightness due to the magnification. In this context, for the aforementioned reasons, the magnification of the source would just be given by the ratio of the image area to the source area. For each $\theta_{\text{\tiny I}}$ solution of Eq.\,(\ref{eq:lens_eq}), this magnification ($\mathfrak{m}_{\text{\tiny I}}$) may be expressed as
\begin{equation}\label{eq:mag_exp}
\mathfrak{m}_{\text{\tiny I}}=\frac{\theta_{\text{\tiny I}}}{\theta_{\text{\tiny S}}}\dv{\theta_{\text{\tiny I}}}{\theta_{\text{\tiny S}}} \; .
\end{equation}
The total magnification ($\mathfrak{m}$) obtained in the detector is then just the sum of magnification produced by all the images; obtained as solutions to Eq.\,(\ref{eq:lens_eq})
\begin{equation}\label{eq:total_mag}
\mathfrak{m}=\sum_\text{\tiny I} \abs{\mathfrak{m}_{\text{\tiny I}}}\; .
\end{equation}

Let us now proceed to discuss the main results of the study in the context of Q-ball configurations that may exist in the universe.
    
\section{Astrophysical Q-ball structures}\label{sec:results}
In this section, we discuss some of the theoretical considerations for Q-balls, when they may form extended astrophysical structures, followed by discussions of their gravitational lensing characteristics and constraints on astrophysical Q-balls from microlensing surveys.

\subsection{Theoretical bounds from existence, stability and viability}\label{sec:thbounds}

As we mentioned earlier, by Q-balls, we strictly mean here the pure field theoretical non-topological soliton solutions obtained from Eq.\eqref{eq:eom}, ignoring any role of gravity. Obviously, the latter regime puts constraints on how dense these astrophysical Q-ball structures could be and are constrained by relevant Jeans' criterion\,\cite{Jeans:1902fpv}. We will be considering diffuse Q-balls in the low compactness regime.

When a Q-ball of total mass $M_\text{\tiny {Q}}$ is confined within a characteristic radius $R_{\text{\tiny {Q}}}$ and is slightly compressed, there is a propensity for gravity to compress the system, thereby attempting to decrease the gravitational potential energy, while in contrast, the Q-matter inside would try to resist this compression by virtue of outward internal pressure. Assuming a homogeneous medium with almost constant density, which is a good approximation for large Q-ball solutions, we may now formulate the relevant Jeans' criterion. 

The characteristic speed of the outward pressure waves is the sound speed in the homogeneous medium ($v^\text{\tiny s}_\text{\tiny Q}$). The relevant time scales are, therefore, the sound crossing time in Q-matter and the corresponding gravitational free-fall time, which is given by
\begin{equation}\label{ff_time}
t^{\text{\tiny {g}}}_{\text{\tiny {Q}}}\simeq\frac{1}{\sqrt{G_{\text{\tiny N}}\, \rho_{\text{\tiny Q}}^{\text{\tiny M}}}}\;.
\end{equation} 
Here, $G_\text{\tiny N}$ is Newton's gravitational constant and $\rho_{\text{\tiny Q}}^{\text{\tiny M}}$, as before, is the mass density of the Q-ball. Within this characteristic free-fall time $t^{\text{\tiny {g}}}_{\text{\tiny {Q}}}$ the pressure waves would travel a characteristic length
\begin{equation}\label{eq:qdj}
d^{\text{\tiny J}}_\text{\tiny Q} =v^\text{\tiny s}_\text{\tiny Q} \, t^{\text{\tiny {g}}}_{\text{\tiny {Q}}} \simeq \frac{v^\text{\tiny s}_\text{\tiny Q}}{\sqrt{G_{\text{\tiny N}}\, \rho_{\text{\tiny Q}}^{\text{\tiny M}}}}\;.
\end{equation}
This characteristic length scale $d^{\text{\tiny J}}_{\text{\tiny Q}}$ is the Jeans' length for the Q-ball configuration. 

The basic idea is that if Q-matter is extended beyond $d^{\text{\tiny J}}_{\text{\tiny Q}}$ then the pressure waves resisting gravity will not have enough time to travel the length scale and counter the external compressive disturbance due to gravity. In terms of the Lagrangian parameters, assuming to a good approximation almost homogeneous internal densities, we have
\begin{equation}\label{eq:Q_pert_speed}
v^{\text{\tiny s}\,2}_\text{\tiny Q}\equiv \frac{dP_{\text{\tiny Q}}}{d\rho_{\text{\tiny Q}}^{\text{\tiny M}}} =\frac{\lambda^2}{4m^2\zeta}\; .
\end{equation}
We note that as a direct consequence of the existence condition for Q-balls given in Eq.\,\eqref{eq:lambda_zeta_rel}, one automatically satisfies the causality criterion $v^\text{\tiny s}_\text{\tiny Q} \leqslant 1$ in the Q-matter.

From Eq.\eqref{eq:Qball_radius}, we have for the radius of a Q-ball 
\begin{equation}\label{eq:radius_exp}
R_{\text{\tiny {Q}}}=\frac{\sqrt{\lambda^2/4\zeta}}{(\omega^2-\omega_\text{\tiny Q}^2)}\;.
\end{equation}
Since $\omega$ has an upper bound, as proscribed by Eq.\eqref{eq:exist_cond}, from Eqs.\,\eqref{eq:omegaqexp} and \eqref{eq:radius_exp} one obtains a minimum radius for the Q-ball solution
\begin{equation}\label{eq:rmin_qball}
R_{\text{\tiny {Q}},\,\text{\tiny min}}^{\,\omega<m}\equiv \sqrt{\frac{4 \zeta}{\lambda^2}}\;.
\end{equation}
The existence of such a minimum radius was also recently pointed out to be valid even in general polynomial potentials\,\cite{Heeck:2022iky}. This agrees with a recently proposed conjecture\,\cite{Freivogel:2019mtr} that all bound states in a theory must be greater than the Compton wavelength $\lambda_\phi^{\text{\tiny C}}\equiv1/m$. In the case of Q-balls we note particularly that $R_{\text{\tiny {Q}},\,\text{\tiny min}}^{\,\omega<m}/\lambda_\phi^{\text{\tiny C}}=\sqrt{4\zeta m^2/\lambda^2}\geqslant1$ by virtue of the Q-ball existence condition Eq.\,\eqref{eq:lambda_zeta_rel}.

From Eqs.\,\eqref{eq:qdj}, \eqref{eq:Q_pert_speed}, and  \eqref{eq:energy_den_TW}, we may also calculate the corresponding Jeans' length explicitly. This gives
\begin{equation}\label{eq:qball_dj}
d^{\text{\tiny J}}_\text{\tiny Q}= \sqrt{\frac{\lvert \lambda \rvert}{4 m^4 G_{\text{\tiny N}}\left( 1-\frac{\lambda^2}{4 m^2 \zeta}\right)} } \equiv R_{\text{\tiny {Q}},\,\text{\tiny max}}^{\text{\tiny J}}\; .
\end{equation}
For the pure non-topological soliton-like Q-ball solution to be valid, the radius of the corresponding stable Q-ball must be less than this Jeans length. 

The above equation, therefore, furnishes an upper bound on the radius of astrophysically viable Q-balls. Otherwise, the initial Q-ball will be unstable to gravitational collapse and potentially reconfigure into a different field configuration, whose features may now depend on the effects of gravity as well.
The criterion
\begin{equation}\label{eq:rad_jeans_cond}
R_{\text{\tiny {Q}}}~<~d^{\text{\tiny J}}_\text{\tiny Q}\equiv R_{\text{\tiny {Q}},\,\text{\tiny max}}^{\text{\tiny J}}\; ,
\end{equation}
 for stability with respect to gravitational collapse and continued validity of the pure Q-ball solution, then gives a condition on the Lagrange multiplier $\omega$ as
\begin{equation}\label{eq:omega_jeanslimit}
\omega>\omega_{\text{\tiny Q}}\left[1+\left\{\frac{\abs{\lambda}G_{\text{\tiny N}}/\zeta}{1-\frac{\lambda^2}{4m^2\zeta}}\right\}^{1/2}\right]^{1/2}\equiv\omega^{\text{\tiny J}}_{\text{\tiny min}}\;.
\end{equation}
At $\omega=\omega^{\text{\tiny J}}_{\text{\tiny min}}$ the Q-ball will become unstable to collapse.

Therefore, we note that there exists both a minimum and maximum limit on the radius of astrophysical Q-balls. Of course, for physically viable astrophysical Q-ball solutions to actually exist, we must then obviously require
\begin{equation}
R_{\text{\tiny {Q}},\,\text{\tiny min}}^{\,\omega<m} < R_{\text{\tiny {Q}},\,\text{\tiny max}}^{\text{\tiny J}}\;.
\end{equation}
As emphasised by the superscripts, the lower and upper bound respectively come from the restriction on the Lagrangian parameter $\omega$, for the existence of Q-ball solutions, and Jeans' criterion, signifying stability to gravitational collapse and continued validity of the flat-spacetime Q-ball solutions. From Eqs.\,\eqref{eq:lambda_zeta_rel}, \eqref{eq:rmin_qball} and \eqref{eq:qball_dj} this gives the constraint
\begin{equation}\label{eq:jeans_cond}
\left\{\left(1-\frac{\lambda^2}{4m^2\zeta}\right)\frac{G_{\text{\tiny N}} \abs{\lambda}}{\zeta}\right\}^{1/2} < \frac{\lambda^2}{4m^2\zeta}\leqslant1\;.
\end{equation}
Note that here the first part of the inequality comes from Jeans' criteria, and the second part of the inequality results from the existence condition for Q-balls.

A measure of the extent to which gravity influences the global properties of an astrophysical object is quantified by the compactness parameter, which is defined as $\mathcal{C}\equiv{G_{\text{\tiny N}}M}/{R_{\text{\tiny {Q}}}}$. For large Q-balls, we can express this in terms of the Lagrangian parameters and Lagrange multiplier $\omega$ as follows,
\begin{equation}\label{eq:comp_qball_exp}
\mathcal{C}_{\text{\tiny Q}}=\frac{8\pi G_{\text{\tiny N}}}{3}\left(\frac{\abs{\lambda}}{2\zeta}\right)\left(m^2-\frac{\lambda^2}{4\zeta}\right)\frac{\lambda^2/4\zeta}{\left[\omega^2-\left(m^2-\frac{\lambda^2}{4\zeta}\right) \right]^2}\;.
\end{equation}
The maximum compactness of the profiles that we work with is $\mathcal{O}(10^{-5})$. Hence we can safely say that gravity has very little role to play in determining the global properties of such Q-balls.

Let us examine another perspective to think about the non-gravitational limit and the validity of the flat spacetime Q-ball solutions in an astrophysical context. Let us start from the case where any effects due to gravity are negligible and a scalar field potential with non-zero self-interactions (i.e. $\lambda \neq 0$, $\zeta \neq 0$) gives us viable flat spacetime Q-ball configurations; as a solution to Eq.\,\eqref{eq:eom}. Let us now consider how slowly incorporating any effects due to gravity may start changing the solutions.

Towards this, let us change the flat spacetime metric in the scalar field equation of motion to include gravity. In the weak-field limit this may be done through the substitution
\begin{equation}\label{}
\eta_{\mu\nu}\rightarrow g_{\mu\nu}(x)=(1+2\varphi^{\text{\tiny G}},-1,-1,-1) \;.
\end{equation}
$\varphi^{\text{\tiny G}}$, as before, is the Newtonian gravitational potential. Let us express the scalar field $\Phi$ as
\begin{equation}\label{}
\Phi(r,t)=(2E/N)^{-1/2}\psi(r)e^{-iEt}\;.
\end{equation}
Here, $N$ is the total number of particles in the configuration, $E\approx m+2 E_B$ is the energy of the ground state, and $E_B \ll m$ is the binding energy. The equation of motion takes the form
\begin{equation}\label{eq:Qball_GP_eq}
E_B\, \psi=-\dfrac{1}{2m}\nabla^2\psi+m\varphi^{\text{\tiny G}}\psi+\dfrac{N\lambda}{2m^2}\psi^3+\dfrac{3\zeta N^2}{8m^3}\psi^5 \; .
\end{equation}
The second and third term on the right-hand side leads to attraction between bosons; due to gravity and attractive self-interactions ($\lambda<0$). The last term on the right-hand side leads to repulsive interactions among the bosons. 

After inspecting the attractive components in Eq.\,\eqref{eq:Qball_GP_eq}, we discover that the effect of gravitational interaction is weaker than that due to the attractive self-interaction among the fields when
\begin{equation}\label{eq:nongrav_cond_terms}
	\vert m\varphi^{\text{\tiny G}}\rvert <\frac{N\abs{\lambda}}{2m^2}\psi^2\simeq\frac{M_{\text{\tiny {Q}}}\abs{\lambda}}{2\omega m^2}\psi^2\;.
\end{equation}

Here, as suggested by Eq.\,(\ref{eq:energyphi}), $M_{\text{\tiny {Q}}}\sim N\omega$ denotes the approximate mass of the Q-ball. Using $\varphi^{\text{\tiny G}}\sim -\,G_{\text{\tiny N}} M_{\text{\tiny {Q}}}/R_{\text{\tiny {Q}}}$ and $\psi^2\sim 1/R^3_{\text{\tiny {Q}}}$,  we may estimate the strength of gravitational and self-interaction terms. 

From Eqs.\,\eqref{eq:nongrav_cond_terms}, \eqref{eq:radius_exp} and \eqref{eq:omegaqexp}, along with the scalings already identified, one finds that the quartic self-interaction among the fields is the dominant attractive interaction when
\begin{equation}\label{eq:nongrav_coupling}
\frac{G_{\text{\tiny N}}m^3\omega\abs{\lambda}}{2\zeta\left(\omega^2-m^2+\frac{\abs{\lambda}^{2}}{4\zeta}\right)^2}<1\;.
\end{equation}

This implies that for the Lagrangian parameters satisfying the above condition, gravity does not play a dominant role when compared to the attractive self-interaction.

An analogous but slightly different reasoning has been used to converge on the non-gravitational limit for generic Boson stars\,\cite{Kling:2017hjm}, starting now from an initial configuration which is non-self-interacting ($\lambda = 0$), and where the scaling, in contrast, is now $R\sim 1/(G_{\text{\tiny N}} M m^2)$.

The above discussions mean that with these constraints the field-theoretic Q-ball solutions of Sec.\,\ref{sec:qballs} are still good approximations for the astrophysical structures under consideration. In our study, we ensure that the Jeans' criterion and subservient gravity conditions, given by Eqs.\,\eqref{eq:rad_jeans_cond} and \eqref{eq:nongrav_coupling} respectively, are well satisfied. This legitimises the consideration of these astrophysical configurations as actual Q-balls and, furthermore, the use of the theoretical expressions and profiles we work with.

For brevity, we will henceforth refer to Q-balls satisfying the criteria of Secs.\,\ref{sec:qballs} and \ref{sec:thbounds} as \textit{astrophysical Q-balls}. These bosonic field configurations may be expected to be very similar in characteristics to ideal flat-spacetime Q-balls while satisfying all the requisite existence, stability and non-gravitational limit qualifications. For our studies, we will require these conditions to ensure the theoretical validity and physical viability of the astrophysical Q-ball structures.

Let us now discuss the various gravitational lensing characteristics of astrophysical Q-ball structures and how they may change for the various types of profiles possible for such objects.

\subsection{Gravitational lensing of point sources by thin-wall and beyond-thin-wall Q-balls}\label{subsec:twbtwlensing}

When an astrophysical Q-ball passes across the field of view of the observer and progressively occludes the source, the gravitational lensing by the Q-ball will cause a waxing and waning of the source brightness. If we have a moving Q-ball lens, then $\theta_{\text{\tiny S}}$ may vary with time. Assuming approximately rectilinear motion of the source for the duration of the lensing transit, the temporal dependence will be given by
\begin{equation}
\theta_{\text{\tiny S}}=\frac{\sqrt{d_{\text{\tiny L}}^2\theta_{\text{\tiny S},\text{min}}^2+v^2 t^2}}{d_{\text{\tiny L}}}\;.
\end{equation}
where $\theta_{\text{\tiny S},\text{min}}$ is the minimum source angle attained during transit. The above relation is just a consequence of the Pythagorean theorem satisfied by the angular diameter distances in the lens plane. As the Q-ball travels across the foreground of the source, gravitational lensing will result in the formation of images, whose positions will be given as a function of time by
\begin{equation}
\frac{\sqrt{d_{\text{\tiny L}}^2\theta_{\text{\tiny S},\text{min}}^2+v^2 t^2}}{d_{\text{\tiny L}}}=\theta_{\text{\tiny I}}-\frac{d_{\text{\tiny LS}}}{d_{\text{\tiny S}}d_{\text{\tiny L}}}\frac{4G_{\text{\tiny N}}\widetilde{M}_\text{\tiny {Q}}(\theta_{\text{\tiny I}})}{c^2\theta_{\text{\tiny I}}}\;,
\end{equation}
following Eq.\,\eqref{eq:lens_eq}. In the microlensing context, the images so formed at each instance of time will contribute to the overall magnification of the source. These, when put together for the transit duration, leads to the full microlensing light curve. 

This gravitational microlensing will depend on the density profile of the spherical Q-ball. Specifically, the maximum magnification and the brightness profile crucially depend on the number and contribution from the images formed and hence on the density profile through $M(\chi)$. Let us explore these aspects now for thin-walled and beyond-thin-walled Q-ball solutions.

From the improved large Q-ball interior, exterior and intermediate region field profiles\,\cite{Heeck:2020bau}, which are in a unified form well approximated by Eq.\,\eqref{global_transition_profile}, the charge and mass density profiles may be computed. We find that they have the respective forms

\begin{widetext}
\begin{equation}\label{eq:Cden_btw}
\rho_{\text{\tiny Q}}^{\text{\tiny C}}(r)=\frac{2\omega\phi_*^2}{1+2\exp\left[2\sqrt{m^2-\omega_\text{\tiny Q}^2}(r-R_{\text{\tiny {Q}}})\right]}\;,
\end{equation}
\begin{equation}\label{eq:Mden_btw}
\rho_{\text{\tiny Q}}^{\text{\tiny M}}(r)=2\phi_*^2\left[\frac{\omega^2}{1+2\exp\left[2\sqrt{m^2-\omega_\text{\tiny Q}^2}(r-R_{\text{\tiny {Q}}})\right]}+\frac{4(m^2-\omega_\text{\tiny Q}^2)}{3}\left(\frac{\exp\left[4\sqrt{m^2-\omega_\text{\tiny Q}^2}(r- R_{\text{\tiny {Q}}})\right]}{\left(1+2\exp\left[2\sqrt{m^2-\omega_\text{\tiny Q}^2}(r- R_{\text{\tiny {Q}}})\right]\right)^{3}}\right)\right].
 \end{equation}
\end{widetext}
 \begin{figure*}[t]
 	\includegraphics[scale=0.9]{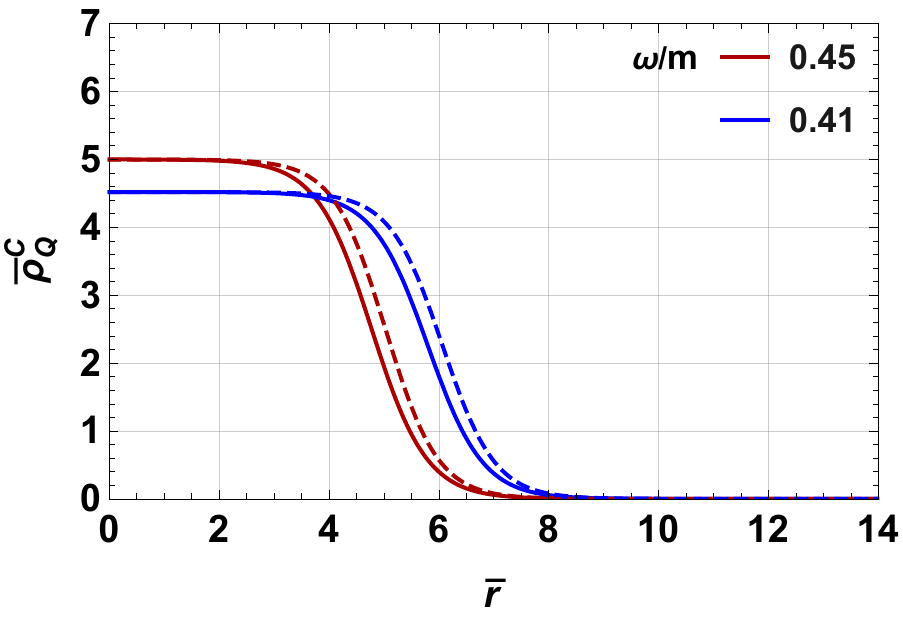}
 	\includegraphics[scale=0.9]{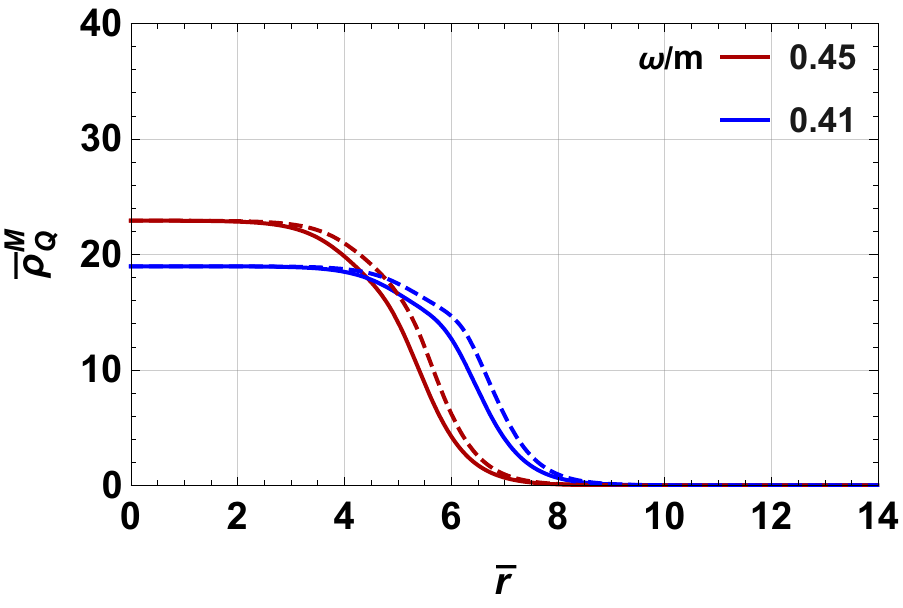}
 	\caption{Plot showing the charge density and energy density profiles calculated using transition profile function (solid) and exact numerical profile (dashed). Here, $\bar{\rho}_{\text{\tiny Q}}^{\text{\tiny C}}\equiv \rho_{\text{\tiny Q}}^{\text{\tiny C}}/2\omega_{\text{\tiny Q}}\phi_{\text{\tiny Q}}^2$, $\bar{\rho}_{\text{\tiny Q}}^{\text{\tiny M}}\equiv \rho_{\text{\tiny Q}}^{\text{\tiny M}}/2\omega_{\text{\tiny Q}}^2\phi_{\text{\tiny Q}}^2$ and $\bar{r}\equiv r\sqrt{m^2-\omega_\text{\tiny Q}^2}$. One sees that the agreement between the exact numerical solution and the transition profile approximation are relatively good across most regions, as well as for the two distinct profiles shown. The two profiles shown have been labelled by their $\omega/m$ ratios. For both cases we have $m/\omega_{\text{\tiny Q}}\sim 10^{-5}$.}
 	\label{fig:coleman_compar}
 \end{figure*}
Here, $\phi_*$ is as defined earlier in Eq.\,\eqref{eq:phiplus}. Note that the quantity $R_{\text{\tiny {Q}}}$ in the above expressions is defined implicitly through Eq.\,\eqref{eq:qball_radii}. For our lensing studies, the mass density $\rho_{\text{\tiny Q}}^{\text{\tiny M}}(r)$ will be of primary interest.

The transition field profile in Eq.\,\eqref{global_transition_profile}, the density profiles of Eqs.\,\eqref{eq:Cden_btw} and \eqref{eq:Mden_btw}, along with the Q-ball radius defined through Eq.\,\eqref{eq:Qball_radius}, all approximate the respective exact numerical profiles very well for large Q-balls\,\cite{Heeck:2020bau}. In Fig.\,\ref{fig:coleman_compar}, we show a comparison of the charge and energy densities calculated using Eqs.\,\eqref{eq:Cden_btw} and \eqref{eq:Mden_btw} with the corresponding numerically computed exact density profiles. One notes that the approximation is relatively good across the radial range and even for distinct profiles differing in their radius. 

The mass profile for a large radius Q-ball may be calculated based on Eq.\,\eqref{eq:Mden_btw}, and is given by
\begin{widetext}
\begin{eqnarray}\label{eq:massprofile}
M_{\text{\tiny Q}}(r)&=&4\pi\int_{0}^{r}drr^2 \rho_{\text{\tiny Q}}^{\text{\tiny M}}(r) \nonumber\\
&=&8\pi\phi_+^2\left[\frac{\omega^2}{4 \vartheta ^3}\left\{2 \vartheta  r \text{Li}_2\left(-\frac{1}{2} e^{2 (R_{\text{\tiny {Q}}}-r) \vartheta }\right)+\text{Li}_3\left(-\frac{1}{2} e^{2 (R_{\text{\tiny {Q}}}-r) \vartheta }\right)\right.\right.\nonumber\\
&&\left.-\text{Li}_3\left(-\frac{1}{2} e^{2 R_{\text{\tiny {Q}}} \vartheta }\right)+\vartheta ^2 r^2 \left(\log (4)-2 \log \left(e^{2 \vartheta  (R_{\text{\tiny {Q}}}-r)}+2\right)\right)\right\}\nonumber\\
&&+\frac{1}{24 \vartheta}\left\{-\text{Li}_2\left(-2 e^{2 (r-R_{\text{\tiny {Q}}}) \vartheta }\right)+\text{Li}_2\left(-2 e^{-2 R_{\text{\tiny {Q}}} \vartheta }\right)+\frac{4 \vartheta  r \left(2 e^{4 \vartheta  r} (\vartheta  r+1)+e^{2 \vartheta  (r+R_{\text{\tiny {Q}}})}\right)}{\left(2 e^{2 \vartheta  r}+e^{2 \vartheta  R_{\text{\tiny {Q}}}}\right)^2}\right.\nonumber\\
&&\left.\left.-(2 \vartheta  r+1) \log \left(2 e^{2 \vartheta  (r-R_{\text{\tiny {Q}}})}+1\right)+\log \left(2 e^{-2 \vartheta  R_{\text{\tiny {Q}}}}+1\right)\right\}\right]\;,
\end{eqnarray}
\end{widetext}
where we have defined $\vartheta\equiv\sqrt{m^2-\omega_\text{\tiny Q}^2}=\sqrt{\lambda^2/4\zeta}$, for brevity of terms, and defined
\begin{equation}
\text{Li}_n(z)=\sum_{k=1}^{\infty}\frac{z^k}{k^n}\;.
\end{equation}
The total energy and charge for large Q-balls calculated using the above densities satisfy the theoretical consistency condition Eq.\,(\ref{eq:debydq}) to a good approximation. In the large radius limit, they also satisfy the corresponding expressions in the literature\,\cite{Heeck:2020bau}, to the given order. For the BTW Q-ball transition profiles and parameter values that we work with, additionally, all relevant boundary conditions and exact Q-ball relations\,\cite{Heeck:2020bau} are well-satisfied with any conservative errors being at most $\mathcal{O}(10 \%)$.

Let us note an elementary limit of Eqs.\,\eqref{eq:Cden_btw} and \eqref{eq:Mden_btw}, the so called simple thin-wall limit mentioned earlier. When $\omega\rightarrow \omega_{\text{\tiny Q}}$, we have the thin-wall limit, and the field profiles are well-approximated by Eq.\,\eqref{eq:kzero}, with the charge and energy density profiles also taking a simple form given by
\begin{widetext}
\begin{equation}\label{eq:charge_den_TW}
\rho_{\text{\tiny Q}}^{\text{\tiny C},\text{\tiny TW}}(r)\simeq2\omega_{\text{\tiny Q}}\phi_{\text{\tiny Q}}^2\theta(R_{\text{\tiny {Q}}}-r)=2m\left(\frac{\lvert\lambda \rvert}{2\zeta}\right)\sqrt{1-\frac{\lambda^2}{4m^2\zeta}}~\Theta(R_{\text{\tiny {Q}}}-r)\;,
\end{equation}
\begin{equation}\label{eq:energy_den_TW}
\rho_{\text{\tiny Q}}^{\text{\tiny M},\text{\tiny TW}}(r)\simeq2\omega_{\text{\tiny Q}}^2\phi_{\text{\tiny Q}}^2\theta(R_{\text{\tiny {Q}}}-r)=2m^2\left(\frac{\lvert\lambda \rvert}{2\zeta}\right)\left\{1-\frac{\lambda^2}{4m^2\zeta}\right\}~\Theta(R_{\text{\tiny {Q}}}-r)\;.
\end{equation}
\end{widetext}
Here, $\Theta(x)$ is the Heaviside step function. In this strict limit of approximation, some of the gravitational lensing features will start to coincide with that of uniform density profiles\,\cite{Croon:2020wpr,Bai:2020jfm}.

In order to solve the lens equation given in Eq.\eqref{eq:lens_eq}, for a given Q-ball solution, we need to know the corresponding mass-ratio function $	\widetilde{M}_{\text{\tiny Q}}(\theta_{\text{\tiny I}})/M_{\text{\tiny Q}}$. This may be calculated starting from Eq.\eqref{eq:mproj} as
\begin{equation}\label{eq:massratioTW}
	\frac{\widetilde{M}_{\text{\tiny Q}}(\theta_{\text{\tiny I}})}{M_{\text{\tiny Q}}}=\frac{\int_0^{\theta_{\text{\tiny I}}/\theta_{\text{\tiny E}}} du~ u\int_0^\infty dv~\rho^{\text{\tiny M}}_{\text{\tiny Q}}\left(\sqrt{u^2+v^2}\right)}{\int_0^\infty dw ~w^2\rho^{\text{\tiny M}}_{\text{\tiny Q}}\left(w\right)}\;.
\end{equation}
Here, we have normalised all the coordinates with respect to the Einstein radius $R_{\text{\tiny E}}=d_{\text{\tiny L}} \theta_{\text{\tiny E}}$, as defined in Eq.\,\eqref{eq:ein_rad}, in the following way-- $u\equiv\chi/R_\text{\tiny E}$, $v\equiv z/R_\text{\tiny E}$ and $w\equiv r/R_\text{\tiny E}$. $M_{\text{\tiny Q}}$ is the total mass of the Q-ball lens. Thus, it is through this mass-ratio function that the precise details of the Q-ball density profile will appear in the lens equation, its solutions and the total magnification finally obtained.

Let us start by analysing astrophysical Q-ball solutions very close to the $\omega\rightarrow \omega_{\text{\tiny Q}}$ limit, the so-called TW Q-balls. For the case of TW Q-ball solutions, evaluating Eq.\,\eqref{eq:massratioTW} with Eq.\eqref{eq:energy_den_TW}, one obtains the mass-ratio function

\begin{equation}\label{eq:massrationTWexp}
	\frac{\widetilde{M}_{\text{\tiny Q}}(\theta_{\text{\tiny I}})}{M_{\text{\tiny Q}}}=
	\begin{cases}
		1-\left(1-\frac{(\theta_{\text{\tiny I}}/\theta_{\text{\tiny E}})^2}{(R_{\text{\tiny {Q}}}/R_{\text{\tiny E}})^2}\right)^{3/2} & ; \hspace{0.25cm}  \left(\frac{\theta_{\text{\tiny I}}}{\theta_{\text{\tiny E}}}\right) R_{\text{\tiny E}} < R_{\text{\tiny {Q}}}\\
		1 & ; \hspace{0.25cm} \left(\frac{\theta_{\text{\tiny I}}}{\theta_{\text{\tiny E}}}\right) R_{\text{\tiny E}} \geqslant R_{\text{\tiny {Q}}}\; .
	\end{cases}
\end{equation}

We note from above, with suitable re-scalings, that the dependence on $\omega_{\text{\tiny Q}}$ and $\phi_{\text{\tiny Q}}$ in the lens equation Eq.\,\eqref{eq:lens_eq} comes solely through $R_{\text{\tiny {Q}}}/R_{\text{\tiny E}}$ and $\theta_{\text{\tiny E}}$ for thin-wall Q-balls. The lens equation Eq.\,\eqref{eq:lens_eq} may now be solved with the mass function Eq.\,\eqref{eq:massrationTWexp} to find the image positions. It is found that there are two broad regimes for the lens equation with respect to the obtained solutions, depending on the size of the Q-ball.

As we see from Eq.\,\eqref{eq:massrationTWexp}, for $\theta_{\text{\tiny I}}/\theta_{\text{\tiny E}}<R_{\text{\tiny {Q}}}/R_\text{\tiny E}$, the lens equation as given in Eq.\eqref{eq:lens_eq} is a quintic polynomial in $\theta_{\text{\tiny I}}$. For the special case $\theta_{\text{\tiny S}}=0$, it can be factorised into a product of a quartic and linear polynomial. The linear polynomial gives the trivial solution at the origin (i.e. $\theta_{\text{\tiny I}}=0$), irrespective of the value of $R_{\text{\tiny {Q}}}/R_\text{\tiny E}$. The quartic equation has two real solutions for $0<R_{\text{\tiny {Q}}}/R_\text{\tiny E}<\sqrt{3/2}$. For $R_{\text{\tiny {Q}}}/R_\text{\tiny E}\leq1$ there exits two solutions which are equally spaced from the origin---at $\theta_{\text{\tiny I}}=\pm \theta_{\text{\tiny E}}$---on the Einstein ring. For $1<R_{\text{\tiny {Q}}}/R_\text{\tiny E}<\sqrt{3/2}$, we again have equally spaced solutions which lie at $\lvert \theta_{\text{\tiny I}} \rvert< \theta_{\text{\tiny E}}$, i.e. inside the Einstein ring. Specifically, they are located at, 
\begin{widetext}
\begin{equation}
\theta_{\text{\tiny I}}=\pm \frac{\theta_{\text{\tiny E}}R_{\text{\tiny {Q}}}}{R_\text{\tiny E}\sqrt{2}}\left(3-(R_{\text{\tiny {Q}}}/R_\text{\tiny E})^4-((R_{\text{\tiny {Q}}}/R_\text{\tiny E})^2-1)^{3/2}\sqrt{(R_{\text{\tiny {Q}}}/R_\text{\tiny E})^2+3}\right)^{1/2}\;.
\end{equation}
\end{widetext}
For $R_{\text{\tiny {Q}}}/R_\text{\tiny E}\geqslant\sqrt{3/2}$, all four solutions to the quartic equation becomes imaginary and we are left with only one solution at the origin.

For non zero $\theta_{\text{\tiny S}}$, we can use Descartes' rule of signs (See, for instance,\,\cite{barnard1959higher}) to get the number of positive and negative real roots. From Descartes' rule of signs, we deduce that there will be two negative real roots and one positive real root. Among them the two negative real roots disappear when either $\theta_{\text{\tiny S}}\gg1$ or $R_{\text{\tiny {Q}}}/R_{\text{\tiny E}}>\sqrt{3/2}$. Below we will try to get some analytical form of some of these solutions in particular limits. We will be largely restricted to numerical analysis, though, due to the Abel-Ruffini theorem (for instance, see\,\cite{khovanskii2014topological}).

For a thin-walled Q-ball with a radius satisfying $R_{\text{\tiny {Q}}}>\sqrt{3/2}R_{\text{\tiny E}}$,  Eq.\,\eqref{eq:lens_eq}, with the functional form Eq.\eqref{eq:massrationTWexp}, yields only a single viable solution. The solution may be readily obtained numerically, but in special cases, a semi-analytical expression may be found. If the domain of the solution is such that $\left( \theta_{\text{\tiny I}}/\theta_\text{\tiny E}\right)  R_\text{\tiny E} \ll R_{\text{\tiny {Q}}} \simeq \sqrt{m^2-\omega_{\text{\tiny Q}}^2}/\left(\omega^2-\omega_{\text{\tiny Q}}^2\right)$, which is the case when $\theta_{\text{\tiny S}}/\theta_{\text{\tiny E}}\ll 1$, then in this region we would have
\begin{equation}\label{eq:massfunc_limit}
\frac{\widetilde{M}_{\text{\tiny Q}}(\theta_{\text{\tiny I}})}{M_{\text{\tiny Q}}}\Big \vert_{\left( \frac{\theta_{\text{\tiny I}}}{\theta_\text{\tiny E}}\right)  R_\text{\tiny E} \ll R_{\text{\tiny {Q}}}} \simeq \frac{3}{2} \left( \frac{\theta_{\text{\tiny I}}R_{\text{\tiny E}}}{\theta_{\text{\tiny E}}R_{\text{\tiny {Q}}}}\right)^2 \; ,
\end{equation}
and Eq.\,\eqref{eq:lens_eq} may be solved analytically to obtain an approximate solution
\begin{equation}\label{eq:theta3}
\theta_{\text{\tiny I}}^{(0)}\simeq \frac{\theta_{\text{\tiny S}}}{1-\left(3 R_{\text{\tiny E}}^2/2 R_{\text{\tiny {Q}}}^2\right)}  \; .
\end{equation}
One may calculate the magnification in this regime by taking the derivative of Eq.\,\eqref{eq:massrationTWexp}, with the limit $\left( \theta_{\text{\tiny I}}/\theta_\text{\tiny E}\right)  R_\text{\tiny E} \ll R_{\text{\tiny {Q}}}$ and using the solution in Eq.\,\eqref{eq:theta3}. In this regime, from the single viable solution, one obtains a magnification
\begin{widetext}
\begin{equation}\label{eq:mag3TW}
\mathfrak{m}\left({\theta_{\text{\tiny I}}^{(0)}}\right) \Big\vert_{\left( \frac{\theta_{\text{\tiny I}}}{\theta_\text{\tiny E}}\right)  R_\text{\tiny E} \ll R_{\text{\tiny {Q}}}}=\frac{{\theta_{\text{\tiny I}}^{(0)}}}{{\theta_{\text{\tiny S}}}}\frac{d\theta_{\text{\tiny I}}^{(0)}}{d \theta_{\text{\tiny S}}}=\left(1-\frac{3}{2(R_{\text{\tiny {Q}}}/R_\text{\tiny E})^2}\right)^{-2}\left[1-\frac{3(\theta_{\text{\tiny S}}/\theta_{\text{\tiny E}})^2}{2(R_{\text{\tiny {Q}}}/R_\text{\tiny E})^4}\left(1-\frac{3}{2(R_{\text{\tiny {Q}}}/R_\text{\tiny E})^2}\right)^{-3}\right]
\end{equation}
\end{widetext}
It is straightforward to deduce by looking at Eq.\,\eqref{eq:mag3TW} that as $\theta_{\text{\tiny S}}\rightarrow 0$, we get a finite maximum magnification. Hence we conclude that for Q-balls with a radius $R_{\text{\tiny {Q}}}>\sqrt{3/2}R_{\text{\tiny E}}$ the maximum magnification produced is finite and only depends on the ratio $R_{\text{\tiny {Q}}}/R_{\text{\tiny E}}$.

Let us now consider the other regime in Q-ball sizes. For Q-balls satisfying $R_{\text{\tiny {Q}}}<\sqrt{3/2}R_{\text{\tiny E}}$, there may be three viable solutions for small $\theta_{\text{\tiny S}}/\theta_\text{\tiny E}$, two of which disappear as we move towards larger values. Again, one may numerically solve the lens equation to find the solutions, but in special cases, semi-analytic expressions may be obtained. 

Two out of the three solutions are found in the region satisfying $\frac{\theta_{\text{\tiny I}}}{\theta_{\text{\tiny E}}}\geqslant\frac{R_{\text{\tiny {Q}}}}{R_{\text{\tiny E}}}$. Here, in fact, we can analytically solve the lens equation, which is now only quadratic in $\theta_{\text{\tiny I}}/\theta_{\text{\tiny E}}$. The solutions obtained are
\begin{equation}\label{eq:pointsolTW}
	\theta_{\text{\tiny I}}^{(\pm)}=\frac{\theta_{\text{\tiny S}}}{2}\left[1\pm\sqrt{1+\frac{4\theta_{\text{\tiny E}}^2}{\theta_{\text{\tiny S}}^2}}\right] \; .
\end{equation}
\begin{figure}[h!]
	\begin{center}
		\includegraphics[scale=0.9]{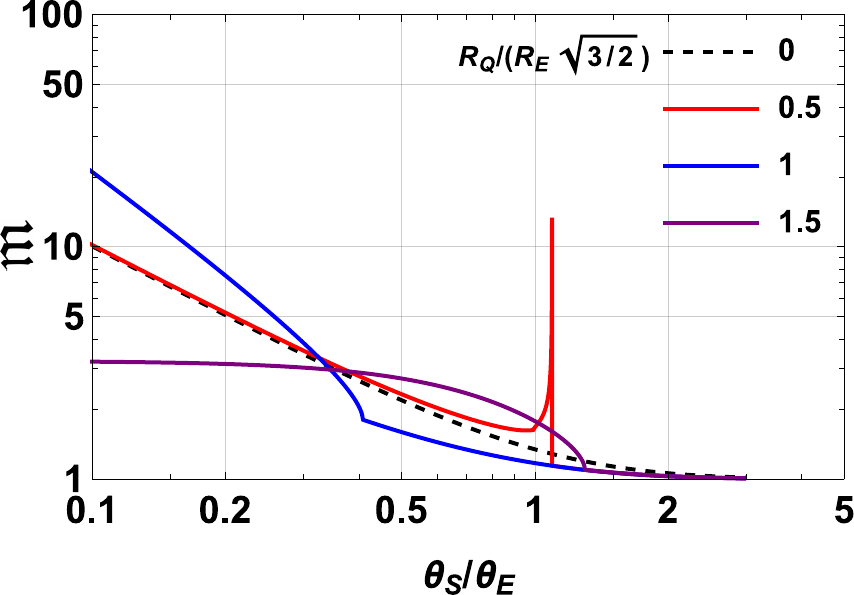}	
	\end{center}
	\caption{Plot showing total magnification produced with changing $\theta_{\text{\tiny S}}$ for the case of TW astrophysical Q-balls. Here the legend indicates the ratio $R_{\text{\tiny {Q}}}/(R_{\text{\tiny E}}\sqrt{3/2})$, where $\sqrt{3/2}$ denotes the critical ratio of $R_{\text{\tiny {Q}}}/R_{\text{\tiny E}}$ which separates the regimes with different number of solutions. The dashed line refers to the total magnification produced due to a point lens.}
	\label{fig:mag_vs_impact_TW}
\end{figure}
These are observed to just coincide with image solutions for a point-like lens. The magnification of each of these images is given by
\begin{widetext}
\begin{equation}
\mathfrak{m}\left({\theta_{\text{\tiny I}}^{(\pm)}}\right) \Big\vert_{\left( \frac{\theta_{\text{\tiny I}}}{\theta_\text{\tiny E}}\right)  R_\text{\tiny E} \geqslant R_{\text{\tiny {Q}}}}=\frac{{\theta_{\text{\tiny I}}^{(\pm)}}}{{\theta_{\text{\tiny S}}}}\frac{d\theta_{\text{\tiny I}}^{(\pm)}}{d \theta_{\text{\tiny S}}}=\pm \frac{(\theta_{\text{\tiny S}}/\theta_{\text{\tiny E}})^2+2}{2\,\theta_{\text{\tiny S}}/\theta_{\text{\tiny E}}\sqrt{(\theta_{\text{\tiny S}}/\theta_{\text{\tiny E}})^2+4}}+\frac{1}{2} \; ,
\end{equation}
\end{widetext}

leading to a contribution to the total magnification from these two solutions to be

\begin{equation}\label{eq:totmagpointTW}
\mathfrak{m}\Big\vert^{(\pm)}_{\left( \frac{\theta_{\text{\tiny I}}}{\theta_\text{\tiny E}}\right)  R_\text{\tiny E} \geqslant R_{\text{\tiny {Q}}}}=\abs{\mathfrak{m}_{\text{\tiny I}}^{(+)}}+\abs{\mathfrak{m}_{\text{\tiny I}}^{(-)}}=\frac{(\theta_{\text{\tiny S}}/\theta_{\text{\tiny E}})^2+2}{\theta_{\text{\tiny S}}/\theta_{\text{\tiny E}}\sqrt{(\theta_{\text{\tiny S}}/\theta_{\text{\tiny E}})^2+4}}\; .
\end{equation}
In this regime of Q-ball sizes, the contribution from two of the solutions, therefore, just coincides with the analogous contribution from point-like lenses. Note that in the above expressions, the characteristics of the Q-ball under consideration are still present, entering through the Einstein angle $\theta_{\text{\tiny E}}$. We also note here that the maximum magnification produced due to these images formally diverges as $\theta_{\text{\tiny S}}\rightarrow 0$. In actuality, this apparent divergence will be mitigated by finite source effects coming into play as we approach the limit.

An analytic form for the third solution may also be sought if it lies in the region $\left( \theta_{\text{\tiny I}}/\theta_\text{\tiny E}\right)  R_\text{\tiny E} \ll R_{\text{\tiny {Q}}}$. It yields an expression identical to that given in Eq.\,\eqref{eq:theta3}, and therefore enhances the image by the same magnification factor as that given in Eq.\,\eqref{eq:mag3TW}. As we discussed before, we can see from Eq.\,\eqref{eq:mag3TW}, that as we decrease $\theta_{\text{\tiny S}}/\theta_{\text{\tiny E}}$ the total magnification increases. The maximum magnification is produced for $\theta_{\text{\tiny S}}=0$ with a magnitude
\begin{equation}\label{eq:mag3maxTW}
	\mathfrak{m}_3^{\text{max}}\approx \left(1-\frac{3}{2(R_{\text{\tiny {Q}}}/R_\text{\tiny E})^2}\right)^{-2} \; .
\end{equation}
\begin{figure*}
\begin{center}
	\includegraphics[scale=0.9]{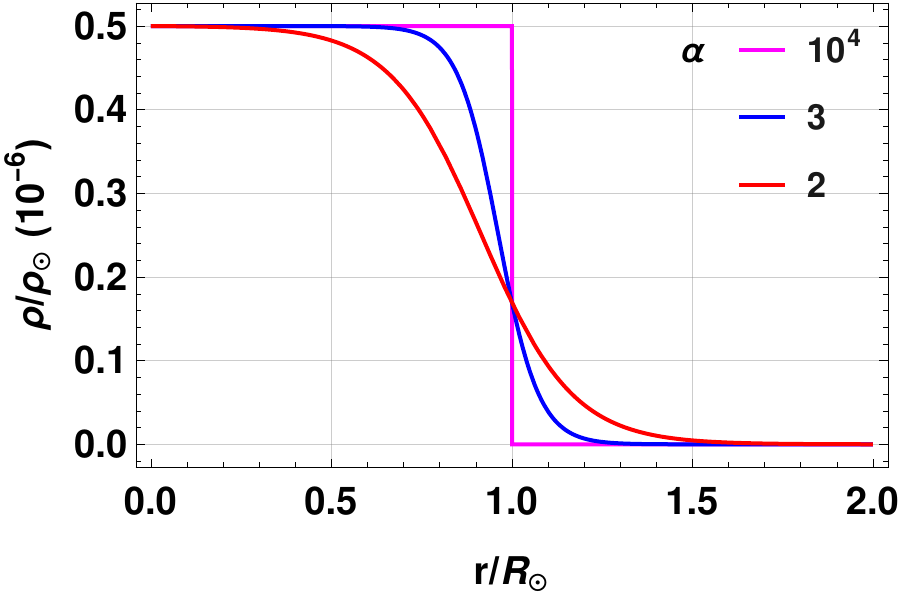}
	\includegraphics[scale=0.9]{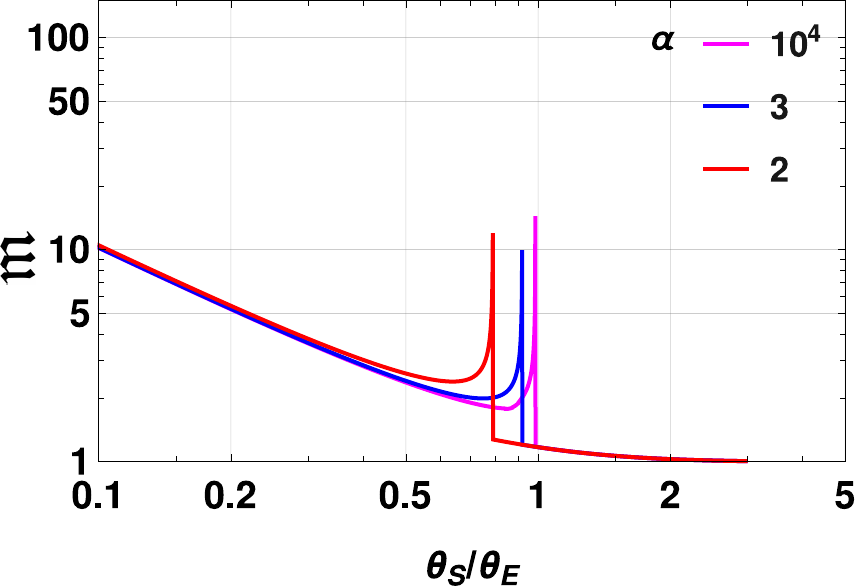}	
\end{center}
	\caption{In the above figures, we display the density profiles for three different astrophysical Q-ball profiles with approximately equal radii and average densities and compare their relative magnification profiles. We define $\alpha\equiv\sqrt{(m^2-\omega_{\text{\tiny Q}}^2)/(\omega^2-\omega_{\text{\tiny Q}}^2)}$ to quantify the different Q-ball profiles. \textbf{Top:} Density profile for a TW and two BTW Q-balls are shown, all of which have an equal radius, around $1\,R_{\odot}$. Here, we have normalised the radial coordinate with respect to the solar radius, $R_{\odot}=6.9\times10^8\,\mathrm{m}$, and the density is normalised to the average solar density, $\rho_\odot=1.4\times10^3\,\mathrm{kg}/\mathrm{m}^3$. \textbf{Bottom:} The figure shows the dependence of the total magnification on the source position $\theta_{\text{\tiny S}}$ for the same set of Q-ball profiles as in the adjoining figure. Though the magnification profiles have similar shapes overall, the position of the caustic progressively shifts towards lower $\theta_{\text{\tiny S}}$ values for thicker transition regions. A larger magnification is also seen in the small $\theta_{\text{\tiny S}}$ domain for profiles with a thicker transition region.}
\label{fig:comp_TW_BTW}
\end{figure*}

For intermediate values of $R_{\text{\tiny {Q}}}/R_{\text{\tiny E}}$, analytical expressions for the solutions and magnification are challenging to obtain in general. In our subsequent studies, we will, therefore, numerically compute all the solutions to the lens equation. 

The numerically computed total magnification produced by thin-wall Q-balls are shown in Fig.\,\ref{fig:mag_vs_impact_TW}, for different values of $R_{\text{\tiny {Q}}}/R_{\text{\tiny E}}$. From Fig.\,\ref{fig:mag_vs_impact_TW}, we see that for TW Q-balls having $R_{\text{\tiny {Q}}}>\sqrt{3/2}R_{\text{\tiny E}}$, decreasing the value of $\theta_{\text{\tiny S}}$ to zero leads to a finite maximum value for the magnification. This is quantified by the analytical expression given in Eq.\eqref{eq:mag3TW}. 

This may be contrasted with the $R_{\text{\tiny {Q}}}\leqslant \sqrt{3/2}R_{\text{\tiny E}}$ case where there is formally no upper bound on the total magnification, though in reality it will be regulated when finite source sizes are correctly accounted for, as we mentioned earlier. When the size of TW Q-balls are of the order of their corresponding Einstein radius i.e. $R_{\text{\tiny {Q}}}\approx R_{\text{\tiny E}}$, the magnification profile displays some interesting features---such as the peak for the case $R_{\text{\tiny {Q}}}/R_{\text{\tiny E}}=0.5\times\sqrt{3/2}$. Such a peak is termed a `caustic', and is the value of $\theta_{\text{\tiny S}}$ at which the number of solutions changes discontinuously. The sudden divergence of the magnification comes from the fact that the derivative $d\theta_{\text{\tiny S}}/d\theta_{\text{\tiny I}}$ vanishes, and by virtue of Eq.\eqref{eq:mag_exp} makes the magnification formally diverge. 

Intuitively, it may also be understood that Q-balls which are not in the vicinity of the line of sight joining the source and the observer should not affect the source brightness significantly. Therefore we see from Fig.\,\ref{fig:mag_vs_impact_TW} that for large values of $\theta_{\text{\tiny S}}$, Q-balls of all sizes fail to produce any magnification and therefore the magnification factor $\mathfrak{m}\rightarrow 1$ irrespective of the size of the Q-balls.

Having looked at the salient features of lensing due to thin-wall Q-balls, we now move on to investigate the gravitational lensing by BTW Q-balls, where $\omega$ may take more general values. Here again, in order to solve the lens equation given in Eq.\eqref{eq:lens_eq}, we need to calculate the mass ratio given in Eq.\eqref{eq:massratioTW}, but now using the full mass density profile for BTW Q-balls given in Eq.\,\eqref{eq:Mden_btw}, with arbitrary $\omega$. This does not yield a simple analytical form, except in very special cases. So, for the case of BTW Q-balls, we will rely on numerical analysis for finding out solutions to lens equation as well as calculating the total magnification. We also ensure that all the requisite criteria for viable astrophysical Q-balls, as encapsulated in Secs.\,\ref{sec:qballs} and \ref{sec:thbounds} are satisfied. Note that the mass profile Eq.\,\eqref{eq:massprofile} implicitly appears in Eq.\,\eqref{eq:mproj}, for the definition of $\widetilde{M}_{\text{\tiny Q}}(\chi')$, through the identification $r=\sqrt{\chi'^2+z^2}$, for a fixed transverse distance $\chi'$.

The major difference in TW and BTW Q-balls comes from the possibility of a distinct overall functional form and thicker transition region for the latter's field profiles and mass density profiles. It is natural to suspect that such a feature in the mass density profile may have a different overall impact on the microlensing signature, depending on the actual functional form, even for Q-balls which have roughly the same mass and size. One of the astrophysical observables to help contrast the distinct characteristics of TW and BTW Q-balls is to compare their respective magnification profiles. 

For this particular reason, in Fig.\,\ref{fig:comp_TW_BTW}, we plot the magnification profile for a thin wall Q-ball and two BTW Q-balls. All three Q-ball profiles have equal radii and roughly the same average density. We can see from the figure that qualitatively the magnification profiles look the same, but the position of the caustic progressively shifts towards lower $\theta_{\text{\tiny S}}$ values for profiles with thicker transition regions. Also, for smaller values of $\theta_{\text{\tiny S}}$, the profile with the thicker region of transition produces a larger magnification. For large $\theta_{\text{\tiny S}}$ values, all three Q-balls lens cases converge to a magnification of unity, which is true for any generic astrophysical object. 

As we commented earlier, when a lens passes through the line of sight of the observer and the source, it magnifies the background source. Whether this is observed or detected will depend on the microlensing survey instrument's sensitivity to this waxing of brightness. Usually, for categorising an event to be a microlensing event, we define a threshold magnification value ($\mathfrak{m}^*$) above which an instrument is able to detect it as a viable event. This threshold magnification ($\mathfrak{m}^*$) is conventionally defined as the magnification generated by a point lens when $\theta_{\text{\tiny S}}=\theta_{\text{\tiny E}}$. Utilising Eq.\,\eqref{eq:totmagpointTW}, this gives the threshold magnification as $\mathfrak{m}^*=1.34$. The idea is that, with the above prototypical magnification threshold assumption for a survey, all point lenses which are positioned with $\theta_{\text{\tiny S}}<\theta_{\text{\tiny S}}^*$ will produce magnification $\mathfrak{m}>\mathfrak{m}^*$, and hence will be detected.

\begin{figure}[]
\begin{center}
\includegraphics[scale=0.9]{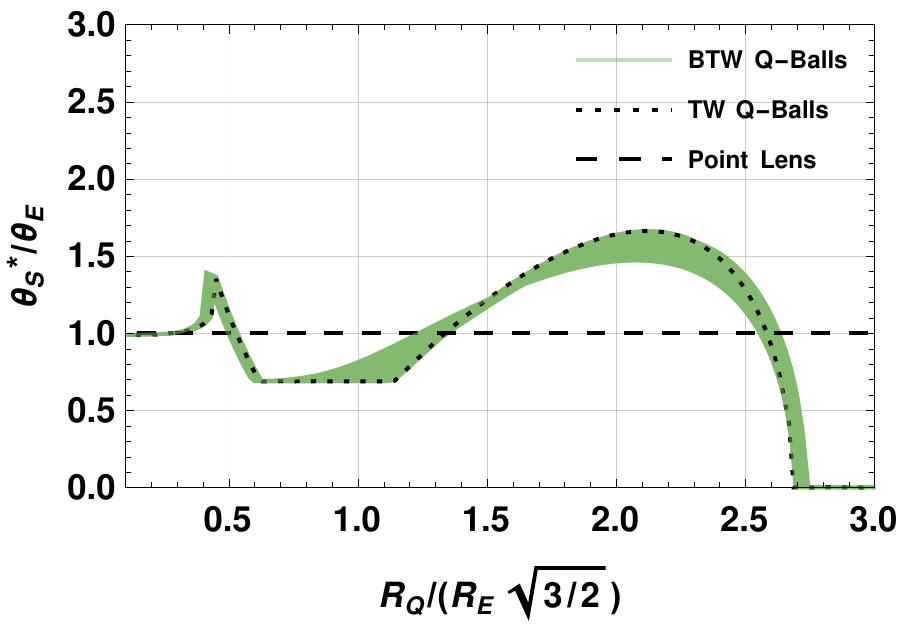}	
\end{center}
\caption{The threshold value of the source position (denoted by $\theta_{\text{\tiny S}}^*$), below which one obtains a total magnification $\mathfrak{m}\geqslant \mathfrak{m}^*$, is shown as a function of $R_{\text{\tiny {Q}}}/R_{\text{\tiny E}}$ .  Here, a range bounded by $\mathfrak{m}^*=1.34\pm0.01$ is the assumed threshold magnification band (corresponding to the magnification due to a point mass lens when $\theta_{\text{\tiny S}}\rightarrow\theta_{\text{\tiny E}}$) delimiting what may be detectable by a microlensing survey. The dashed and dotted black curves indicate point lens and TW Q-ball profiles, respectively.  
The green band is composed of various distinct astrophysical Q-ball profiles with the restriction $\sqrt{(m^2-\omega_{\text{\tiny Q}}^2)/(\omega^2-\omega_{\text{\tiny Q}}^2)}\lesssim 0.5$, so as to comfortably satisfy the field-theoretic stability constraints. For the $R_{\text{\tiny {Q}}}/R_{\text{\tiny E}}$ values shown, we have verified the existence of astrophysical Q-balls, satisfying the criteria of Secs.\,\ref{sec:qballs} and \ref{sec:thbounds}.}
	\label{fig:size_vs_thres_impact}
\end{figure}
For extended distributions of matter like Q-balls, due to the non-trivial density profiles, the value of $\theta_{\text{\tiny S}}$ that produces a magnification of $\mathfrak{m}^*$ may not be the same as $\theta_{\text{\tiny E}}$, and may even vary depending on the relative size of the Q-ball as compared to its Einstein radius. 

\begin{figure}[]
	\begin{center}
		\includegraphics[scale=0.9]{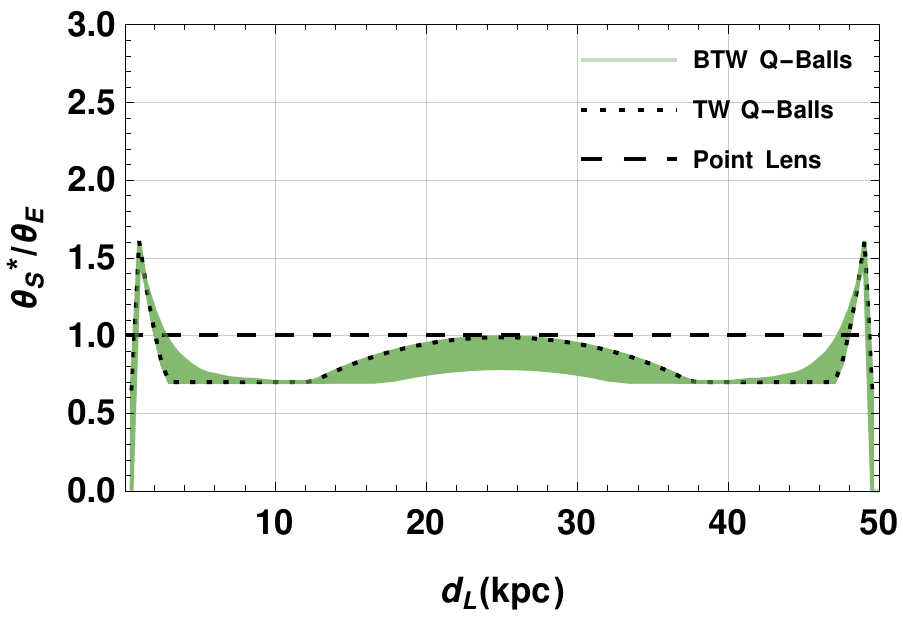}
		\includegraphics[scale=0.9]{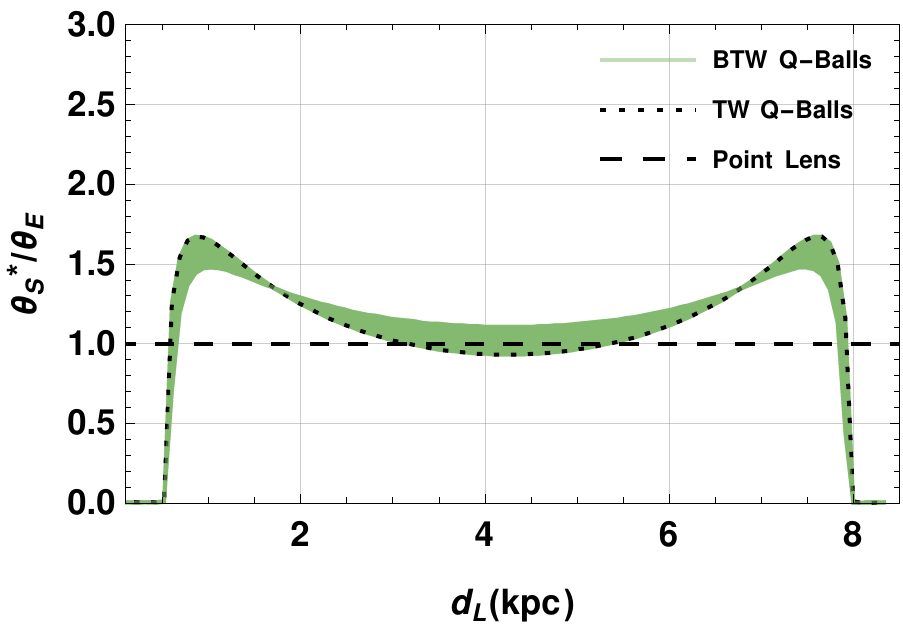}
	\end{center}	
	\caption{The threshold $\theta_{\text{\tiny S}}^*$ is shown now as a function of the lens distance $d_{\text{\tiny L}}$. The lensing source objects are assumed to be the Large Magellanic Cloud (top figure) and the Milky Way bulge (bottom figure). The point lens and TW Q-balls are again represented by dashed and dotted black curves, respectively.  The green band, as in Fig.\,\,\ref{fig:size_vs_thres_impact}, represents various astrophysical Q-ball configurations, all with similar mass and radii but distinct density profiles. The plots have been made for $M_\text{\tiny {Q}}\simeq 5 \times 10^{-6}~ \solarmass $  and $R_\text{\tiny {Q}}= 1 ~\solarrad $. We have again adopted $\mathfrak{m}^*\in [1.33, 1.35]$ and have restricted $\sqrt{(m^2-\omega_{\text{\tiny Q}}^2)/(\omega^2-\omega_{\text{\tiny Q}}^2)}\lesssim 0.5$ based on field-theoretic stability.}
	\label{fig:thres_imp_vs_dl}
\end{figure}

The threshold characteristics may be understood more clearly from Fig.\,\ref{fig:size_vs_thres_impact}, where we have computed and plotted how the threshold value of $\theta_{\text{\tiny S}}$, which gives magnification in the range $\mathfrak{m}^*=1.34\pm0.01$,  changes with the ratio $R_{\text{\tiny {Q}}}/R_{\text{\tiny E}}$.  In Fig.\,\ref{fig:size_vs_thres_impact}, the green band depicts a range of astrophysical Q-balls, with distinct density profiles. The band is obtained by sifting through numerous Q-ball profiles with the requirement that they satisfy the criteria of Secs.\,\ref{sec:qballs} and \ref{sec:thbounds}. The band is not an artifact of the narrow range ($\pm0.01$) we have assumed for $\mathfrak{m}^*$. From Fig.\,\ref{fig:size_vs_thres_impact}, it is particularly obvious that when $R_{\text{\tiny {Q}}}/R_{\text{\tiny E}}\rightarrow 0$, for all Q-balls, the corresponding $\theta_{\text{\tiny S}}$ value goes to the point lens value (i.e. $1$), as should be expected. Additionally, the band gets terminated at $R_{\text{\tiny {Q}}}/(R_{\text{\tiny E}}\sqrt{3/2})\sim 2.7 $, because above it the magnification is always smaller than $1.34$. When $R_{\text{\tiny {Q}}}/(R_{\text{\tiny E}}\sqrt{3/2})$ is near $0.4$ or in the range $[1.4,\,2.5]$ the BTW Q-ball lenses are more effective in magnifying sources than their point-lens counterparts. One also notes that there is a peak near $R_{\text{\tiny {Q}}}/(R_{\text{\tiny E}}\sqrt{3/2}) \sim 0.4$. In the range $[0.5,\,1.3]$ the astrophysical Q-ball lenses are weaker lenses than point lenses.

In Fig.\,\ref{fig:thres_imp_vs_dl}, we further plot the variation of $\theta_{\text{\tiny S}}^*$ as a function of the lens position $d_{\text{\tiny L}}$ for the Large Magellanic Cloud (LMC) and Milky Way bulge (MW) sources; top and bottom figures, respectively. The LMC is located at $50$ kpc and the MW at $8.5$ kpc from earth. As we shall see in the next section, akin to the data encapsulated in Fig.\,\ref{fig:thres_imp_vs_dl}, a dictionary of $\theta_{\text{\tiny S}}^*$ for various astrophysical Q-ball configurations is critical for determining microlensing event rates and physically viable astrophysical Q-ball populations. Here, the  band indicates various astrophysical Q-balls with nearly identical masses and radii but differing density profiles. For Fig.\,\ref{fig:thres_imp_vs_dl}, as in Fig.\,\ref{fig:comp_TW_BTW}, the mass and radius of the Q-ball are taken to be about $ 5 \times 10^{-6}\,\solarmass $  and $ 1\,\solarrad $. Clearly, for the MW source, the Q-ball lens is more effective compared to LMC, manifesting as higher values for $\theta_{\text{\tiny S}}^*$. This will reflect in the microlensing event rates and hence in the limits on viable astrophysical Q-ball populations. 

In the next subsection, we generalise the Q-ball gravitational lensing study also to incorporate scenarios where finite source size effects need to be accounted for. 
\begin{figure}[h]
	\begin{center}
		\includegraphics[scale=0.5]{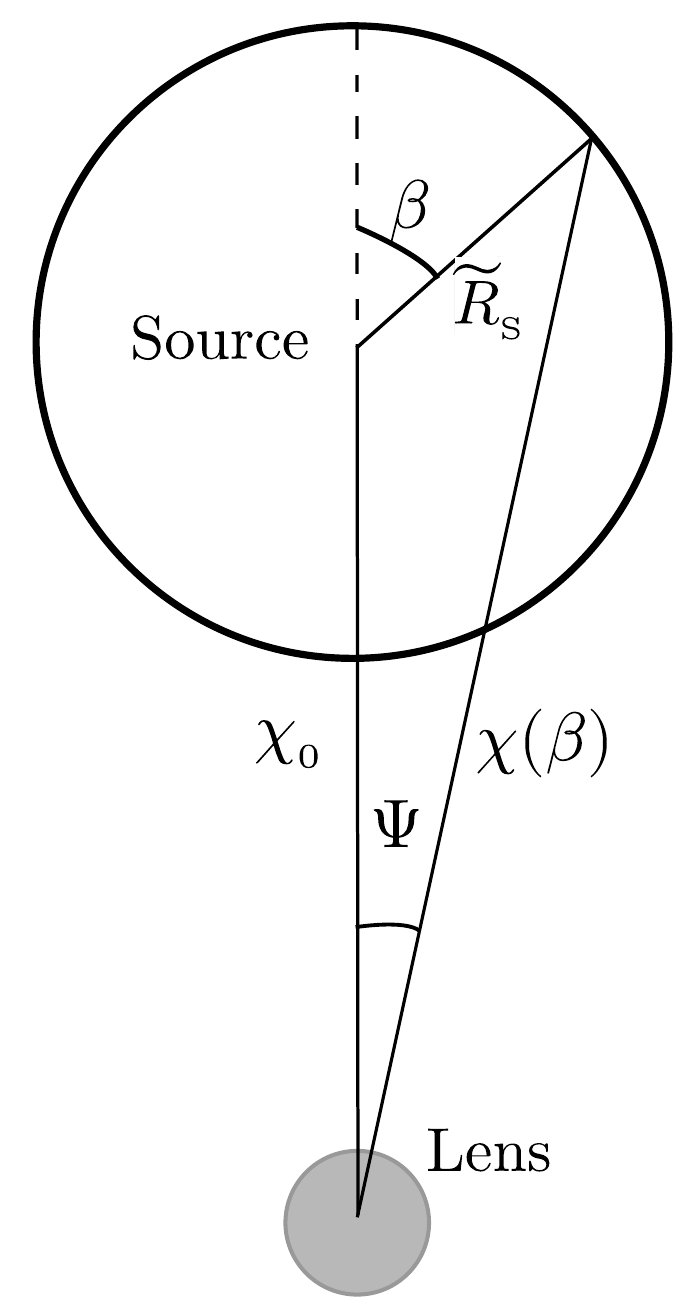}
	\end{center}	
	\caption{A diagrammatic representation of a circular source projected onto the lens plane. The scales are slightly exaggerated for clarity. The center of the source is at impact parameter $\chi_0$ and the edge of the source is parameterized by $\beta $ and is at impact parameter $\chi(\beta)$.}
	\label{fig:fin_src_geometry}
\end{figure}
\begin{figure*}
\begin{center}
\includegraphics[scale=0.8]{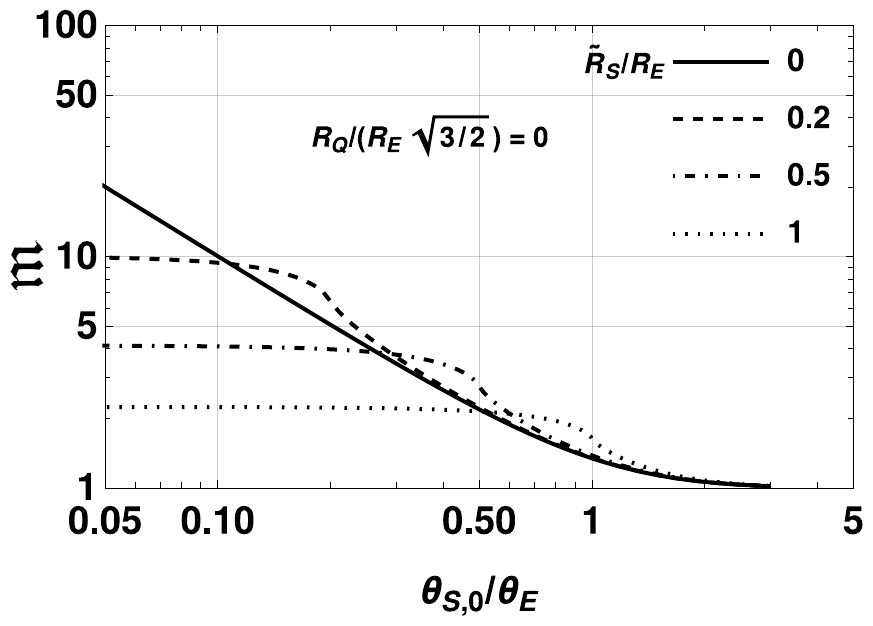}
\includegraphics[scale=0.8]{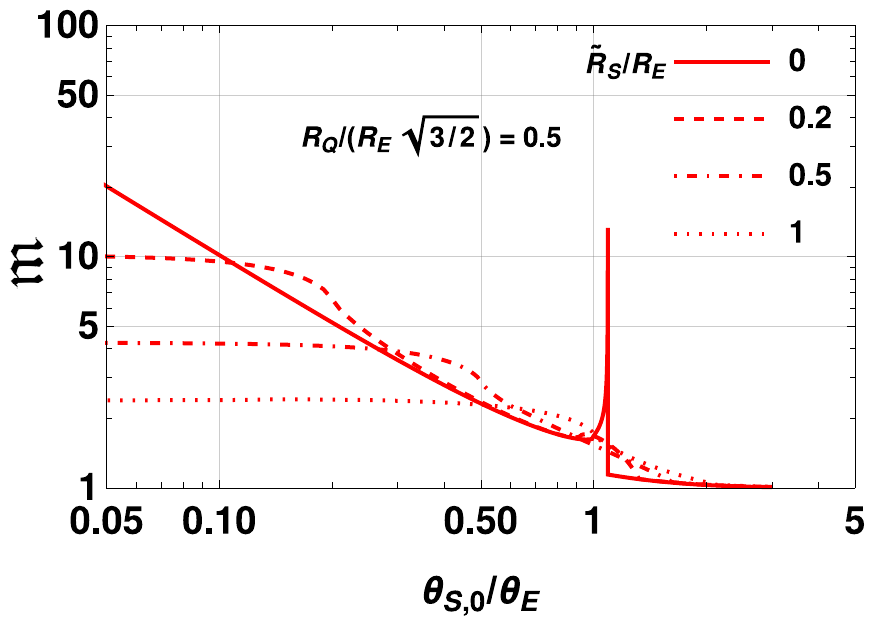}
\includegraphics[scale=0.8]{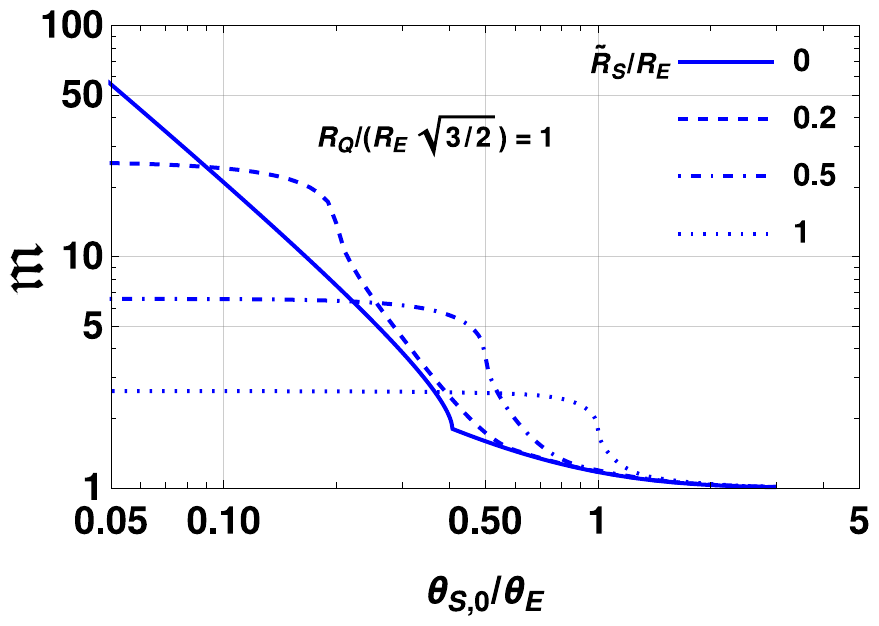}
\includegraphics[scale=0.8]{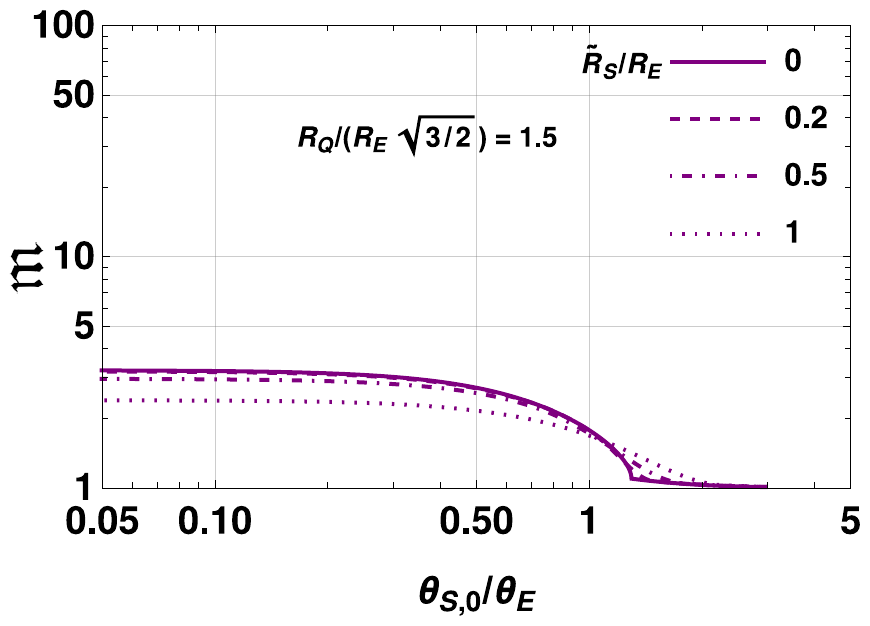}
\end{center}	
\caption{ The plot shows total magnification produced as a function of $\theta_{\text{\tiny S,0}}$ for the case of TW astrophysical Q-balls for a finite-size source. Here, black, red, blue and purple curves represent the ratio $R_{\text{\tiny {Q}}}/(R_{\text{\tiny E}}\sqrt{3/2})=\{0~(\text{Point lens}),~ 0.5, ~1,~ 1.5\}$. Solid, dashed, dashed-dotted and dotted lines represent $\widetilde{R}_{\text{\tiny {S}}}/R_{\text{\tiny E}}=$ \{0 (\text{Point source }),~0.2,~0.5,~1\}. For small $\theta_{\text{\tiny S,0}}/\theta_{\text{\tiny E}}$, the magnification due to finite source is regulated and is smaller than the point source. Additionally, for a given source size and large $\theta_{\text{\tiny S,0}}/\theta_{\text{\tiny E}}$, the magnification matches with the corresponding point lens values (top-left figure).}
\label{fig:fin_src_mag_vs_imp}
\end{figure*}

\subsection{Gravitational lensing of extended sources by thin-wall and beyond-thin-wall Q-balls}\label{subsubsec:twbtwfinsrclensing}
We have so far explored gravitational lensing by  TW and BTW Q-balls for point sources.   However, when the angular diameter of the source is comparable with the Einstein radius, the point source assumption fails, and we obtain different magnification characteristics\,\cite{Witt:1994}. Finite-size source effects become particularly important when we are considering lower-mass lenses with correspondingly small Einstein radii. Furthermore, as lens size approaches the wavelength of visible light, wave effects in gravitational lensing become significant\,\cite{Gould_1997,Nakamura}. In our study, we investigate the Q-balls with astrophysical sizes, for which the wave effects on gravitational lensing are negligible. In this subsection, we will discuss gravitational lensing of extended sources by astrophysical Q-balls. We will follow and adapt accordingly the methodologies in\,\cite{Witt:1994,Croon:2020ouk, Niikura:2017zjd, Sugiyama:2021xqg} for our cases of interest.

Consider a circular source with a radius $R_{\text{\tiny S}}$ and project it onto the lens plane. The source radius in the lens plane is $\widetilde{R}_{\text{\tiny S}}\equiv d_{\text{\tiny L}}R_{\text{\tiny S}}/d_{\text{\tiny S}}$. As shown in Fig.\,\ref{fig:fin_src_geometry}, the impact parameter of the center of the source star is $\chi_{\text{\tiny 0}}$.   An arbitrary point at the source's edge may be parameterised by angle $\beta$ and has an angular position
\begin{equation}\label{eq:fin_src_lens_plane_src_dist}
\theta_{\text{\tiny S}}(\beta)=\sqrt{\theta_{\text{\tiny S,0}}^2+(\widetilde{R}_{\text{\tiny S}}/d_{\text{\tiny L}})^2+2\theta_{\text{\tiny S,0}}(\widetilde{R}_{\text{\tiny S}}/d_{\text{\tiny L}})\cos\beta}\;,
\end{equation}
where $\theta_{\text{\tiny S,0}}\equiv\chi_{\text{\tiny 0}}/d_{\text{\tiny L}}$, is the angular position of the center of the source star. 

\begin{figure*}
\begin{center}
\includegraphics[scale=0.7]{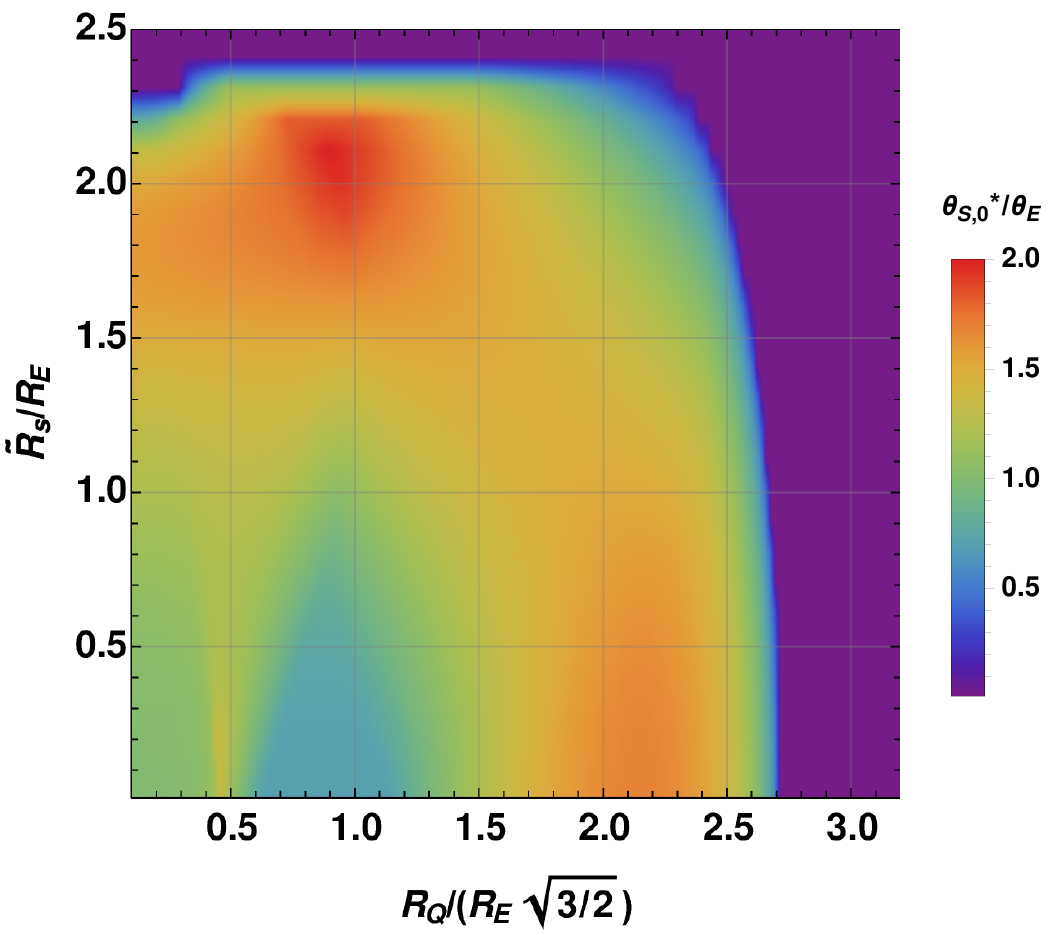} \\	
\includegraphics[scale=0.7]{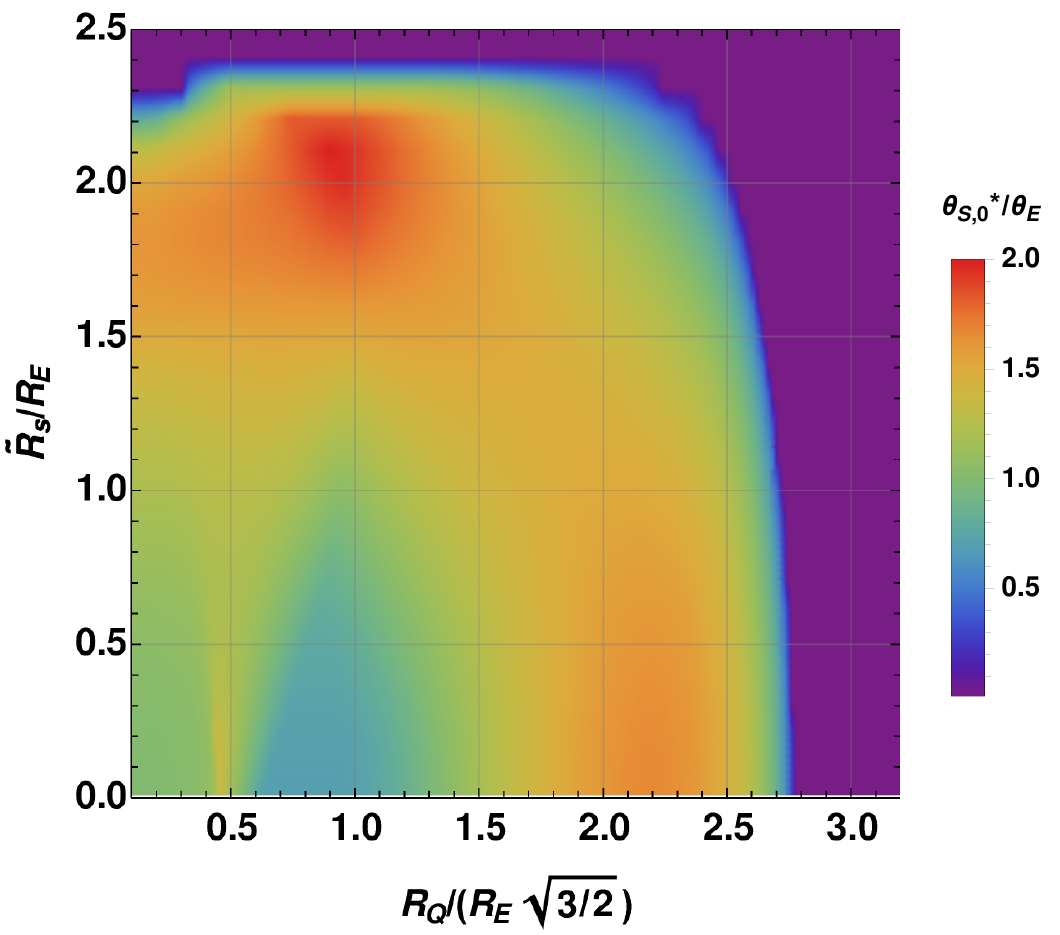}	
\includegraphics[scale=0.7]{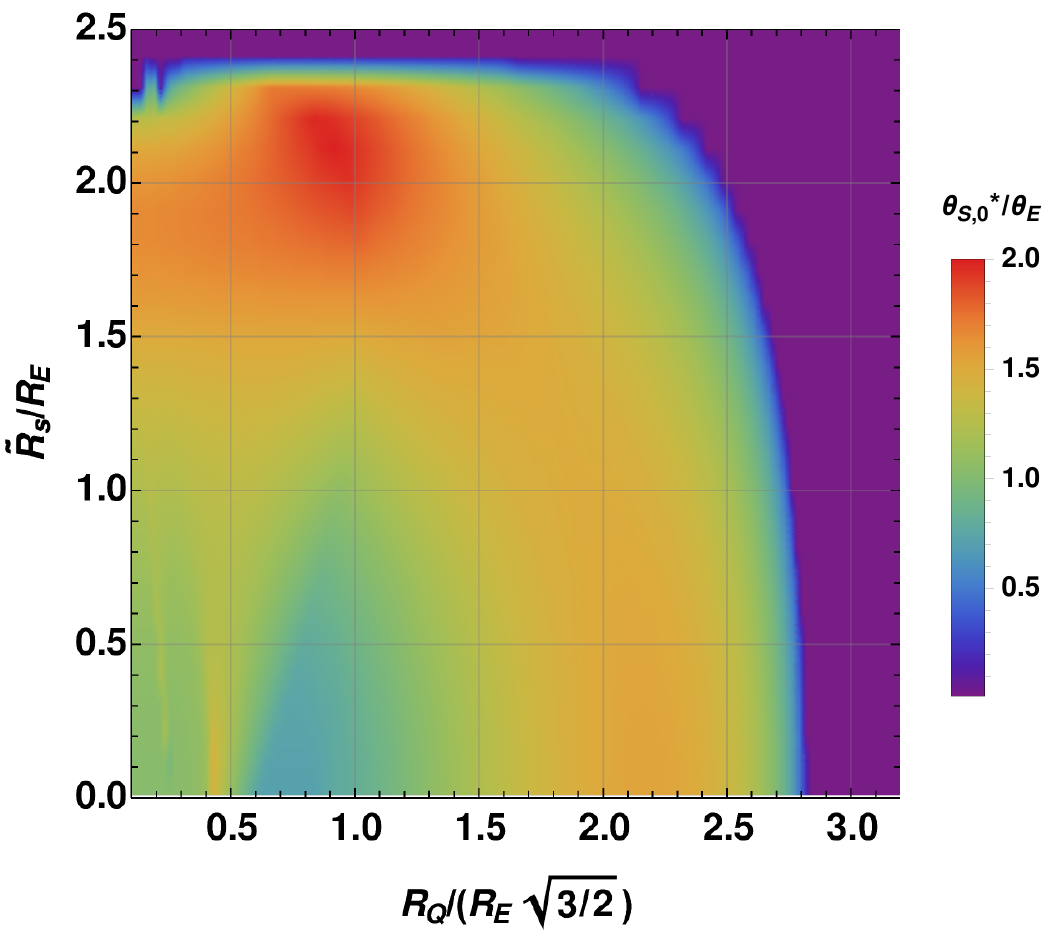}	
\end{center}
\caption{Density plots showing the threshold source positions ($\theta_{\text{\tiny S,0}}^*/\theta_{\text{\tiny E}}$), below which one obtains a total magnification $\mathfrak{m}\geqslant \mathfrak{m}^*$, as a function of $R_{\text{\tiny {Q}}}/R_{\text{\tiny E}}$ and $\widetilde{R}_{\text{\tiny {S}}}/R_{\text{\tiny E}}$. As before, a range bounded by $\mathfrak{m}^*=1.34\pm0.01$ is the assumed threshold magnification. The plots are made for TW Q-balls profiles (top), and  BTW  profiles with $\alpha =$ 3 (bottom left), and 2 (bottom right). For the $R_{\text{\tiny {Q}}}/R_{\text{\tiny E}}$ values shown, we have verified the existence of astrophysical Q-balls, satisfying the criteria of Secs.\,\ref{sec:qballs} and \ref{sec:thbounds}.}
\label{fig:fin_src_size_vs_thres_impact}
\end{figure*}
Using the lens equation Eq.\;\eqref{eq:lens_eq}, the angular position of the image of an arbitrary point on the edge of the circle can be determined. The lens equation in this scenario is given by
\begin{equation}\label{key}
\theta_{\text{\tiny S}}(\beta)=\theta_{\text{\tiny I}}(\beta)-\frac{d_{\text{\tiny LS}}}{d_{\text{\tiny S}}d_{\text{\tiny L}}}\frac{4G_{\text{\tiny N}}\widetilde{M}_{\text{\tiny Q}}(\theta_{\text{\tiny I}}(\beta))}{c^2\theta_{\text{\tiny I}}(\beta)}\;.
\end{equation}
We will numerically solve the lens equation for  $0\le\beta<2\pi$ for the given source position to obtain the image of the edge of the source and, like in the case of the point lens case, we can have multiple images. 

Assuming uniform intensity across the source, the magnification produced by the image is the ratio of the image area and the source area
\begin{equation}\label{eq:fin_src_mag_exp}
\mathfrak{m}_{\text{\tiny I}}=\frac{(-1)^P d_{\text{\tiny L}}^2}{2\pi \widetilde{R}_{\text{\tiny S}}^2}\int_{\beta=0}^{\beta=2 \pi}d\Psi(\beta )~\ \theta_{\text{\tiny I}}(\beta)^2 \; .
\end{equation}
Here,  $\Psi(\beta)=\tan^{-1}\left(\widetilde{R}_{\text{\tiny S}}\sin\beta/(d_{\text{\tiny L}}\theta_{\text{\tiny S,0}}+\widetilde{R}_{\text{\tiny S}}\cos\beta)\right)$, and $(-1)^P=\text{sign}(d^2\theta_{\text{\tiny I}}/d^2\theta_{\text{\tiny S}})\vert_{\beta=\pi} $ is the parity of the image. The total magnification is sum of individual magnification by each image. 
	
We numerically compute the total magnification produced by thin-wall Q-balls, and the characteristic curves are as shown in Fig.\,\ref{fig:fin_src_mag_vs_imp}. For $\theta_{\text{\tiny S,0}} \to 0$, we get finite magnification for the extended source, which decreases as the source size increases, as anticipated.  For $\theta_{\text{\tiny S,0}}/\theta_{\text{\tiny E}}\le \widetilde{R}_{\text{\tiny S}}/ R_{\text{\tiny E}}$, the magnification gradually decreases, after which there is steeper decrease in magnification. Also, for large $\theta_{\text{\tiny S,0}}/\theta_{\text{\tiny E}}$ and given source size, the magnification coincides with the corresponding point lens values. Furthermore, unlike the point source scenario, due to the finite extent of the source the number of solutions for images changes continuously as we increase the source position. Hence, there is no "caustic" in the magnification curve. BTW Q-ball magnification curves are also discovered to exhibit similar qualitative characteristics.
\begin{figure*}
	\begin{center}		\includegraphics[scale=0.945]{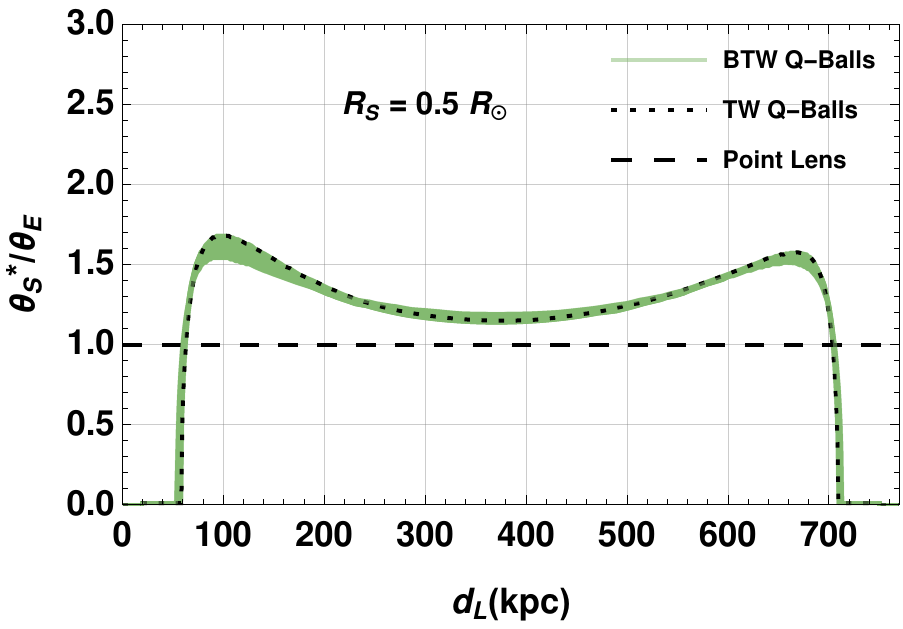}
		\includegraphics[scale=0.945]{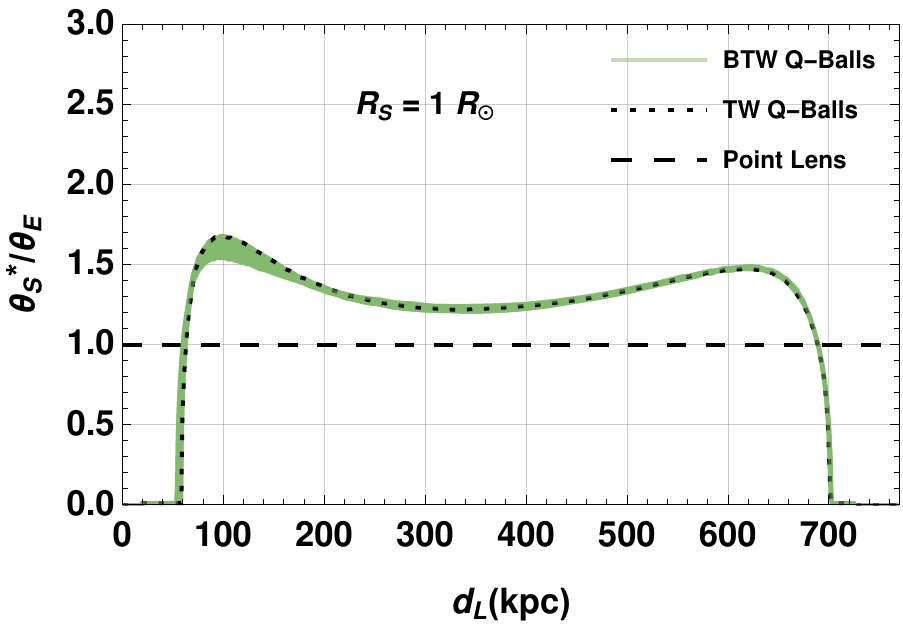}
		\includegraphics[scale=0.945]{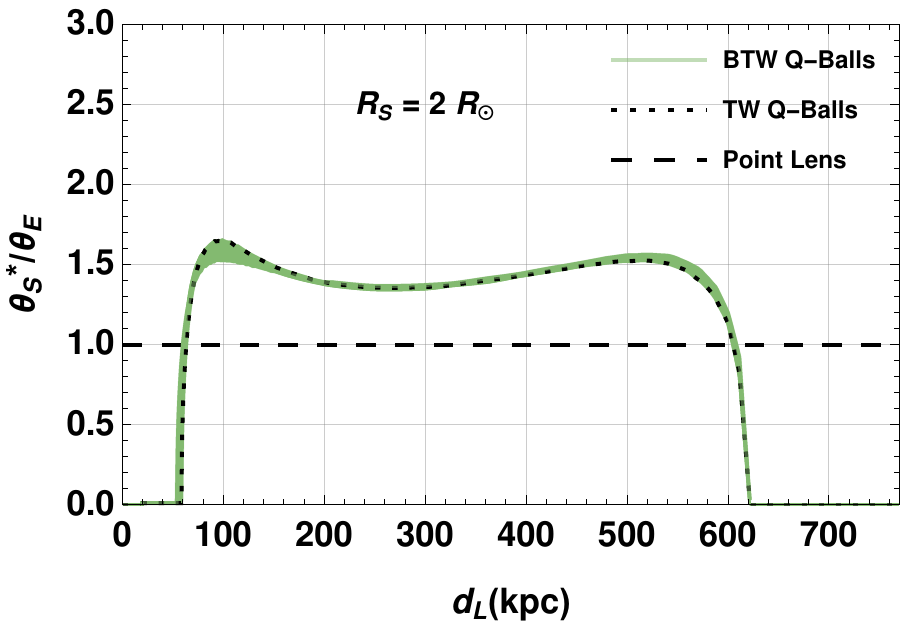}
	\end{center}	
	\caption{The threshold $\theta_{\text{\tiny S,0}}^*$   are shown as a function of the lens distance $d_{\text{\tiny L}}$. The lensing source objects are located in Andromeda galaxy (M31), with the source size $\widetilde{R}_{\text{\tiny S}}=0.5\,R_\odot,\,1\,R_\odot$ and $2\,R_\odot$. The plots have been made for $M_\text{\tiny {Q}}=10^{-8}~ \solarmass $  and $R_\text{\tiny {Q}}= 1 ~\solarrad $. The point lens and TW Q-balls are again represented by dashed and dotted black curves, respectively.  The green band, as in Fig.\,\,\ref{fig:size_vs_thres_impact}, represents various astrophysical Q-ball configurations; all with similar mass and radii but distinct density profiles.  We have again taken $\mathfrak{m}^*\in [1.33, 1.35]$ and have restricted $\sqrt{(m^2-\omega_{\text{\tiny Q}}^2)/(\omega^2-\omega_{\text{\tiny Q}}^2)}\lesssim 0.5$ based on field-theoretic stability.}
	\label{fig:fin_src_thres_imp_vs_dl}
\end{figure*}

For the finite source case, the threshold values of the source positions are shown in   Fig.\,\ref{fig:fin_src_size_vs_thres_impact}. Here, similar to the point source case, we have computed and plotted the characteristic value of $\theta_{\text{\tiny S,0}}$, which would give a magnification in the range $\mathfrak{m}^*=1.34\pm0.01$; now as a function of the ratio $R_{\text{\tiny {Q}}}/R_{\text{\tiny E}}$ and $\widetilde{R}_{\text{\tiny {S}}}/R_{\text{\tiny E}}$. The plots are shown for the TW Q-Ball profile and the BTW profile with $\alpha =\{3,~2\} $, satisfying the criteria of Secs.\,\ref{sec:qballs} and \ref{sec:thbounds}. For $\widetilde{R}_{\text{\tiny {S}}}/R_{\text{\tiny E}}\to0$, we have verified that the threshold behavior reproduces the point source. Additionally, for $R_{\text{\tiny {Q}}}/(R_{\text{\tiny E}} \sqrt{3/2})\lesssim 1.5$ as $\widetilde{R}_{\text{\tiny {S}}}/R_{\text{\tiny E}}$ increases, value of $\theta_{\text{\tiny S,0}}^*/\theta_{\text{\tiny E}}$ also increases. This happens because for a given $\theta_{\text{\tiny S,0}}/\theta_{\text{\tiny E}}$, the lower part of the finite source is more magnified than the upper part, leading to an overall larger magnification which increases as source size increases. This aspect may be observed from Fig.\,\ref{fig:fin_src_mag_vs_imp}. For large   $R_{\text{\tiny {Q}}}/(R_{\text{\tiny E}}\sqrt{3/2})  $  and $\widetilde{R}_{\text{\tiny {S}}}/R_{\text{\tiny E}}$, the magnification produced by lens is always less than $1.34$, and the corresponding region is colored purple in Fig.\,\ref{fig:fin_src_size_vs_thres_impact}. This is expected because, for large $R_{\text{\tiny {Q}}}/R_{\text{\tiny E}}$, the Q-ball lens is too diffused and for large $\widetilde{R}_{\text{\tiny {S}}}/R_{\text{\tiny E}}$ the Q-ball lens is too small to produce a threshold magnification.

For the Andromeda galaxy (M31) located $770$ kpc from earth, we plot the variation of $\theta_{\text{\tiny S}}^*$ as a function of the lens position $d_{\text{\tiny L}}$.  This is shown in Fig.\,\ref{fig:fin_src_thres_imp_vs_dl}. As mentioned before in the context of point sources, in Sec.\,\ref{subsec:twbtwlensing}, the information on $\theta_{\text{\tiny S}}^*$ is crucial for estimating microlensing event rates for physically feasible astrophysical Q-ball populations; for various astrophysical Q-ball configurations as well as source sizes. In Fig.\,\ref{fig:fin_src_thres_imp_vs_dl}, as before, the band represents various viable astrophysical Q-ball configurations satisfying the criteria of Secs.\,\ref{sec:qballs} and \ref{sec:thbounds}, with identical masses and radii but differing density profiles. The mass and radius of the Q-ball have been taken to be about $ 10^{-8}\,\solarmass $  and $ 1\,\solarrad $. The plots are computed for $\widetilde{R}_{\text{\tiny S}}=0.5\,R_\odot ,\,1\,R_\odot$ and $2\,R_\odot$. Unlike the point source case in Fig.\,\ref{fig:thres_imp_vs_dl}, the band is not symmetric around the midpoint of the source distance. This is happening because of the linear dependence of $\widetilde{R}_{\text{\tiny S}}$ with the lens position, which breaks the symmetry. Furthermore, when lens is too close to the source or the source size is large, $\widetilde{R}_{\text{\tiny S}}/ {R}_{\text{\tiny E}}$ becomes huge and it is unable able to generate the requisite threshold magnification.

In the next section, we place constraints on the population of astrophysical Q-balls, assuming them to form a component of the dark matter sector, by leveraging the observations from microlensing surveys like EROS-2, OGLE-IV, HSC-Subaru,  and WFIRST (proposed, future survey). 

\subsection{Constraints on astrophysical Q-balls from microlensing surveys}\label{subsec:microlensing_bounds}

If astrophysical Q-balls exist and may be forming a small fraction of the missing matter in the universe, they may be detected or at least constrained by gravitational microlensing surveys. In this section, we wish to study bounds on the fraction of dark matter that may, in principle, be in the form of Q-balls. 

The number of microlensing events induced by astrophysical Q-balls will be directly proportional to the number density in the line of sight of the observer and the source. Therefore, the idea is that depending on the number of `anomalous' events observed in a microlensing survey, we may hope to put crude upper bounds on the population of astrophysical Q-balls in the vicinity of our galaxy. We will adapt the methodologies expounded in \,\cite{Croon:2020wpr,Bai:2020jfm} for studying our particular cases of interest. If the dark matter density at a distance $d_{\text{\tiny L}}$ from observer is $\rho_{\text{DM}}(d_{\text{\tiny L}})$, then we may write the fraction of that density contained in Q-balls as $\rho^{\text{\tiny {Q}}}_{\text{DM}}(d_{\text{\tiny L}})=f_{\text{DM}}\,\rho_{\text{DM}}(d_{\text{\tiny L}})$. Here, $f_{\text{DM}}$ is the fraction of total dark matter contained in the form of Q-balls. This is the quantity we hope to put constraints on.

For a single background source, we may quantify the rate of events, assuming unit exposure time, from the differential event rate expression\,\cite{Griest:1990vu}
\begin{equation}\label{eq:diff_event_rate}
\frac{d^2\Gamma}{d\gamma d\tau}=\frac{2d_{\text{\tiny S}}e(\tau)}{v_\odot^2M_{\text{\tiny {Q}}}}f_{\text{DM}}\rho_{\text{DM}}(\gamma)v_\text{\tiny {Q}}^4(\gamma)e^{-v_\text{\tiny {Q}}^2(\gamma)/v_\odot^2}\; .
\end{equation}
We have defined $\gamma\equiv d_{\text{\tiny L}}/d_{\text{\tiny S}}$ and $v_\odot=2.2\times 10^5 \mathrm{m\,s}^{-1}$, which is the circular speed of the solar system around the galactic centre. In the above expression, $\tau$ is the transit time for a Q-ball to pass through the ``lensing tube''. The latter is defined as the volume of space between the observer and the source,
constructed by cylinders of infinitesimal thickness and radius $\theta_{\text{\tiny S}}^*d_{\text{\tiny L}}$. This basically means that as long as the Q-ball resides inside the lensing tube, it produces a magnification greater than the threshold value $\mathfrak{m}^*$. Here, $\theta_{\text{\tiny S}}^*$ is the threshold source position as a function of $d_\text{\tiny L}/d_\text{\tiny S}$ , similar to Fig.\,\ref{fig:thres_imp_vs_dl}. $e(\tau)$ is the efficiency of detection, $M_{\text{\tiny {Q}}}$ is the mass of the astrophysical Q-ball and $v_{\text{\tiny {Q}}}\equiv 2\theta_{\text{\tiny S}}^*(\gamma)R_{\text{\tiny E}}(\gamma)/\theta_{\text{\tiny E}}\tau$ is the velocity of the astrophysical Q-ball. The above expression is derived assuming that dark matter and astrophysical Q-ball velocities follow a simple Maxwell-Boltzmann distribution\,\cite{Griest:1990vu}. For the dark matter distribution concerned, we assume an isothermal profile\,\cite{Cirelli:2010xx}
\begin{equation}\label{}
\rho_{\text{\tiny DM}}(r)=\frac{\rho_c}{1+(r/r_c)^2}
\end{equation}
where $\rho_c=1.4\,\mathrm{GeV}/\mathrm{cm}^3$ is the dark matter core density and $r_c=4.4\,\mathrm{kpc}$ quantifies the size of the dark matter core. For the HSC-Subaru case, we consider the contribution of the M31 halo dark matter distribution. Assuming again an isothermal profile, we take $\rho_c=0.66\,\mathrm{GeV}/\mathrm{cm}^3$ and $r_c=6.3\,\mathrm{kpc}$\,\cite{Boshkayev:2022vpn}.
\begin{figure*}
	\begin{center}
	\includegraphics[scale=1]{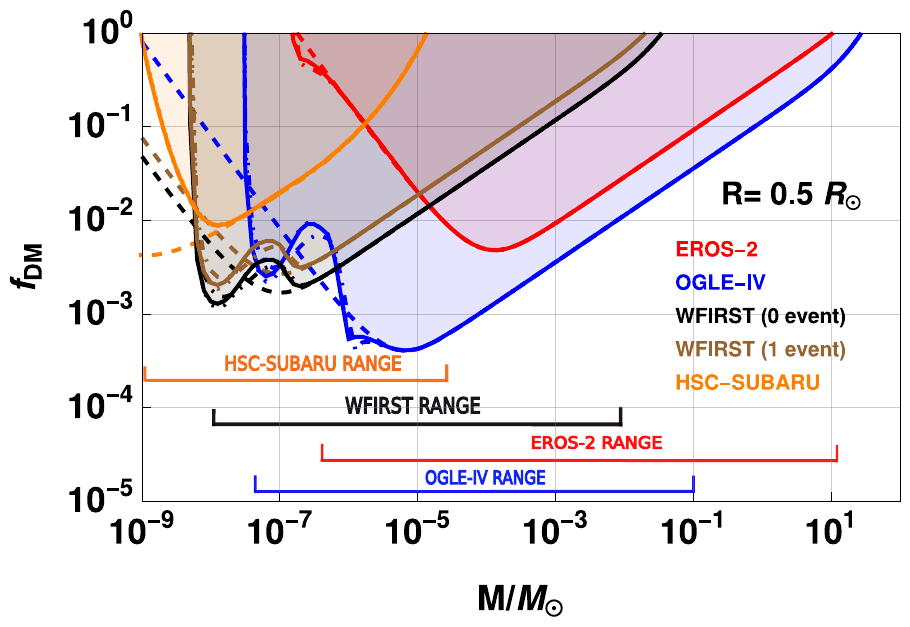}
	\includegraphics[scale=1]{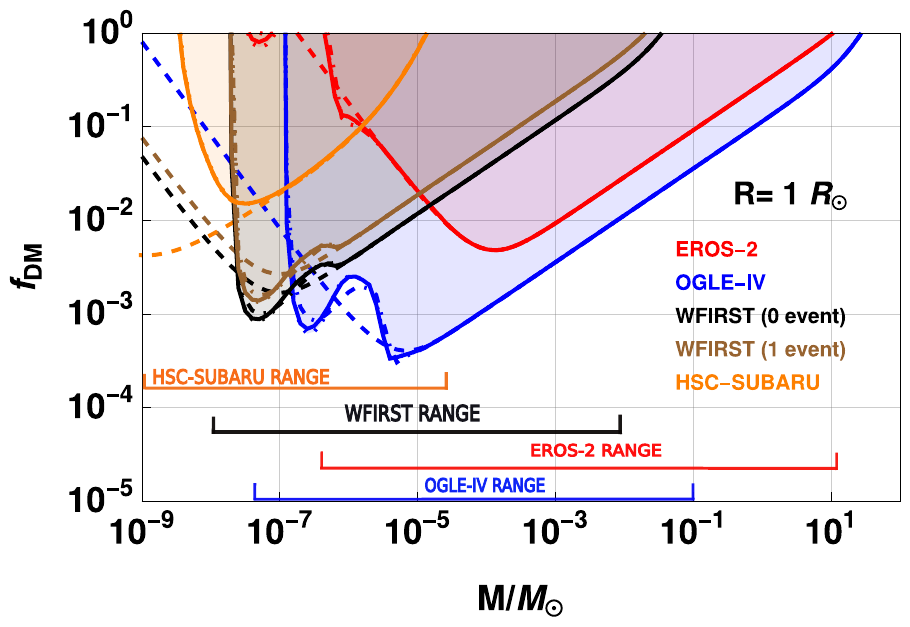}
	\end{center}
	\begin{center}
	\includegraphics[scale=1]{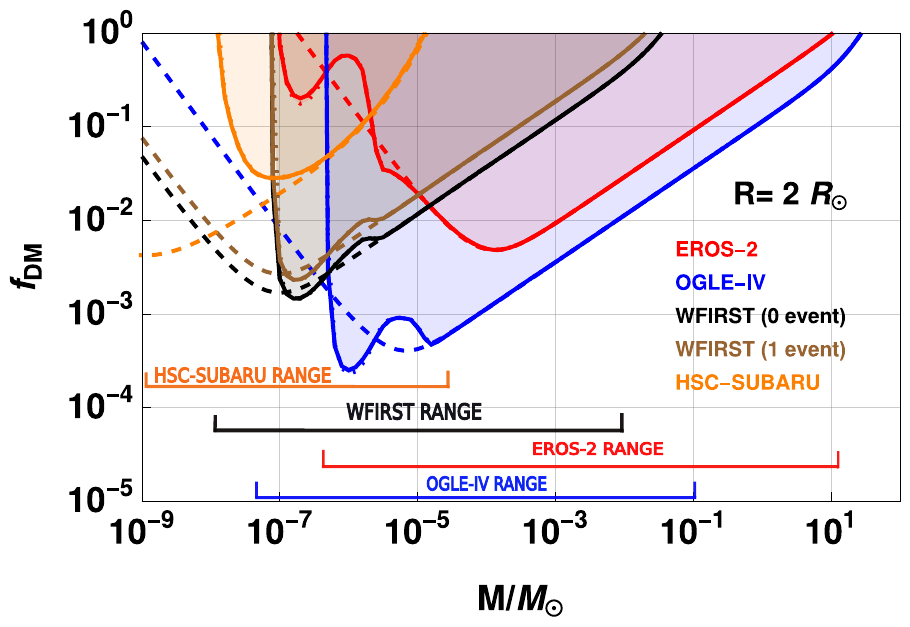}	
	\end{center}
	\caption{Bounds on the fraction of dark matter comprised of astrophysical Q-balls are shown from various microlensing surveys. The point lens and TW astrophysical Q-balls are shown by the dashed and dotted lines, respectively. While the solid and dashed-dotted lines represent the two BTW Q-balls. The three profiles have the same radius $R_{\text{\tiny Q}}$ in each of the cases. As in Fig.\,\ref{fig:comp_TW_BTW} we have taken $\alpha=\{10^4,3,2\}$. From top to bottom, the constraints are given for astrophysical Q-balls of sizes $0.5\,R_\odot, 1\,R_\odot$ and $2\,R_\odot$ respectively; again with the solar radius $R_{\odot}=6.9\times10^8\,\mathrm{m}$. We note that inspite of having quantitatively disntinct individual magnification signatures, the BTW Q-balls and TW Q-ball have almost the same $f_{\text{DM}}$ bound. The $M/M_\odot$ ranges adopted are accommodating regions where we have verified the existence of astrophysical Q-balls, following the conditions derived in Secs.\,\ref{sec:qballs} and \ref{sec:thbounds}. We have taken $M_\odot=1.98\times10^{30}\,\mathrm{kg}$.}
	\label{fig:fdm_cons}
\end{figure*}

We may now integrate Eq.\eqref{eq:diff_event_rate}, along the line joining the source and the observer and between a range of astrophysical Q-ball transit times. This gives the total number of Q-ball microlensing events registered ($\eta$) for a single source and for unit observation time
\begin{equation}\label{eq:tot_event_rate}
\eta=\int_0^1 d\gamma
\int_{\tau_{\text{min}}}^{\tau_{\text{max}}}d\tau\frac{d^2\Gamma}{d\gamma d\tau}\;.
\end{equation}
In the HSC-Subaru microlensing case, the events rates may be calculated using
\begin{equation}\label{eq:tot_event_rate_Sub}
	\eta_{\text{\tiny Sub.}}=\int_0^1d\gamma \int  \frac{dn_{\text{\tiny S}}}{dR_{\text{\tiny S}}}
	\int_{\tau_{\text{min}}}^{\tau_{\text{max}}}d\tau\frac{d^2\Gamma}{d\gamma d\tau}\;.
\end{equation}
Here, we have used the normalised stellar radius distribution ($dn_{\text{\tiny S}}/dR_{\text{\tiny S}}$) of the source stars in M31 derived in \cite{Smyth:2019whb}, using the Panchromatic Hubble Andromeda Treasury star catalogue\,\cite{Williams_2014,Dalcanton_2012} and the Mesa Isochrones and Stellar Tracks stellar evolution package\,\cite{choi2016mesa,dotter2016mesa}. 

Multiplying $\eta$ with the total number of sources in a survey ($N_{s}$) and the total observation time ($T_{o}$), one obtains the total number of expected astrophysical Q-ball events
\begin{equation}\label{eq:constraint_cond}
N_{\text{\tiny exp}}=\eta N_{s} T_{o} \;.
\end{equation}
Here, we emphasize that $\eta$ indirectly contains information about the astrophysical Q-ball density profile distribution through $\theta_{\text{\tiny S}}^*(\gamma)$. For an astrophysical Q-ball with fixed radius and mass, $\theta_{\text{\tiny S}}^*/\theta_{\text{\tiny E}}$ still varies with $\gamma$; as we already observed in Fig.\,\ref{fig:thres_imp_vs_dl}, for instance. This may be partly understood from Eq.\,\eqref{eq:ein_rad}. This information goes into Eq.\,\eqref{eq:diff_event_rate} through $v_{\text{\tiny {Q}}}$ and therefore in putting the constraints for Q-balls of different radii and mass. Specifically, the dependence is through relations similar to those encapsulated in Fig.\,\ref{fig:thres_imp_vs_dl}.

We will leverage the gravitational microlensing surveys EROS-2\,\cite{Survey_EROS} and OGLE-IV\,\cite{Survey_OGLEIV_1,Survey_OGLEIV_2}, HSC-Subaru\,\cite{Aihara:2017tri}, and the proposed survey WFIRST\,\cite{Green:2012mj, Survey_WFIRST} to put bounds on the astrophysical Q-ball fraction ($f_{\text{DM}}$). We will place limits on the population of astrophysical Q-balls in the vicinity of our Milky Way based on the number of anomalous events detected (for EROS-2 and OGLE-IV) or assuming future detection (for WFIRST).

EROS-2\,\cite{Survey_EROS} is pointed at the Large and Small Magellanic Clouds, which are located at a distance of $50\,\mathrm{kpc}$ and $60\,\mathrm{kpc}$ from the Sun, respectively. The total number of source stars in the survey is $N_s=5.49\times 10^6$\,\cite{Wenger:2000sw}. The total observation time was $T_o=2500\,\mathrm{days}$, with the accommodated transit times ranging from a day up to a thousand days. We have taken the efficiency $e(\tau)=0.24$, which is the time average over all transit periods. It has detected one event that cannot be fully explained by background modeling. We have placed constraints on the astrophysical Q-ball constituents in dark matter, taking LMC as the source field.

The centre of the Milky Way contains a large number of source stars, around $N_{s}=4.88\times 10^7$. OGLE-IV\,\cite{Survey_OGLEIV_1} takes its source field as the Milky Way bulge, which is at a distance of $8.5\,\mathrm{kpc}$ from the Sun\,\cite{Wenger:2000sw}. OGLE-IV has observed more than $2500$ events in its observation period of $T_o=1826\,\mathrm{days}$. We have taken the efficiency $e(\tau)=0.1$, which is the time average over all transit periods. Assuming these events to be due to stellar structure, we put conservative constraints on the population of astrophysical Q-balls in the Milky Way.

HSC-Subaru\,\cite{Aihara:2017tri} is looking at our neighbor Andromeda(M31) galaxy, located at distance $770$ kpc. It hosts $N_{s}=8.7 \times 10^7$ number of source stars for the survey. The total observation time is taken as $T_{o}= 7~ \mathrm{hour}$. One anomalous event\;\cite{Niikura:2017zjd} has been detected by the survey. Taking the efficiency $e(\tau)=0.5$, which is the time average over all transit periods, we put conservative constraints on the population of astrophysical Q-balls, present across Milky way till the center of the Andromeda galaxy. 

Proposed, future microlensing surveys, like WFIRST, will enable us to look for astrophysical Q-balls in ever smaller mass ranges. The proposed WFIRST mission expects to adopt the Milky Way Bulge as well as Magellanic clouds as its source field. It will scan the sky in intervals spanning $72\,\mathrm{days}$\,\cite{Survey_WFIRST}. We have taken the efficiency $e(\tau)=0.5$. As we mentioned, it will be most sensitive to very low sub-solar mass range objects. We have put constraints on the population of low-mass Q-balls---assuming it detects one or zero anomalous events---taking LMC as the source field.

While counting events, we assume  Poisson statistics for the distribution of microlensing events. Therefore, if the actual number of anomalous events detected by one of these surveys is $N_{\text{obs}}$, say, then assuming Poisson distribution for these events, we may calculate the expected number $N_{\text{exp}}$ as
\begin{equation}\label{Poisson cond}
\sum_{k=0}^{N_{\text{obs}}}P(k,N_{\text{exp}})=0.05\;.
\end{equation}  
Here, $P(k,N_{\text{exp}})=N_{\text{exp}}^k e^{-N_{\text{exp}}}/k!$ denotes the probability of observing $k$ events for a fixed $N_{\text{exp}}$. So if we have already observed $N_{\text{obs}}$ events then by virtue of Eq.\eqref{Poisson cond} we can exclude the values of $N_{\text{exp}}$ such that $\sum_{k=0}^{N_{\text{obs}}}P(k,N_{\text{exp}})<0.05$ with $95\%$ C.L. This puts an upper bound on $N_{\text{exp}}$ which can be used to put the upper bound on $f_{\text{DM}}$ as function of the lens mass through Eq.\eqref{eq:constraint_cond}. We have performed this analysis for the surveys mentioned above and our results are shown in Fig.\,\ref{fig:fdm_cons}.

The limits on three astrophysical Q-ball profiles, the same as those displayed in Fig.\,\ref{fig:comp_TW_BTW}, are shown in Fig.\,\ref{fig:fdm_cons}. As a further check of our methodology and analysis pipeline, we have verified that it reproduce the limits for some of the representative density profiles in\,\cite{Croon:2020wpr,Croon:2020ouk}. In each case that we show, the astrophysical Q-ball radii are the same. The profiles are chosen to have $\alpha=\{10^4,3,2\}$, where $\alpha\equiv\sqrt{(m^2-\omega_{\text{\tiny Q}}^2)/(\omega^2-\omega_{\text{\tiny Q}}^2)}$. It is worth noting that for sub-earth mass astrophysical Q-balls, very minimal variance exists in the constrained $f_{\text{DM}}$, for various BTW Q-ball profiles. This is a consequence of the fact that the area under the curve in Fig.\,\ref{fig:thres_imp_vs_dl} is almost the same for the different profiles. This renders the contribution from the lensing tube integration to the number of microlensing events to be almost equal. There are also no significant differences in the $f_{\text{DM}}$ bounds for higher-mass TW and BTW astrophysical Q-balls. This is because, for large-mass Q-balls (at fixed radii), the ratio $R_\text{\tiny {Q}}/R_{\text{\tiny E}}\ll1$ and hence all Q-ball profiles start showing point lens behaviour. This is clearly visible from Fig.\,\ref{fig:size_vs_thres_impact}, for instance. Therefore, one concludes that despite their quantitatively unique individual magnification characteristics, as far as the $f_{\text{DM}}$ bounds are concerned, the various astrophysical Q-balls generate almost identical bounds.

 We observe from Fig.\,\ref{fig:fdm_cons} that the strictest limits on solar mass astrophysical Q-balls may be placed from OGLE-IV. Also, for $R_{\text{\tiny {Q}}}\sim R_\odot$, we see that for the mass range $10^{-5}-10^{-4}M_\odot$, the astrophysical Q-ball population is restricted to being within about $0.1\%$ of the total dark matter. As we move away from $10^{-5}M_\odot$, the $f_{\text{DM}}$ constraints weaken. For low mass Q-balls within a mass range $10^{-9} M_\odot<M_{\text{\tiny {Q}}}<10^{-7}M_\odot$, and $R_{\text{\tiny Q}}\lesssim R_\odot$, HSC-Subaru  and the proposed WFIRST place the best constraints on $f_{\text{DM}}$. Furthermore, for mass range $10^{-7} M_\odot<M_{\text{\tiny {Q}}}<10^{-5}M_\odot$, we observe that WFIRST and OGLE-IV put stricter bounds than EROS-2 and HSC-Subaru when $R_{\text{\tiny {Q}}}<R_\odot$. This trend could also have been already speculated based on the characteristics observed in Fig.\,\ref{fig:thres_imp_vs_dl}. For $R_{\text{\tiny {Q}}}>R_\odot$, in the vicinity of $10^{-7} M_\odot$, HSC-Subaru currently gives better limits than EROS-2, and OGLE-IV. WFIRST can potentially constrain this region very strongly in the future. For mass range $10^{-5}M_\odot-10^{-2}M_\odot$, OGLE-IV and EROS-2 put more strict bounds on $f_{\text{DM}}$ as compared to WFIRST. This is expected as WFIRST is designed to detect earth-mass astrophysical objects. Furthermore, for $R_{\text{\tiny Q}}\sim R_{\odot}$, HSC-Subaru doesn't place any significant bounds for mass ranges greater than $ 10^{-5} \solarmass$. We also observe that for astrophysical Q-ball mass ranges $10^{-1}\,M_\odot-10\,M_\odot$, one cannot put any really strong restrictive bounds on $f_{\text{DM}}$, from the microlensing surveys considered. In these regions, therefore, astrophysical Q-balls, if they exist, could abound.

\section{Summary and conclusions}
\label{sec:summary}
In this paper, we have explored a few aspects of non-topological solitonic Q-ball solutions when such configurations may form astrophysically viable structures in the universe. We focused on theoretical aspects related to their viability, as well as investigated their gravitational lensing features for different astrophysical Q-ball profiles. Speculating on what fraction of dark matter may be comprised in Q-balls, we derived constraints from microlensing surveys in the viable ranges. Apart from Figs.\,\ref{fig:mag_vs_impact_TW}-\ref{fig:fdm_cons}, let us summarise below some of the main results.

In the context of the theoretical bounds, we deduced novel limits on the size of astrophysical Q-balls in Eqs.\,\eqref{eq:rmin_qball} and \eqref{eq:qball_dj}. These were obtained from the analytic form of the Q-ball radius and by considerations of gravitational instability, respectively. In the same context, while exploring the Jeans' criteria for astrophysical Q-balls, we found a new lower bound in Eq.\,\eqref{eq:omega_jeanslimit} on the Lagrange multiplier, which is akin to a chemical potential in the Q-ball context; as may be motivated from Eq.\,\eqref{eq:debydq}. Also, these limits on the astrophysical Q-ball radii provided new inequalities for the Lagrangian parameters, as encapsulated by Eq.\,\eqref{eq:jeans_cond}.

Through the study of gravitational lensing features by different Q-ball profiles, we noted distinct characteristics as the profiles were varied. We observed, for instance, in Fig.\,\ref{fig:comp_TW_BTW} that thin-wall and beyond-thin-wall Q-balls, having roughly the same mass and size, exhibit different magnification profiles. This suggests that the microlensing light curve will also be distinct and may help differentiate profiles, provided sufficient anomalous events are observed in the future, if astrophysical Q-balls indeed exist in the universe. In Figs.\,\ref{fig:size_vs_thres_impact},\,\ref{fig:thres_imp_vs_dl},\,\ref{fig:fin_src_size_vs_thres_impact}, and \ref{fig:fin_src_thres_imp_vs_dl},  we explored threshold angular source positions, leading to potentially detectable magnifications in surveys, for beyond-thin-wall Q-balls.

Finally, we derived gravitational microlensing constraints on astrophysical Q-balls---both for the case of thin wall and beyond-thin-wall profiles---using data from EROS-2, OGLE-IV, HSC-Subaru, and projections for the proposed WFIRST survey. The limits on $f_{\text{DM}}$, the fraction of dark matter that may be in the form of astrophysical Q-balls, are shown in Fig.\,\ref{fig:fdm_cons}. For the intermediate masses that we consider, it is seen that this fraction may at most be $\mathcal{O}(0.1\%)$, while for higher and lower masses the fraction may be much higher.

\begin{acknowledgments}
We would like to thank Ranjan Laha for discussions. AT would like to thank the organisers of the `Horizons in Accelerators, Particle/Nuclear Physics and Laboratory-based Quantum Sensors for HEP/NP' and `Particle Physics: Phenomena, Puzzles, Promises' ICTS conferences, as well as ICTS, Bengaluru, for their kind hospitality during the completion of parts of this work. AA and LB acknowledge support from a Senior Research Fellowship, granted by the Human Resource Development Group, Council of Scientific and Industrial Research, Government of India. AT acknowledges support from an Early Career Research award, from the Department of Science and Technology, Government of India.
\end{acknowledgments}

\bibliographystyle{unsrtnat}
\bibliography{AstrophysicalQBalls} 

\begin{thebibliography}{112}
\providecommand{\natexlab}[1]{#1}
\providecommand{\url}[1]{\texttt{#1}}
\expandafter\ifx\csname urlstyle\endcsname\relax
  \providecommand{\doi}[1]{doi: #1}\else
  \providecommand{\doi}{doi: \begingroup \urlstyle{rm}\Url}\fi

\bibitem[Bertone et~al.(2005)Bertone, Hooper, and Silk]{Bertone:2004pz}
Gianfranco Bertone, Dan Hooper, and Joseph Silk.
\newblock {Particle dark matter: Evidence, candidates and constraints}.
\newblock \emph{Phys. Rept.}, 405:\penalty0 279--390, 2005.
\newblock \doi{10.1016/j.physrep.2004.08.031}.

\bibitem[Feng(2010)]{Feng:2010gw}
Jonathan~L. Feng.
\newblock {Dark Matter Candidates from Particle Physics and Methods of
  Detection}.
\newblock \emph{Ann. Rev. Astron. Astrophys.}, 48:\penalty0 495--545, 2010.
\newblock \doi{10.1146/annurev-astro-082708-101659}.

\bibitem[Bergstrom(2009)]{Bergstrom:2009ib}
Lars Bergstrom.
\newblock {Dark Matter Candidates}.
\newblock \emph{New J. Phys.}, 11:\penalty0 105006, 2009.
\newblock \doi{10.1088/1367-2630/11/10/105006}.

\bibitem[Steffen(2009)]{Steffen:2008qp}
Frank~Daniel Steffen.
\newblock {Dark Matter Candidates - Axions, Neutralinos, Gravitinos, and
  Axinos}.
\newblock \emph{Eur. Phys. J. C}, 59:\penalty0 557--588, 2009.
\newblock \doi{10.1140/epjc/s10052-008-0830-0}.

\bibitem[Workman et~al.(2022)]{ParticleDataGroup:2022pth}
R.~L. Workman et~al.
\newblock {Review of Particle Physics}.
\newblock \emph{PTEP}, 2022:\penalty0 083C01, 2022.
\newblock \doi{10.1093/ptep/ptac097}.

\bibitem[Henriksen and Widrow(1995)]{Henriksen:1994ep}
Richard~N. Henriksen and Lawrence~M. Widrow.
\newblock {Hydrogen clouds and the MACHO / EROS events}.
\newblock \emph{Astrophys. J.}, 441:\penalty0 70, 1995.
\newblock \doi{10.1086/175336}.

\bibitem[Gerhard and Silk(1996)]{Gerhard:1995ff}
Ortwin Gerhard and Joseph Silk.
\newblock {Baryonic dark halos: MACHOs and cold gas?}
\newblock \emph{Astrophys. J.}, 472:\penalty0 34--45, 1996.
\newblock \doi{10.1086/178039}.

\bibitem[Bertone and Hooper(2018)]{Bertone:2016nfn}
Gianfranco Bertone and Dan Hooper.
\newblock {History of dark matter}.
\newblock \emph{Rev. Mod. Phys.}, 90\penalty0 (4):\penalty0 045002, 2018.
\newblock \doi{10.1103/RevModPhys.90.045002}.

\bibitem[Erickcek and Sigurdson(2011)]{Erickcek:2011us}
Adrienne~L. Erickcek and Kris Sigurdson.
\newblock {Reheating Effects in the Matter Power Spectrum and Implications for
  Substructure}.
\newblock \emph{Phys. Rev. D}, 84:\penalty0 083503, 2011.
\newblock \doi{10.1103/PhysRevD.84.083503}.

\bibitem[Fan et~al.(2013{\natexlab{a}})Fan, Katz, Randall, and
  Reece]{Fan:2013tia}
JiJi Fan, Andrey Katz, Lisa Randall, and Matthew Reece.
\newblock {Dark-Disk Universe}.
\newblock \emph{Phys. Rev. Lett.}, 110\penalty0 (21):\penalty0 211302,
  2013{\natexlab{a}}.
\newblock \doi{10.1103/PhysRevLett.110.211302}.

\bibitem[Fan et~al.(2013{\natexlab{b}})Fan, Katz, Randall, and
  Reece]{Fan:2013yva}
JiJi Fan, Andrey Katz, Lisa Randall, and Matthew Reece.
\newblock {Double-Disk Dark Matter}.
\newblock \emph{Phys. Dark Univ.}, 2:\penalty0 139--156, 2013{\natexlab{b}}.
\newblock \doi{10.1016/j.dark.2013.07.001}.

\bibitem[Barenboim and Rasero(2014)]{Barenboim:2013gya}
Gabriela Barenboim and Javier Rasero.
\newblock {Structure Formation during an early period of matter domination}.
\newblock \emph{JHEP}, 04\penalty0 (0):\penalty0 138, 2014.
\newblock \doi{10.1007/JHEP04(2014)138}.

\bibitem[Fan et~al.(2014)Fan, \"Ozsoy, and Watson]{Fan2014}
JiJi Fan, Ogan \"Ozsoy, and Scott Watson.
\newblock Nonthermal histories and implications for structure formation.
\newblock \emph{Phys. Rev. D}, 90:\penalty0 043536, Aug 2014.
\newblock \doi{10.1103/PhysRevD.90.043536}.
\newblock URL \url{https://link.aps.org/doi/10.1103/PhysRevD.90.043536}.

\bibitem[Wise and Zhang(2014)]{Wise:2014jva}
Mark~B. Wise and Yue Zhang.
\newblock {Stable Bound States of Asymmetric Dark Matter}.
\newblock \emph{Phys. Rev. D}, 90\penalty0 (5):\penalty0 055030, 2014.
\newblock \doi{10.1103/PhysRevD.90.055030}.
\newblock [Erratum: Phys.Rev.D 91, 039907 (2015)].

\bibitem[Graham et~al.(2016)Graham, Mardon, and Rajendran]{Graham:2015rva}
Peter~W. Graham, Jeremy Mardon, and Surjeet Rajendran.
\newblock {Vector Dark Matter from Inflationary Fluctuations}.
\newblock \emph{Phys. Rev. D}, 93\penalty0 (10):\penalty0 103520, 2016.
\newblock \doi{10.1103/PhysRevD.93.103520}.

\bibitem[Fairbairn et~al.(2017)Fairbairn, Marsh, and
  Quevillon]{Fairbairn:2017dmf}
Malcolm Fairbairn, David J.~E. Marsh, and Jérémie Quevillon.
\newblock {Searching for the QCD Axion with Gravitational Microlensing}.
\newblock \emph{Phys. Rev. Lett.}, 119\penalty0 (2):\penalty0 021101, 2017.
\newblock \doi{10.1103/PhysRevLett.119.021101}.

\bibitem[Gresham et~al.(2018)Gresham, Lou, and Zurek]{Gresham:2017cvl}
Moira~I. Gresham, Hou~Keong Lou, and Kathryn~M. Zurek.
\newblock {Early Universe synthesis of asymmetric dark matter nuggets}.
\newblock \emph{Phys. Rev. D}, 97\penalty0 (3):\penalty0 036003, 2018.
\newblock \doi{10.1103/PhysRevD.97.036003}.

\bibitem[Dror et~al.(2018)Dror, Kuflik, Melcher, and Watson]{Dror:2017gjq}
Jeff~A. Dror, Eric Kuflik, Brandon Melcher, and Scott Watson.
\newblock {Concentrated dark matter: Enhanced small-scale structure from
  codecaying dark matter}.
\newblock \emph{Phys. Rev. D}, 97\penalty0 (6):\penalty0 063524, 2018.
\newblock \doi{10.1103/PhysRevD.97.063524}.

\bibitem[Grabowska et~al.(2018)Grabowska, Melia, and
  Rajendran]{Grabowska:2018lnd}
Dorota~M. Grabowska, Tom Melia, and Surjeet Rajendran.
\newblock {Detecting Dark Blobs}.
\newblock \emph{Phys. Rev. D}, 98\penalty0 (11):\penalty0 115020, 2018.
\newblock \doi{10.1103/PhysRevD.98.115020}.

\bibitem[Curtin and Setford(2020)]{Curtin:2019ngc}
David Curtin and Jack Setford.
\newblock {Signatures of Mirror Stars}.
\newblock \emph{JHEP}, 03:\penalty0 041, 2020.
\newblock \doi{10.1007/JHEP03(2020)041}.

\bibitem[Fox et~al.(2023)Fox, Weiner, and Xiao]{Fox:2023aat}
Patrick~J. Fox, Neal Weiner, and Huangyu Xiao.
\newblock {Recurrent Axinovae and their Cosmological Constraints}.
\newblock 2 2023.

\bibitem[Ruffini and Bonazzola(1969)]{Ruffini:1969qy}
Remo Ruffini and Silvano Bonazzola.
\newblock {Systems of selfgravitating particles in general relativity and the
  concept of an equation of state}.
\newblock \emph{Phys. Rev.}, 187:\penalty0 1767--1783, 1969.
\newblock \doi{10.1103/PhysRev.187.1767}.

\bibitem[Kling and Rajaraman(2017)]{Kling:2017mif}
Felix Kling and Arvind Rajaraman.
\newblock {Towards an Analytic Construction of the Wavefunction of Boson
  Stars}.
\newblock \emph{Phys. Rev. D}, 96\penalty0 (4):\penalty0 044039, 2017.
\newblock \doi{10.1103/PhysRevD.96.044039}.

\bibitem[Kling and Rajaraman(2018)]{Kling:2017hjm}
Felix Kling and Arvind Rajaraman.
\newblock {Profiles of boson stars with self-interactions}.
\newblock \emph{Phys. Rev. D}, 97\penalty0 (6):\penalty0 063012, 2018.
\newblock \doi{10.1103/PhysRevD.97.063012}.

\bibitem[Cardoso and Pani(2019)]{Cardoso:2019rvt}
Vitor Cardoso and Paolo Pani.
\newblock {Testing the nature of dark compact objects: a status report}.
\newblock \emph{Living Rev. Rel.}, 22\penalty0 (1):\penalty0 4, 2019.
\newblock \doi{10.1007/s41114-019-0020-4}.

\bibitem[Cardoso et~al.(2022)Cardoso, Macedo, Maeda, and
  Okawa]{Cardoso:2021ehg}
Vitor Cardoso, Caio F.~B. Macedo, Kei-ichi Maeda, and Hirotada Okawa.
\newblock {ECO-spotting: looking for extremely compact objects with bosonic
  fields}.
\newblock \emph{Class. Quant. Grav.}, 39\penalty0 (3):\penalty0 034001, 2022.
\newblock \doi{10.1088/1361-6382/ac41e7}.

\bibitem[Visinelli(2021)]{Visinelli:2021uve}
Luca Visinelli.
\newblock {Boson stars and oscillatons: A review}.
\newblock \emph{Int. J. Mod. Phys. D}, 30\penalty0 (15):\penalty0 2130006,
  2021.
\newblock \doi{10.1142/S0218271821300068}.

\bibitem[Bo\v{s}kovi\'c and Barausse(2022)]{Boskovic:2021nfs}
Mateja Bo\v{s}kovi\'c and Enrico Barausse.
\newblock {Soliton boson stars, Q-balls and the causal Buchdahl bound}.
\newblock \emph{JCAP}, 02\penalty0 (02):\penalty0 032, 2022.
\newblock \doi{10.1088/1475-7516/2022/02/032}.

\bibitem[Del~Grosso et~al.(2023)Del~Grosso, Franciolini, Pani, and
  Urbano]{DelGrosso:2023trq}
L.~Del~Grosso, G.~Franciolini, P.~Pani, and A.~Urbano.
\newblock {Fermion soliton stars}.
\newblock 1 2023.

\bibitem[Croon et~al.(2020{\natexlab{a}})Croon, McKeen, and Raj]{Croon:2020wpr}
Djuna Croon, David McKeen, and Nirmal Raj.
\newblock {Gravitational microlensing by dark matter in extended structures}.
\newblock \emph{Phys. Rev. D}, 101\penalty0 (8):\penalty0 083013,
  2020{\natexlab{a}}.
\newblock \doi{10.1103/PhysRevD.101.083013}.

\bibitem[Bai et~al.(2020)Bai, Long, and Lu]{Bai:2020jfm}
Yang Bai, Andrew~J. Long, and Sida Lu.
\newblock {Tests of Dark MACHOs: Lensing, Accretion, and Glow}.
\newblock \emph{JCAP}, 09:\penalty0 044, 2020.
\newblock \doi{10.1088/1475-7516/2020/09/044}.

\bibitem[Croon et~al.(2020{\natexlab{b}})Croon, McKeen, Raj, and
  Wang]{Croon:2020ouk}
Djuna Croon, David McKeen, Nirmal Raj, and Zihui Wang.
\newblock {Subaru-HSC through a different lens: Microlensing by extended dark
  matter structures}.
\newblock \emph{Phys. Rev. D}, 102\penalty0 (8):\penalty0 083021,
  2020{\natexlab{b}}.
\newblock \doi{10.1103/PhysRevD.102.083021}.

\bibitem[Fujikura et~al.(2021)Fujikura, Hertzberg, Schiappacasse, and
  Yamaguchi]{Fujikura:2021omw}
Kohei Fujikura, Mark~P. Hertzberg, Enrico~D. Schiappacasse, and Masahide
  Yamaguchi.
\newblock {Microlensing constraints on axion stars including finite lens and
  source size effects}.
\newblock \emph{Phys. Rev. D}, 104\penalty0 (12):\penalty0 123012, 2021.
\newblock \doi{10.1103/PhysRevD.104.123012}.

\bibitem[Giudice et~al.(2016)Giudice, McCullough, and Urbano]{Giudice:2016zpa}
Gian~F. Giudice, Matthew McCullough, and Alfredo Urbano.
\newblock {Hunting for Dark Particles with Gravitational Waves}.
\newblock \emph{JCAP}, 10:\penalty0 001, 2016.
\newblock \doi{10.1088/1475-7516/2016/10/001}.

\bibitem[Croon et~al.(2022)Croon, Ipek, and McKeen]{Croon:2022tmr}
Djuna Croon, Seyda Ipek, and David McKeen.
\newblock {Gravitational wave constraints on extended dark matter structures}.
\newblock 5 2022.

\bibitem[Marsh and Hoof(2022)]{Marsh:2022dqv}
David J.~E. Marsh and Sebastian Hoof.
\newblock \emph{{Astrophysical Searches and Constraints}}, pages 73--122.
\newblock 3 2022.
\newblock \doi{10.1007/978-3-030-95852-7_3}.

\bibitem[Drlica-Wagner et~al.(2022)]{Drlica-Wagner:2022lbd}
Alex Drlica-Wagner et~al.
\newblock {Report of the Topical Group on Cosmic Probes of Dark Matter for
  Snowmass 2021}.
\newblock 9 2022.

\bibitem[Green et~al.(2022)]{Green:2022hhj}
Daniel Green et~al.
\newblock {Snowmass Theory Frontier: Astrophysics and Cosmology}.
\newblock 9 2022.

\bibitem[Frieman et~al.(1988)Frieman, Gelmini, Gleiser, and
  Kolb]{Frieman:1988ut}
Joshua~A. Frieman, G.~B. Gelmini, Marcelo Gleiser, and Edward~W. Kolb.
\newblock {Solitogenesis: Primordial Origin of Nontopological Solitons}.
\newblock \emph{Phys. Rev. Lett.}, 60:\penalty0 2101, 1988.
\newblock \doi{10.1103/PhysRevLett.60.2101}.

\bibitem[Kusenko(1997)]{Kusenko:1997ad}
Alexander Kusenko.
\newblock {Small Q balls}.
\newblock \emph{Phys. Lett. B}, 404:\penalty0 285, 1997.
\newblock \doi{10.1016/S0370-2693(97)00582-0}.

\bibitem[Kusenko and Shaposhnikov(1998)]{Kusenko:1997si}
Alexander Kusenko and Mikhail~E. Shaposhnikov.
\newblock {Supersymmetric Q balls as dark matter}.
\newblock \emph{Phys. Lett. B}, 418:\penalty0 46--54, 1998.
\newblock \doi{10.1016/S0370-2693(97)01375-0}.

\bibitem[Lee and Pang(1992)]{Lee:1991ax}
T.~D. Lee and Y.~Pang.
\newblock {Nontopological solitons}.
\newblock \emph{Phys. Rept.}, 221:\penalty0 251--350, 1992.
\newblock \doi{10.1016/0370-1573(92)90064-7}.

\bibitem[Rosen(1968{\natexlab{a}})]{Rosen:1968mfz}
Gerald Rosen.
\newblock {Particlelike Solutions to Nonlinear Complex Scalar Field Theories
  with Positive-Definite Energy Densities}.
\newblock \emph{J. Math. Phys.}, 9:\penalty0 996, 1968{\natexlab{a}}.
\newblock \doi{10.1063/1.1664693}.

\bibitem[Rosen(1968{\natexlab{b}})]{Rosen:1968zwl}
Gerald Rosen.
\newblock {Charged Particlelike Solutions to Nonlinear Complex Scalar Field
  Theories}.
\newblock \emph{J. Math. Phys.}, 9\penalty0 (7):\penalty0 999--1002,
  1968{\natexlab{b}}.
\newblock \doi{10.1063/1.1664694}.

\bibitem[Coleman(1985)]{Coleman:1985ki}
Sidney~R. Coleman.
\newblock {Q-balls}.
\newblock \emph{Nucl. Phys. B}, 262\penalty0 (2):\penalty0 263, 1985.
\newblock \doi{10.1016/0550-3213(86)90520-1}.
\newblock [Addendum: Nucl.Phys.B 269, 744 (1986)].

\bibitem[Arias and Schaposnik(2014)]{Arias:2014tka}
Paola Arias and Fidel~A. Schaposnik.
\newblock {Vortex solutions of an Abelian Higgs model with visible and hidden
  sectors}.
\newblock \emph{JHEP}, 12:\penalty0 011, 2014.
\newblock \doi{10.1007/JHEP12(2014)011}.

\bibitem[Vachaspati(2009)]{Vachaspati:2009jx}
Tanmay Vachaspati.
\newblock {Dark Strings}.
\newblock \emph{Phys. Rev. D}, 80:\penalty0 063502, 2009.
\newblock \doi{10.1103/PhysRevD.80.063502}.

\bibitem[Long and Vachaspati(2014)]{Long:2014lxa}
Andrew~J. Long and Tanmay Vachaspati.
\newblock {Cosmic Strings in Hidden Sectors: 2. Cosmological and Astrophysical
  Signatures}.
\newblock \emph{JCAP}, 12:\penalty0 040, 2014.
\newblock \doi{10.1088/1475-7516/2014/12/040}.

\bibitem[Long et~al.(2014)Long, Hyde, and Vachaspati]{Long:2014mxa}
Andrew~J. Long, Jeffrey~M. Hyde, and Tanmay Vachaspati.
\newblock {Cosmic Strings in Hidden Sectors: 1. Radiation of Standard Model
  Particles}.
\newblock \emph{JCAP}, 09:\penalty0 030, 2014.
\newblock \doi{10.1088/1475-7516/2014/09/030}.

\bibitem[Hyde et~al.(2014)Hyde, Long, and Vachaspati]{Hyde:2013fia}
Jeffrey~M. Hyde, Andrew~J. Long, and Tanmay Vachaspati.
\newblock {Dark Strings and their Couplings to the Standard Model}.
\newblock \emph{Phys. Rev. D}, 89:\penalty0 065031, 2014.
\newblock \doi{10.1103/PhysRevD.89.065031}.

\bibitem[Hindmarsh et~al.(2015)Hindmarsh, Kirk, No, and
  West]{Hindmarsh:2014zba}
Mark Hindmarsh, Russell Kirk, Jose~Miguel No, and Stephen~M. West.
\newblock {Dark Matter with Topological Defects in the Inert Doublet Model}.
\newblock \emph{JCAP}, 05:\penalty0 048, 2015.
\newblock \doi{10.1088/1475-7516/2015/05/048}.

\bibitem[Hartmann and Arbabzadah(2009)]{Hartmann:2009ki}
Betti Hartmann and Farhad Arbabzadah.
\newblock {Cosmic strings interacting with dark strings}.
\newblock \emph{JHEP}, 07:\penalty0 068, 2009.
\newblock \doi{10.1088/1126-6708/2009/07/068}.

\bibitem[Forg\'acs and Luk\'acs(2016)]{Forgacs:2016iva}
P\'eter Forg\'acs and \'Arp\'ad Luk\'acs.
\newblock {Vortices and magnetic bags in Abelian models with extended scalar
  sectors and some of their applications}.
\newblock \emph{Phys. Rev. D}, 94\penalty0 (12):\penalty0 125018, 2016.
\newblock \doi{10.1103/PhysRevD.94.125018}.

\bibitem[Forg\'acs and Luk\'acs(2020)]{Forgacs:2019tbn}
P\'eter Forg\'acs and \'Arp\'ad Luk\'acs.
\newblock {Electroweak strings with dark scalar condensates and their
  stability}.
\newblock \emph{Phys. Rev. D}, 102\penalty0 (2):\penalty0 023009, 2020.
\newblock \doi{10.1103/PhysRevD.102.023009}.

\bibitem[Brihaye and Hartmann(2009)]{Brihaye:2009fs}
Yves Brihaye and Betti Hartmann.
\newblock {The Effect of dark strings on semilocal strings}.
\newblock \emph{Phys. Rev. D}, 80:\penalty0 123502, 2009.
\newblock \doi{10.1103/PhysRevD.80.123502}.

\bibitem[Babeanu and Hartmann(2012)]{Babeanu:2011ie}
Alexandru Babeanu and Betti Hartmann.
\newblock {Stability of superconducting strings coupled to cosmic strings}.
\newblock \emph{Phys. Rev. D}, 85:\penalty0 023518, 2012.
\newblock \doi{10.1103/PhysRevD.85.023518}.

\bibitem[Cohen et~al.(1986)Cohen, Coleman, Georgi, and Manohar]{Cohen:1986ct}
Andrew~G. Cohen, Sidney~R. Coleman, Howard Georgi, and Aneesh Manohar.
\newblock {The Evaporation of $Q$ Balls}.
\newblock \emph{Nucl. Phys. B}, 272:\penalty0 301--321, 1986.
\newblock \doi{10.1016/0550-3213(86)90004-0}.

\bibitem[Kusenko and Steinhardt(2001)]{Kusenko:2001vu}
Alexander Kusenko and Paul~J. Steinhardt.
\newblock {Q ball candidates for selfinteracting dark matter}.
\newblock \emph{Phys. Rev. Lett.}, 87:\penalty0 141301, 2001.
\newblock \doi{10.1103/PhysRevLett.87.141301}.

\bibitem[Troitsky(2016)]{Troitsky:2015mda}
Sergey Troitsky.
\newblock {Supermassive dark-matter Q-balls in galactic centers?}
\newblock \emph{JCAP}, 11:\penalty0 027, 2016.
\newblock \doi{10.1088/1475-7516/2016/11/027}.

\bibitem[Enqvist and McDonald(1998)]{Enqvist:1997si}
Kari Enqvist and John McDonald.
\newblock {Q balls and baryogenesis in the MSSM}.
\newblock \emph{Phys. Lett. B}, 425:\penalty0 309--321, 1998.
\newblock \doi{10.1016/S0370-2693(98)00271-8}.

\bibitem[Krylov et~al.(2013)Krylov, Levin, and Rubakov]{Krylov:2013qe}
E.~Krylov, A.~Levin, and V.~Rubakov.
\newblock {Cosmological phase transition, baryon asymmetry and dark matter
  Q-balls}.
\newblock \emph{Phys. Rev. D}, 87\penalty0 (8):\penalty0 083528, 2013.
\newblock \doi{10.1103/PhysRevD.87.083528}.

\bibitem[Kleihaus et~al.(2005)Kleihaus, Kunz, and List]{Kleihaus:2005me}
Burkhard Kleihaus, Jutta Kunz, and Meike List.
\newblock {Rotating boson stars and Q-balls}.
\newblock \emph{Phys. Rev. D}, 72:\penalty0 064002, 2005.
\newblock \doi{10.1103/PhysRevD.72.064002}.

\bibitem[Copeland et~al.(2014)Copeland, Saffin, and Zhou]{Copeland:2014qra}
Edmund~J. Copeland, Paul~M. Saffin, and Shuang-Yong Zhou.
\newblock {Charge-Swapping Q-balls}.
\newblock \emph{Phys. Rev. Lett.}, 113\penalty0 (23):\penalty0 231603, 2014.
\newblock \doi{10.1103/PhysRevLett.113.231603}.

\bibitem[Xie et~al.(2021)Xie, Saffin, and Zhou]{Xie:2021glp}
Qi-Xin Xie, Paul~M. Saffin, and Shuang-Yong Zhou.
\newblock {Charge-Swapping Q-balls and Their Lifetimes}.
\newblock \emph{JHEP}, 07:\penalty0 062, 2021.
\newblock \doi{10.1007/JHEP07(2021)062}.

\bibitem[Hou et~al.(2022)Hou, Saffin, Xie, and Zhou]{Hou:2022jcd}
Si-Yuan Hou, Paul~M. Saffin, Qi-Xin Xie, and Shuang-Yong Zhou.
\newblock {Charge-swapping Q-balls in a logarithmic potential and Affleck-Dine
  condensate fragmentation}.
\newblock \emph{JHEP}, 07:\penalty0 060, 2022.
\newblock \doi{10.1007/JHEP07(2022)060}.

\bibitem[Zhou(2015)]{Zhou:2015yfa}
Shuang-Yong Zhou.
\newblock {Gravitational waves from Affleck-Dine condensate fragmentation}.
\newblock \emph{JCAP}, 06:\penalty0 033, 2015.
\newblock \doi{10.1088/1475-7516/2015/06/033}.

\bibitem[Kasuya et~al.(2022)Kasuya, Kawasaki, and Murai]{Kasuya:2022cko}
Shinta Kasuya, Masahiro Kawasaki, and Kai Murai.
\newblock {Enhancement of second-order gravitational waves at Q-ball decay}.
\newblock 12 2022.

\bibitem[Saffin et~al.(2022)Saffin, Xie, and Zhou]{Saffin:2022tub}
Paul~M. Saffin, Qi-Xin Xie, and Shuang-Yong Zhou.
\newblock {Q-ball Superradiance}.
\newblock 12 2022.

\bibitem[Loiko and Shnir(2022)]{Loiko:2022noq}
Victor Loiko and Yakov Shnir.
\newblock {Q-ball stress stability criterion in U(1) gauged scalar theories}.
\newblock \emph{Phys. Rev. D}, 106\penalty0 (4):\penalty0 045021, 2022.
\newblock \doi{10.1103/PhysRevD.106.045021}.

\bibitem[Loginov(2022)]{Loginov:2022okj}
A.~Yu. Loginov.
\newblock {Scattering of fermions on a one-dimensional Q-ball}.
\newblock \emph{Nucl. Phys. B}, 984:\penalty0 115964, 2022.
\newblock \doi{10.1016/j.nuclphysb.2022.115964}.

\bibitem[Bai et~al.(2022)Bai, Lu, and Orlofsky]{Bai:2021mzu}
Yang Bai, Sida Lu, and Nicholas Orlofsky.
\newblock {Q-monopole-ball: a topological and nontopological soliton}.
\newblock \emph{JHEP}, 01:\penalty0 109, 2022.
\newblock \doi{10.1007/JHEP01(2022)109}.

\bibitem[Rosen(1969)]{Rosen:1969ay}
G.~Rosen.
\newblock {Dilatation covariance and exact solutions in local relativistic
  field theories}.
\newblock \emph{Phys. Rev.}, 183:\penalty0 1186--1188, 1969.
\newblock \doi{10.1103/PhysRev.183.1186}.

\bibitem[Theodorakis(2000)]{Theodorakis:2000bz}
Stavros Theodorakis.
\newblock {Analytic Q ball solutions in a parabolic-type potential}.
\newblock \emph{Phys. Rev. D}, 61:\penalty0 047701, 2000.
\newblock \doi{10.1103/PhysRevD.61.047701}.

\bibitem[Gulamov et~al.(2013)Gulamov, Nugaev, and Smolyakov]{Gulamov:2013ema}
I.~E. Gulamov, E.~Ya. Nugaev, and M.~N. Smolyakov.
\newblock {Analytic $Q$-ball solutions and their stability in a piecewise
  parabolic potential}.
\newblock \emph{Phys. Rev. D}, 87\penalty0 (8):\penalty0 085043, 2013.
\newblock \doi{10.1103/PhysRevD.87.085043}.

\bibitem[Paccetti~Correia and Schmidt(2001)]{PaccettiCorreia:2001wtt}
F.~Paccetti~Correia and M.~G. Schmidt.
\newblock {Q balls: Some analytical results}.
\newblock \emph{Eur. Phys. J. C}, 21:\penalty0 181--191, 2001.
\newblock \doi{10.1007/s100520100710}.

\bibitem[MacKenzie and Paranjape(2001)]{MacKenzie:2001av}
R.~B. MacKenzie and Manu~B. Paranjape.
\newblock {From Q walls to Q balls}.
\newblock \emph{JHEP}, 08:\penalty0 003, 2001.
\newblock \doi{10.1088/1126-6708/2001/08/003}.

\bibitem[Ioannidou and Vlachos(2003)]{Ioannidou:2003ev}
T.~A. Ioannidou and N.~D. Vlachos.
\newblock {On the Q ball profile function}.
\newblock \emph{J. Math. Phys.}, 44:\penalty0 3562--3568, 2003.
\newblock \doi{10.1063/1.1586792}.

\bibitem[Ioannidou et~al.(2005{\natexlab{a}})Ioannidou, Kuiroukidis, and
  Vlachos]{Ioannidou:2005hk}
T.~A. Ioannidou, A.~Kuiroukidis, and N.~D. Vlachos.
\newblock {An analytic approach to Q-balls}.
\newblock \emph{Theor. Math. Phys.}, 144:\penalty0 1171--1175,
  2005{\natexlab{a}}.
\newblock \doi{10.1007/s11232-005-0147-1}.

\bibitem[Ioannidou et~al.(2005{\natexlab{b}})Ioannidou, Kouiroukidis, and
  Vlachos]{Ioannidou:2004vr}
T.~A. Ioannidou, A.~Kouiroukidis, and N.~D. Vlachos.
\newblock {Universality in a class of Q-ball solutions: An Analytic approach}.
\newblock \emph{J. Math. Phys.}, 46:\penalty0 042306, 2005{\natexlab{b}}.
\newblock \doi{10.1063/1.1851972}.

\bibitem[Heeck et~al.(2021{\natexlab{a}})Heeck, Rajaraman, Riley, and
  Verhaaren]{Heeck:2020bau}
Julian Heeck, Arvind Rajaraman, Rebecca Riley, and Christopher~B. Verhaaren.
\newblock {Understanding Q-Balls Beyond the Thin-Wall Limit}.
\newblock \emph{Phys. Rev. D}, 103\penalty0 (4):\penalty0 045008,
  2021{\natexlab{a}}.
\newblock \doi{10.1103/PhysRevD.103.045008}.

\bibitem[Heeck et~al.(2021{\natexlab{b}})Heeck, Rajaraman, Riley, and
  Verhaaren]{Heeck:2021zvk}
Julian Heeck, Arvind Rajaraman, Rebecca Riley, and Christopher~B. Verhaaren.
\newblock {Mapping Gauged Q-Balls}.
\newblock \emph{Phys. Rev. D}, 103\penalty0 (11):\penalty0 116004,
  2021{\natexlab{b}}.
\newblock \doi{10.1103/PhysRevD.103.116004}.

\bibitem[Heeck and Sokhashvili(2022)]{Heeck:2022iky}
Julian Heeck and Mikheil Sokhashvili.
\newblock {Q-balls in polynomial potentials}.
\newblock 10 2022.

\bibitem[Dine and Kusenko(2003)]{Dine:2003ax}
Michael Dine and Alexander Kusenko.
\newblock {The Origin of the matter - antimatter asymmetry}.
\newblock \emph{Rev. Mod. Phys.}, 76:\penalty0 1, 2003.
\newblock \doi{10.1103/RevModPhys.76.1}.

\bibitem[Enqvist and Mazumdar(2003)]{Enqvist:2003gh}
Kari Enqvist and Anupam Mazumdar.
\newblock {Cosmological consequences of MSSM flat directions}.
\newblock \emph{Phys. Rept.}, 380:\penalty0 99--234, 2003.
\newblock \doi{10.1016/S0370-1573(03)00119-4}.

\bibitem[Derrick(1964)]{Derrick:1964ww}
G.~H. Derrick.
\newblock {Comments on nonlinear wave equations as models for elementary
  particles}.
\newblock \emph{J. Math. Phys.}, 5:\penalty0 1252--1254, 1964.
\newblock \doi{10.1063/1.1704233}.

\bibitem[Lee et~al.(1989)Lee, Stein-Schabes, Watkins, and Widrow]{Lee:1988ag}
Ki-Myeong Lee, Jaime~A. Stein-Schabes, Richard Watkins, and Lawrence~M. Widrow.
\newblock {Gauged q Balls}.
\newblock \emph{Phys. Rev. D}, 39:\penalty0 1665, 1989.
\newblock \doi{10.1103/PhysRevD.39.1665}.

\bibitem[{Einstein}(1936)]{1936Sci....84..506E}
Albert {Einstein}.
\newblock {Lens-Like Action of a Star by the Deviation of Light in the
  Gravitational Field}.
\newblock \emph{Science}, 84\penalty0 (2188):\penalty0 506--507, December 1936.
\newblock \doi{10.1126/science.84.2188.506}.

\bibitem[Narayan and Bartelmann(1996)]{Narayan:1996ba}
Ramesh Narayan and Matthias Bartelmann.
\newblock {Lectures on gravitational lensing}.
\newblock In \emph{{13th Jerusalem Winter School in Theoretical Physics:
  Formation of Structure in the Universe}}, 6 1996.

\bibitem[Jeans(1902)]{Jeans:1902fpv}
J.~H. Jeans.
\newblock {The Stability of a Spherical Nebula}.
\newblock \emph{Phil. Trans. A. Math. Phys. Eng. Sci.}, 199\penalty0
  (312-320):\penalty0 1--53, 1902.
\newblock \doi{10.1098/rsta.1902.0012}.

\bibitem[Freivogel et~al.(2020)Freivogel, Gasenzer, Hebecker, and
  Leonhardt]{Freivogel:2019mtr}
Ben Freivogel, Thomas Gasenzer, Arthur Hebecker, and Sascha Leonhardt.
\newblock {A Conjecture on the Minimal Size of Bound States}.
\newblock \emph{SciPost Phys.}, 8\penalty0 (4):\penalty0 058, 2020.
\newblock \doi{10.21468/SciPostPhys.8.4.058}.

\bibitem[Barnard and Child(1959)]{barnard1959higher}
S.~Barnard and J.M. Child.
\newblock \emph{Higher Algebra}.
\newblock Macmillan and Company, Limited, 1959.
\newblock URL \url{https://books.google.co.in/books?id=9Lm5vQEACAAJ}.

\bibitem[Khovanskii et~al.(2014)Khovanskii, Timorin, Kiritchenko, and
  Kadets]{khovanskii2014topological}
A.~Khovanskii, V.~Timorin, V.~Kiritchenko, and L.~Kadets.
\newblock \emph{Topological Galois Theory: Solvability and Unsolvability of
  Equations in Finite Terms}.
\newblock Springer Monographs in Mathematics. Springer Berlin Heidelberg, 2014.
\newblock ISBN 9783642388712.
\newblock URL \url{https://books.google.co.in/books?id=u\_3HBAAAQBAJ}.

\bibitem[{Witt} and {Mao}(1994)]{Witt:1994}
Hans~J. {Witt} and Shude {Mao}.
\newblock {Can Lensed Stars Be Regarded as Pointlike for Microlensing by
  MACHOs?}
\newblock \emph{\apj}, 430:\penalty0 505, August 1994.
\newblock \doi{10.1086/174426}.

\bibitem[Gould and Gaudi(1997)]{Gould_1997}
Andrew Gould and B.~Scott Gaudi.
\newblock Femtolens imaging of a quasar central engine using a dwarf star
  telescope.
\newblock \emph{The Astrophysical Journal}, 486\penalty0 (2):\penalty0 687, sep
  1997.
\newblock \doi{10.1086/304569}.
\newblock URL \url{https://dx.doi.org/10.1086/304569}.

\bibitem[Nakamura(1998)]{Nakamura}
Takahiro~T. Nakamura.
\newblock Gravitational lensing of gravitational waves from inspiraling
  binaries by a point mass lens.
\newblock \emph{Phys. Rev. Lett.}, 80:\penalty0 1138--1141, Feb 1998.
\newblock \doi{10.1103/PhysRevLett.80.1138}.
\newblock URL \url{https://link.aps.org/doi/10.1103/PhysRevLett.80.1138}.

\bibitem[Niikura et~al.(2019)]{Niikura:2017zjd}
Hiroko Niikura et~al.
\newblock {Microlensing constraints on primordial black holes with Subaru/HSC
  Andromeda observations}.
\newblock \emph{Nature Astron.}, 3\penalty0 (6):\penalty0 524--534, 2019.
\newblock \doi{10.1038/s41550-019-0723-1}.

\bibitem[Sugiyama et~al.(2023)Sugiyama, Takada, and Kusenko]{Sugiyama:2021xqg}
Sunao Sugiyama, Masahiro Takada, and Alexander Kusenko.
\newblock {Possible evidence of axion stars in HSC and OGLE microlensing
  events}.
\newblock \emph{Phys. Lett. B}, 840:\penalty0 137891, 2023.
\newblock \doi{10.1016/j.physletb.2023.137891}.

\bibitem[Griest(1991)]{Griest:1990vu}
Kim Griest.
\newblock {Galactic Microlensing as a Method of Detecting Massive Compact Halo
  Objects}.
\newblock \emph{Astrophys. J.}, 366:\penalty0 412--421, 1991.
\newblock \doi{10.1086/169575}.

\bibitem[Cirelli et~al.(2011)Cirelli, Corcella, Hektor, Hutsi, Kadastik, Panci,
  Raidal, Sala, and Strumia]{Cirelli:2010xx}
Marco Cirelli, Gennaro Corcella, Andi Hektor, Gert Hutsi, Mario Kadastik, Paolo
  Panci, Martti Raidal, Filippo Sala, and Alessandro Strumia.
\newblock {PPPC 4 DM ID: A Poor Particle Physicist Cookbook for Dark Matter
  Indirect Detection}.
\newblock \emph{JCAP}, 03:\penalty0 051, 2011.
\newblock \doi{10.1088/1475-7516/2012/10/E01}.
\newblock [Erratum: JCAP 10, E01 (2012)].

\bibitem[Boshkayev et~al.(2022)Boshkayev, Konysbayev, Kurmanov, Luongo,
  Muccino, Quevedo, and Zhumakhanova]{Boshkayev:2022vpn}
Kuantay Boshkayev, Talgar Konysbayev, Yergali Kurmanov, Orlando Luongo, Marco
  Muccino, Hernando Quevedo, and Gulnur Zhumakhanova.
\newblock {Numerical analyses of M31 dark matter profiles}.
\newblock 12 2022.

\bibitem[Smyth et~al.(2020)Smyth, Profumo, English, Jeltema, McKinnon, and
  Guhathakurta]{Smyth:2019whb}
Nolan Smyth, Stefano Profumo, Samuel English, Tesla Jeltema, Kevin McKinnon,
  and Puragra Guhathakurta.
\newblock {Updated Constraints on Asteroid-Mass Primordial Black Holes as Dark
  Matter}.
\newblock \emph{Phys. Rev. D}, 101\penalty0 (6):\penalty0 063005, 2020.
\newblock \doi{10.1103/PhysRevD.101.063005}.

\bibitem[Williams et~al.(2014)Williams, Lang, Dalcanton, Dolphin, Weisz, Bell,
  Bianchi, Byler, Gilbert, Girardi, and et~al.]{Williams_2014}
Benjamin~F. Williams, Dustin Lang, Julianne~J. Dalcanton, Andrew~E. Dolphin,
  Daniel~R. Weisz, Eric~F. Bell, Luciana Bianchi, Nell Byler, Karoline~M.
  Gilbert, Léo Girardi, and et~al.
\newblock The panchromatic hubble andromeda treasury. x. ultraviolet to
  infrared photometry of 117 million equidistant stars.
\newblock \emph{The Astrophysical Journal Supplement Series}, 215\penalty0
  (1):\penalty0 9, Oct 2014.
\newblock ISSN 1538-4365.
\newblock \doi{10.1088/0067-0049/215/1/9}.
\newblock URL \url{http://dx.doi.org/10.1088/0067-0049/215/1/9}.

\bibitem[Dalcanton et~al.(2012)Dalcanton, Williams, Lang, Lauer, Kalirai, Seth,
  Dolphin, Rosenfield, Weisz, Bell, and et~al.]{Dalcanton_2012}
Julianne~J. Dalcanton, Benjamin~F. Williams, Dustin Lang, Tod~R. Lauer,
  Jason~S. Kalirai, Anil~C. Seth, Andrew Dolphin, Philip Rosenfield, Daniel~R.
  Weisz, Eric~F. Bell, and et~al.
\newblock The panchromatic hubble andromeda treasury.
\newblock \emph{The Astrophysical Journal Supplement Series}, 200\penalty0
  (2):\penalty0 18, May 2012.
\newblock ISSN 1538-4365.
\newblock \doi{10.1088/0067-0049/200/2/18}.
\newblock URL \url{http://dx.doi.org/10.1088/0067-0049/200/2/18}.

\bibitem[Choi et~al.(2016)Choi, Dotter, Conroy, Cantiello, Paxton, and
  Johnson]{choi2016mesa}
Jieun Choi, Aaron Dotter, Charlie Conroy, Matteo Cantiello, Bill Paxton, and
  Benjamin~D Johnson.
\newblock Mesa isochrones and stellar tracks (mist). i. solar-scaled models.
\newblock \emph{The Astrophysical Journal}, 823\penalty0 (2):\penalty0 102,
  2016.

\bibitem[Dotter(2016)]{dotter2016mesa}
Aaron Dotter.
\newblock Mesa isochrones and stellar tracks (mist) 0: Methods for the
  construction of stellar isochrones.
\newblock \emph{The Astrophysical Journal Supplement Series}, 222\penalty0
  (1):\penalty0 8, 2016.

\bibitem[Tisserand et~al.(2007)]{Survey_EROS}
P.~Tisserand et~al.
\newblock {Limits on the Macho Content of the Galactic Halo from the EROS-2
  Survey of the Magellanic Clouds}.
\newblock \emph{Astron. Astrophys.}, 469:\penalty0 387--404, 2007.
\newblock \doi{10.1051/0004-6361:20066017}.

\bibitem[Udalski et~al.(2015)Udalski, Szymański, and
  Szymański]{Survey_OGLEIV_1}
A.~Udalski, M.~K. Szymański, and G.~Szymański.
\newblock Ogle-iv: Fourth phase of the optical gravitational lensing
  experiment, 2015.

\bibitem[Mroz et~al.(2017)Mroz, Udalski, Skowron, Poleski, Kozlowski,
  Szymanski, Soszynski, Wyrzykowski, Pietrukowicz, Ulaczyk, and
  et~al.]{Survey_OGLEIV_2}
Przemek Mroz, Andrzej Udalski, Jan Skowron, Radoslaw Poleski, Szymon Kozlowski,
  Michał~K. Szymanski, Igor Soszynski, Lukasz Wyrzykowski, Pawel Pietrukowicz,
  Krzysztof Ulaczyk, and et~al.
\newblock No large population of unbound or wide-orbit jupiter-mass planets.
\newblock \emph{Nature}, 548\penalty0 (7666):\penalty0 183–186, Jul 2017.
\newblock ISSN 1476-4687.
\newblock \doi{10.1038/nature23276}.
\newblock URL \url{http://dx.doi.org/10.1038/nature23276}.

\bibitem[Aihara et~al.(2018)]{Aihara:2017tri}
Hiroaki Aihara et~al.
\newblock {First Data Release of the Hyper Suprime-Cam Subaru Strategic
  Program}.
\newblock \emph{Publ. Astron. Soc. Jap.}, 70:\penalty0 S8, 2018.
\newblock \doi{10.1093/pasj/psx081}.

\bibitem[Green et~al.(2012)]{Green:2012mj}
J.~Green et~al.
\newblock {Wide-Field InfraRed Survey Telescope (WFIRST) Final Report}.
\newblock 8 2012.

\bibitem[Penny et~al.(2019)Penny, Gaudi, Kerins, Rattenbury, Mao, Robin, and
  Novati]{Survey_WFIRST}
Matthew~T. Penny, B.~Scott Gaudi, Eamonn Kerins, Nicholas~J. Rattenbury, Shude
  Mao, Annie~C. Robin, and Sebastiano~Calchi Novati.
\newblock Predictions of the {WFIRST} microlensing survey. i. bound planet
  detection rates.
\newblock \emph{The Astrophysical Journal Supplement Series}, 241\penalty0
  (1):\penalty0 3, feb 2019.
\newblock \doi{10.3847/1538-4365/aafb69}.
\newblock URL \url{https://doi.org/10.3847%2F1538-4365%2Faafb69}.

\bibitem[Wenger et~al.(2000)]{Wenger:2000sw}
Marc Wenger et~al.
\newblock {The simbad astronomical database}.
\newblock \emph{Astron. Astrophys. Suppl. Ser.}, 143:\penalty0 9, 2000.
\newblock \doi{10.1051/aas:2000332}.

\end{thebibliography}

\end{document}